\documentclass{aastex701}
\usepackage{enumitem}
\shorttitle{Pluto's Atmosphere 2017-2023}
\shortauthors{Sickafoose et al.}
\graphicspath{{./}{}}
\begin{document}

\title{Changes in Pluto's Atmosphere Based on Stellar Occultation Data from 2017 to 2023}

\correspondingauthor{Amanda A. Sickafoose}
\email{asickafoose@psi.edu}

\author[0000-0002-9468-7477]{Amanda A. Sickafoose}
\affiliation{Planetary Science Institute \\
1700 East Fort Lowell, Suite 106 \\
Tucson, AZ 85719, USA}
\email{asickafoose@psi.edu}

\author[0000-0003-0000-0572]{Michael J. Person}
\affiliation{Department of Earth, Atmospheric, and Planetary Sciences \\
Massachusetts Institute of Technology \\
77 Massachusetts Ave. \\
Cambridge, MA 02139, USA}
\email{mjperson@mit.edu}

\author{Carlos A. Zuluaga}
\affiliation{Department of Earth, Atmospheric, and Planetary Sciences \\
Massachusetts Institute of Technology \\
77 Massachusetts Ave. \\
Cambridge, MA 02139, USA}
\email{czuluaga@mit.edu}

\author[0000-0003-4772-528X]{Amanda S. Bosh}
\affiliation{Lowell Observatory \\
1400 West Mars Hill Road \\
Flagstaff, AZ 86001, USA}
\email{amanda@lowell.edu}

\author[0000-0002-1050-3539]{Stephen E. Levine}
\affiliation{Lowell Observatory \\
1400 West Mars Hill Road \\
Flagstaff, AZ 86001, USA}
\email{sel@lowell.edu}

\author{Tim Brothers}
\affiliation{Department of Earth, Atmospheric, and Planetary Sciences \\
Massachusetts Institute of Technology \\
77 Massachusetts Ave. \\
Cambridge, MA 02139, USA}
\email{bro@mit.edu}

\author[0009-0000-7038-0406]{Bastian Knieling}
\affiliation{Institute of Space Systems\\
Universität Stuttgart\\
Pfaffenwaldring 29, 
70569 Stuttgart, Germany}
\affiliation{SOFIA Science Center\\
NASA Ames Research Center\\
Moffett Field, CA 94035, USA}
\email{bastian.knieling@irs.uni-stuttgart.de}

\author[0000-0002-3818-7769]{Timothy A. Lister}
\affiliation{Las Cumbres Observatory \\
6740 Cortona Drive, Suite 102 \\
Goleta, CA 93117, USA}
\email{tlister@lco.global}

\author[0000-0003-0412-9664]{David J. Osip}
\affiliation{Carnegie Observatories \\
813 Santa Barbara Street \\
Pasadena, CA 91101 USA}
\email{dosip@carnegiescience.edu}

\author[0000-0001-7337-2452]{Karsten Schindler}
\affiliation{Institute of Space Systems\\
Universität Stuttgart\\
Pfaffenwaldring 29, 
70569 Stuttgart, Germany}
\affiliation{SOFIA Science Center\\
NASA Ames Research Center\\
Moffett Field, CA 94035, USA}
\email{karsten.schindler@irs.uni-stuttgart.de}

\author{Joe Brimacombe}
\affiliation{Savannah Skies Observatory\\
Chillagoe, QLD 4871, Australia \\}
\email{jbrimaco@bigpond.net.au}

\author[]{Tim Carruthers}
\affiliation{Savannah Skies Observatory\\
Chillagoe, QLD 4871, Australia \\}
\email{tim@astrophoto.com.au}

\author[0009-0001-3422-6900]{Abigail Colclasure}
\affiliation{Department of Earth, Atmospheric, and Planetary Sciences \\
Massachusetts Institute of Technology \\
77 Massachusetts Ave. \\
Cambridge, MA 02139, USA}
\email{acolclas@mit.edu}

\author[0000-0001-6841-8436]{Anja Genade}
\affiliation{Department of Astronomy \\
University of Cape Town\\
Private Bag X3 \\
Rondebosch, 7701, South Africa}
\affiliation{South African Astronomical Observatory \\
1 Observatory Rd. \\
Observatory, 7925 South Africa}
\email{ag@saao.ac.za}

\author[0000-0001-8764-7365]{Petro Janse van Rensburg}
\affiliation{Department of Astronomy \\
University of Cape Town\\
Private Bag X3 \\
Rondebosch, 7701, South Africa}
\affiliation{South African Astronomical Observatory \\
1 Observatory Rd. \\
Observatory, 7925 South Africa}
\email{petro@saao.ac.za}

\author{Stephen B. Potter}
\affiliation{South African Astronomical Observatory \\
1 Observatory Rd. \\
Observatory, 7925 South Africa}
\affiliation{Department of Physics \\
University of Johannesburg \\ 
PO Box 524 \\
Auckland Park  \\
2006 South Africa}
\email{sbp@saao.ac.za}

\author[0000-0002-1607-6443]{Patricio Rojo}
\affiliation{Departmento de Astronomia \\
Universidad de Chile \\
Casilla 36-D\\
Santiago, Chile}
\email{pato@das.uchile.cl}

\received{2026 January 24}
\accepted{2026 May 13}
\submitjournal{PSJ}

\begin{abstract}

 Pluto's tenuous atmosphere has $\mu$bar-level pressure and is composed primarily of ${\text{N}}_2$, with a variable haze. Its eccentric orbit combined with high obliquity leads to significant changes in solar insolation throughout the Plutonian year. The atmosphere is supported by vapor-pressure equilibrium with the surface ices, thus surface changes are coupled with the atmospheric properties. Volatile-transport models have anticipated Pluto’s atmospheric evolution: predictions range from collapse over the coming decades to an atmosphere that remains. Previous work claims that Pluto's atmospheric pressure monotonically increased from 1988 through 2016, that the atmosphere began freezing out in 2018-2019, and that there was a plateau as of 2020. Here, we report results from ten stellar occultations by Pluto between 2017 August and 2023 July. Four events were multi-chord, while six were from single sites. Our results indicate a pressure plateau between the \textit{New Horizons} flyby in 2015 through roughly 2021 and suggest that the atmospheric pressure has started to drop. Between 2015-2021 and 2022, the clear-atmosphere pressure at 1275 km decreased $7\pm6\%$, and it dropped $16\pm2\%$ for pressure at 1215 km when including haze. From 2017-2023, the upper atmospheric structure is consistent, while there is a change in light-curve slope in the lower atmosphere. This change-of-slope is consistent with haze particles settling over yearly or shorter timecales. Spikes in one light curve are indicative of intermittent buoyancy waves. More data are needed to confirm a recent pressure change.

\end{abstract}


%% to the KEY in the \bibitem in the reference list below. Figure to include:\\

\section{Introduction} \label{sec:intro}

Pluto's atmosphere was first detected via stellar occultations in the late 1980s \citep{RN80,RN77,RN1580}. With a ${\rm \mu bar}$-level surface pressure, the atmosphere is composed primarily of ${\text{N}}_2$, with some ${\text{CH}}_4$, CO, and other trace hydrocarbons \citep[e.g.][]{RN3737}. Notably, Pluto's atmosphere is intimately connected to its surface ices through vapor-pressure equilibrium. Pluto has high obliquity, $\sim122^\circ$, and an eccentric orbit ($e\sim0.25$). Over time, this combination leads to significant changes in solar insolation as a function of subsolar latitude on Pluto and causes complex changes in the distribution of surface volatiles as well as atmospheric pressure \citep[e.g.][]{RN3594, RN3738}. Models of volatile transport on Pluto consider ice inventories and properties such as albedo, emissivity, and thermal inertia to predict the future of the atmosphere. Some models have predicted that the atmosphere will collapse as Pluto's seasons progress and it moves away from perihelion, which occurred in 1989 \citep{RN1997,RN3594}. Recent models consistently predict a significant pressure decline within the next decades \citep{RN3485, RN3978,RN3906}, although it appears likely that the atmosphere will be sustained throughout Pluto's orbit due to the ${\text{N}}_2$ reservoir in Sputnik Planitia and the high thermal inertia of the subsurface \citep{RN4100}. 

Studies of Pluto's atmosphere over time lead to a better understanding of the physical characteristics and distribution of surface ices as well as the intertwined surface-atmosphere connections. Pluto's atmosphere also provides insight into other bodies with tenuous atmospheres, such as Triton, Mercury, Io, and Callisto. The surfaces of some trans-Neptunian objects (TNOs) other than Pluto are rich in volatile ices, which could sublimate \citep[e.g][]{RN3194}; therefore, stellar occultations by all large TNOs are carefully analyzed to detect or place upper limits on any global or local atmosphere \citep[e.g.][]{RN3420,RN3492,RN3677,RN3689}. Nonetheless, Pluto is currently the only TNO known to have a global, albeit tenuous, atmosphere. 

NASA's \textit{New Horizons} spacecraft returned detailed information about the Pluto system during the 2015 flyby \citep[e.g][]{RN3645}, with a review of the atmospheric results in \citet{RN3977}. Before and after this snapshot, stellar occultation observations were undertaken to study and monitor Pluto's atmosphere from Earth. Between 2002 and 2016, more than a dozen successful Pluto occultations were reported. The atmospheric pressure doubled between 1988 and 2002 \citep{RN2820,RN2835}, then the pressure held steady through 2006-2007 \citep{RN3171,RN3366,RN3483}. Atmospheric waves were detected in 2007 \citep{RN3286,RN3360}. The presence of haze in Pluto's lower atmosphere was suggested from stellar occultation data \citep[e.g.][]{RN77,RN2820, RN3614, RN3641}: extensive, layered haze up to a few hundred kilometers in altitude was detected in 2015 by \textit{New Horizons} \citep[e.g.][]{RN3742,RN4099}, %>200 from LORRI, 350 km from ALICE occ.
likely composed of hydrocarbon spheres tens of nm in size plus $\mu$m-sized fractal aggregates \citep{RN4098}. From 2013-2015, \citet{RN3641} and \citet{RN3616} reported no significant changes in atmospheric pressure and \citet{RN4101} reported an increase of $5\pm2\%$, with the 2015 occultation occurring just two weeks prior to the arrival of \textit{New Horizons}. By analyzing occultation datasets spanning 2002-2016, \citet{RN3741} found a monotonic increase of pressure from 1988-2016. 

Based on observations in 2018 and 2019, an atmospheric pressure drop was proposed. \citet{RN3823} found a $\sim21\%$ pressure decrease between 2016 and 2019, and a ``freezing out'' of Pluto's atmosphere was reported by \citet{RN3918}. The most recently-published occultation observations are from mid-2020, with the interpretation that the atmosphere was in a "plateau phase" since mid-2015 \citep{RN3947}. 

Here, we use a set of occultation data from 1988 to 2023 that is consistently analyzed to provide the most recent measurements of Pluto's atmospheric pressure and look for trends in Pluto's atmospheric evolution. New occultation datasets from ten different epochs between 2017 and 2023 are presented. Descriptions of the observations are provided in \S~\ref{sec:obs} and Appendix \ref{subsec:appendix}. Light-curve extractions and model fits are described in \S~\ref{sec:analyses}. \S~\ref{sec:results} contains results. A discussion is provided in \S~\ref{sec:discussion}.

\section{Observations} \label{sec:obs}
Star characteristics for each of the stellar occultations presented here are listed in Table~\ref{tab:star}, including predicted geocentric midtimes. The occultations were predicted using Gaia DR2 or DR3 star positions, proper motions, and parallaxes (whichever was the most current version available at the time of the prediction) and our Ephemeris Correction Model (ECM) for Pluto's position. The ECM was developed from many years of measured offsets between Pluto and the JPL Horizons \citep{Giorgini2015} ephemeris, and it has been previously employed for successful occultation observations \citep[e.g.][]{RN3614,RN3469, RN3616,RN3641}.

\begin{figure}[!ht]
\centering
\includegraphics[width=1\textwidth,clip,trim=0mm 0mm 0mm 0mm]{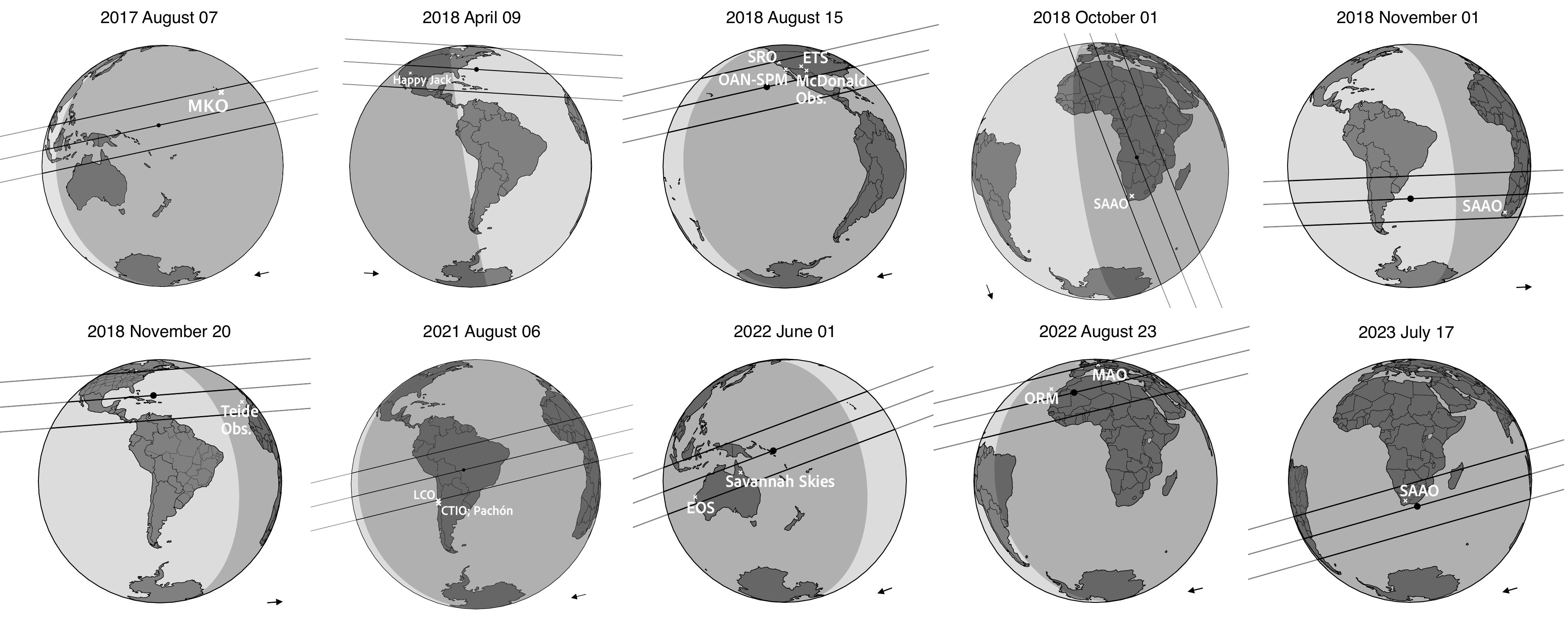}
\caption{Globes with reconstructed shadow paths on Earth and the locations of the observing stations in Table~\ref{tab:sites}. Darker gray portions of the Earth are in darkness at the times of the occultations. The solid lines indicate the top, middle, and bottom of the solid-body shadows, and the arrows show the directions of motion of the shadows. Pluto's diameter is assumed to be 2377~km. The black dots indicate the geocentric closest approaches in the centers of the shadow paths.
% All postdictions from Bastian. SEL notes that the arrows are easy to miss but not sure how to fix.
% Note from Bastian: The RUWE parameter in the Gaia DR3 catalog of the star from the 20180815 event is 1.49, which is greater than 1.4 and thus could be an indicator of a problematic astrometric solution. I don't know if this affects your results, but I just wanted to let you know.
\label{fig:shadowpaths}}
\end{figure}

\begin{figure}[ht!]
\centering
\includegraphics[width=1\textwidth,clip,trim=0mm 0mm 0mm 0mm]{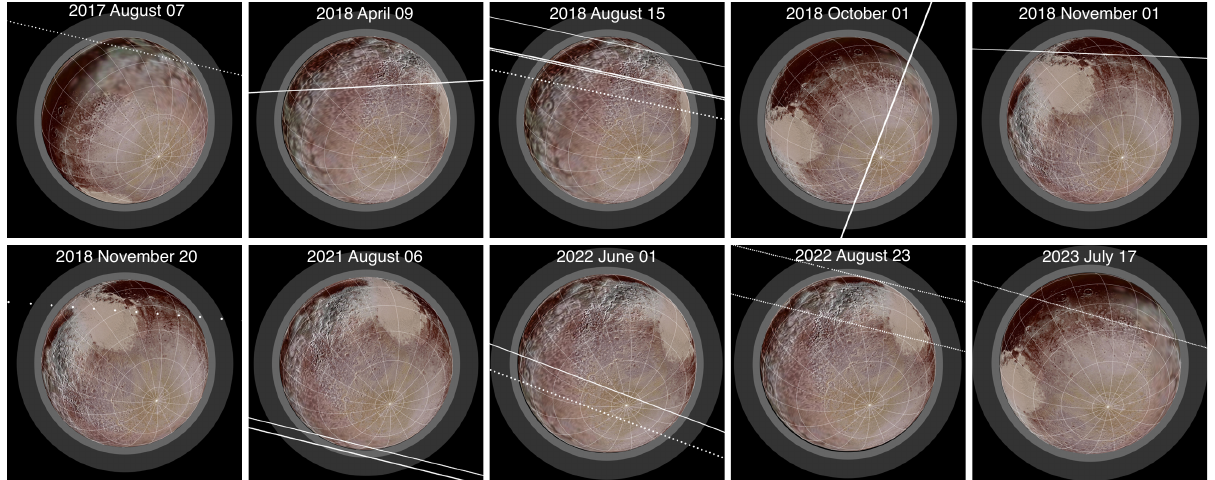}
\caption{Pluto's orientation relative to the Earth at the midtime of each occultation along with sky-plane chords for each successful observation. Sky-plane North is up and East to the left. The white dots indicate the chord locations, with each point representing one image at the fastest cadence from a given site. These chords show the star positions extrapolated behind Pluto: residual starlight observed during the occultations is refracted around Pluto's limb.  Pluto's atmosphere as shown here is represented at a fixed size, with gray shading out to 450 km above the surface and within 100 km shaded more opaquely.
\label{fig:globes}}
\end{figure}

\begin{figure}[ht!]
\includegraphics[width=0.35\textwidth,clip,trim=0mm 0mm 0mm 0mm]{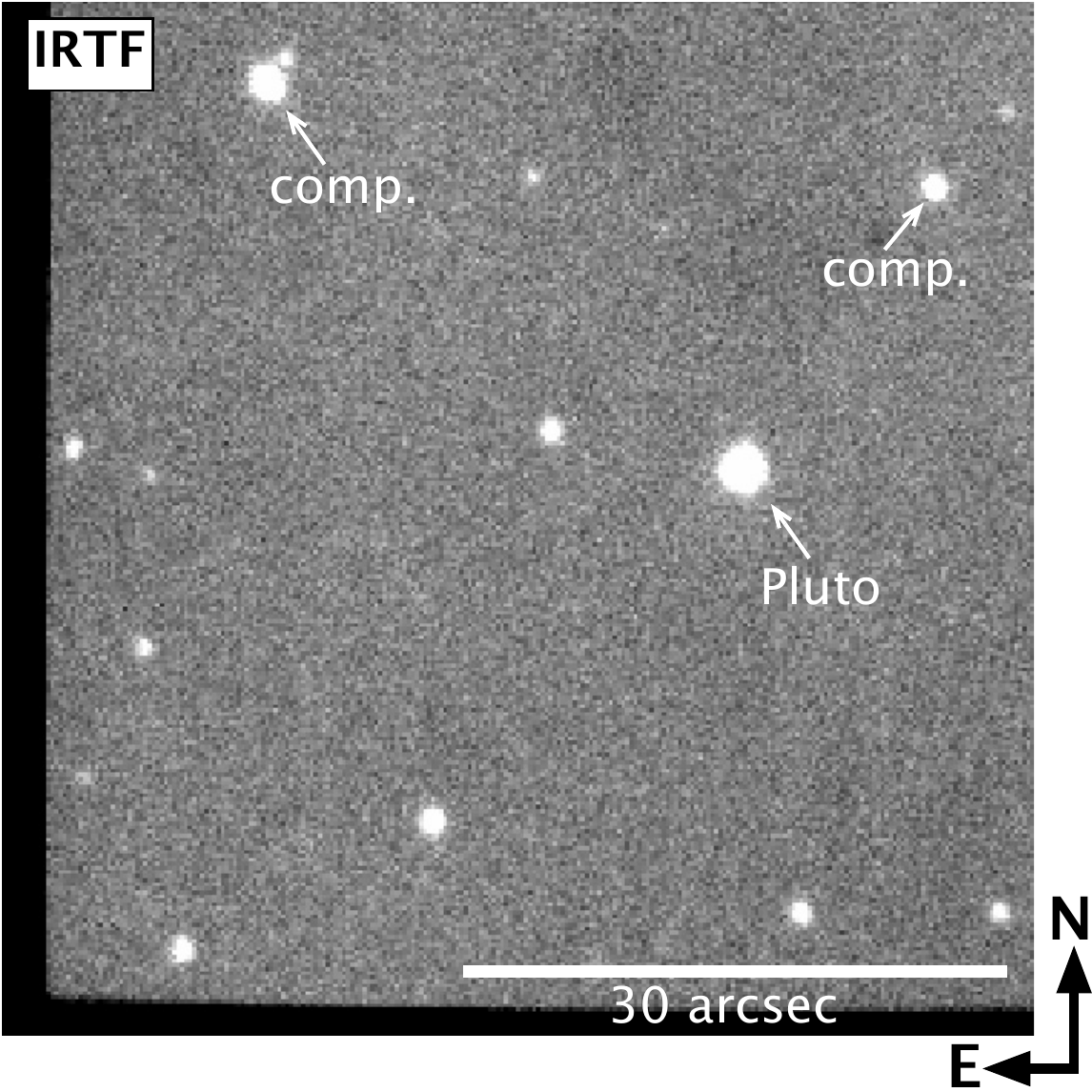}
\centering
%\plottwo can be used for side by side
\caption{Example image from the single successful telescope for the 2017 August 07 event (details are provided in Table \ref{tab:scopes}). The Pluto system and the occultation star (merged) are labeled, as well as the comparison stars used to derive the light curve.  \label{fig:20170807images}}
\end{figure}

\begin{figure}[ht!]
\includegraphics[width=0.35\textwidth,clip,trim=0mm 0mm 0mm 0mm]{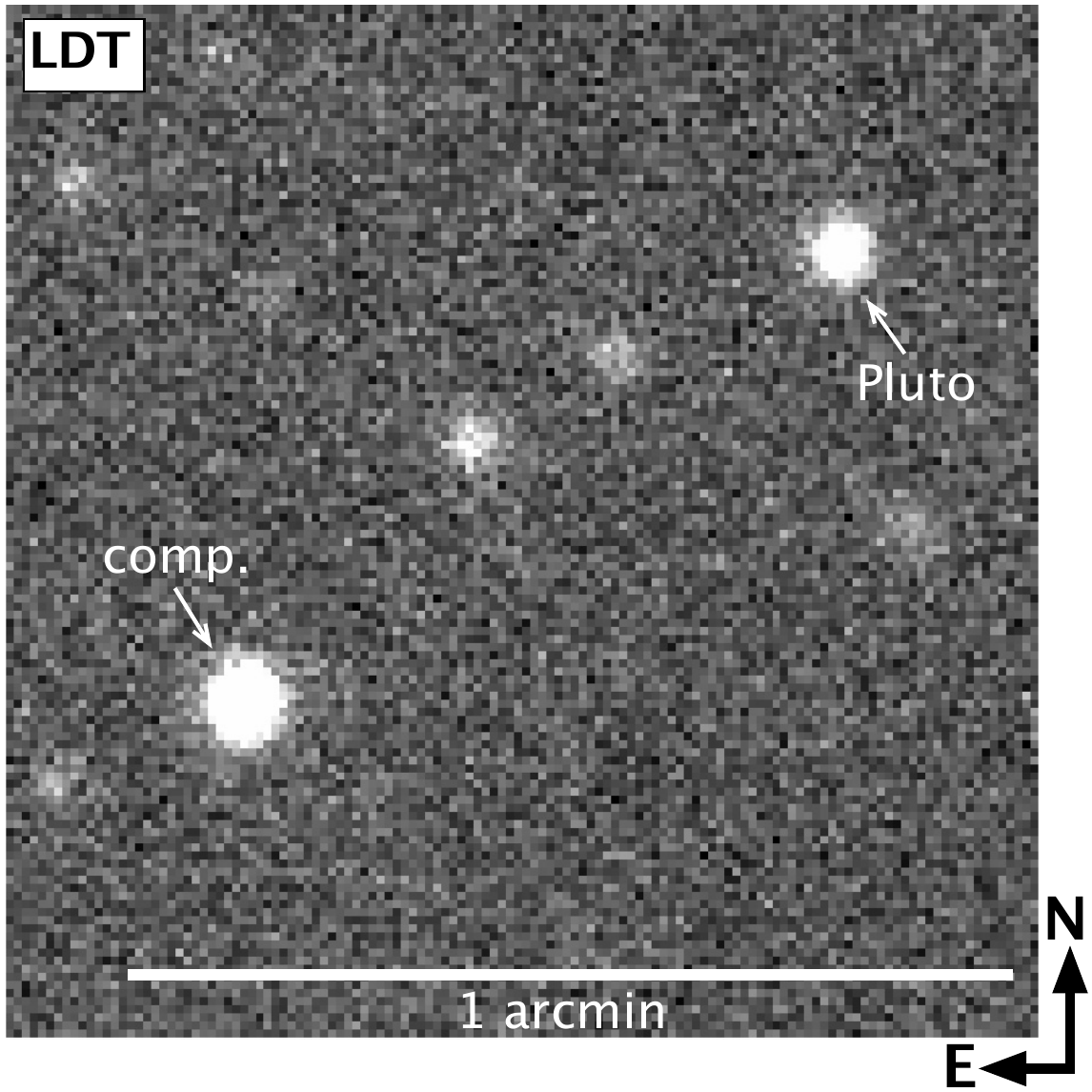}
\centering
%\plottwo can be used for side by side
\caption{Example image from the single successful telescope for the 2018 April 09 event (details are provided in Table \ref{tab:scopes}). The Pluto system and the occultation star (merged) are labeled, as well as the comparison star used to derive the light curve.  \label{fig:20180409images}}
\end{figure}

\begin{figure}[ht!]
\includegraphics[width=0.9\textwidth,clip,trim=0mm 0mm 0mm 0mm]{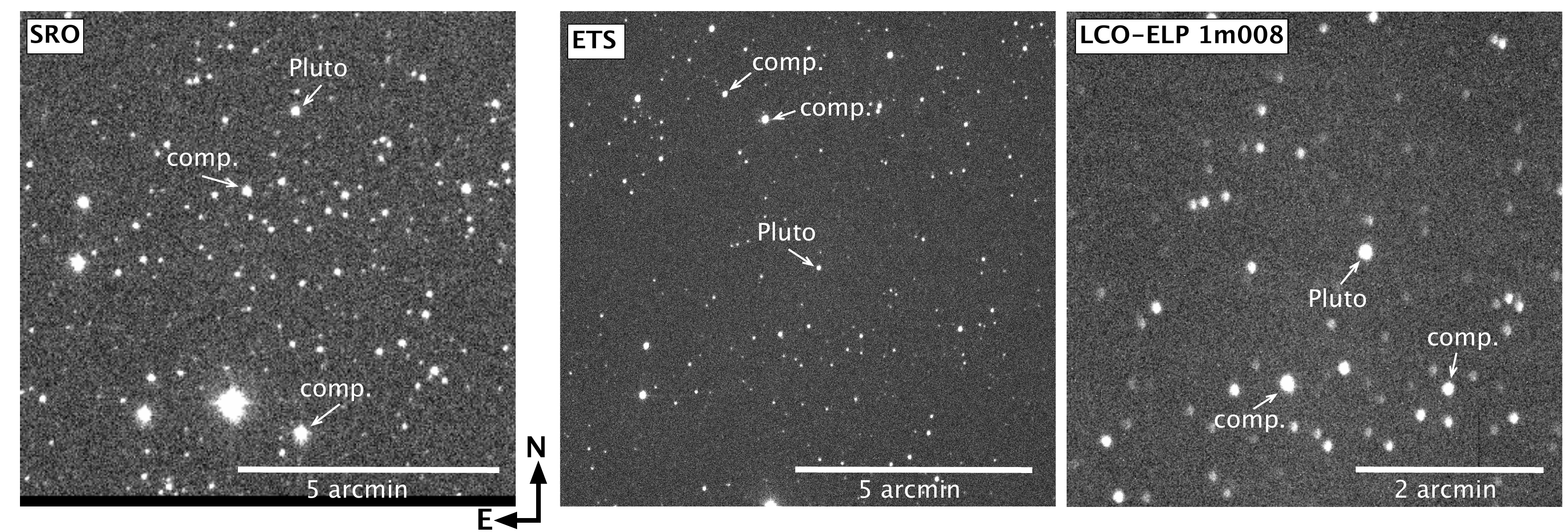}
\centering
%\plottwo can be used for side by side
\caption{Example images from each of the successful sites for the 2018 August 15 event (details are provided in Table~\ref{tab:scopes}). The Pluto system and the occultation star (merged) are labeled, as well as the comparison stars used to derive the light curves.  \label{fig:20180815images}}
\end{figure}

\begin{figure}[ht!]
\includegraphics[width=1\textwidth,clip,trim=0mm 0mm 0mm 0mm]{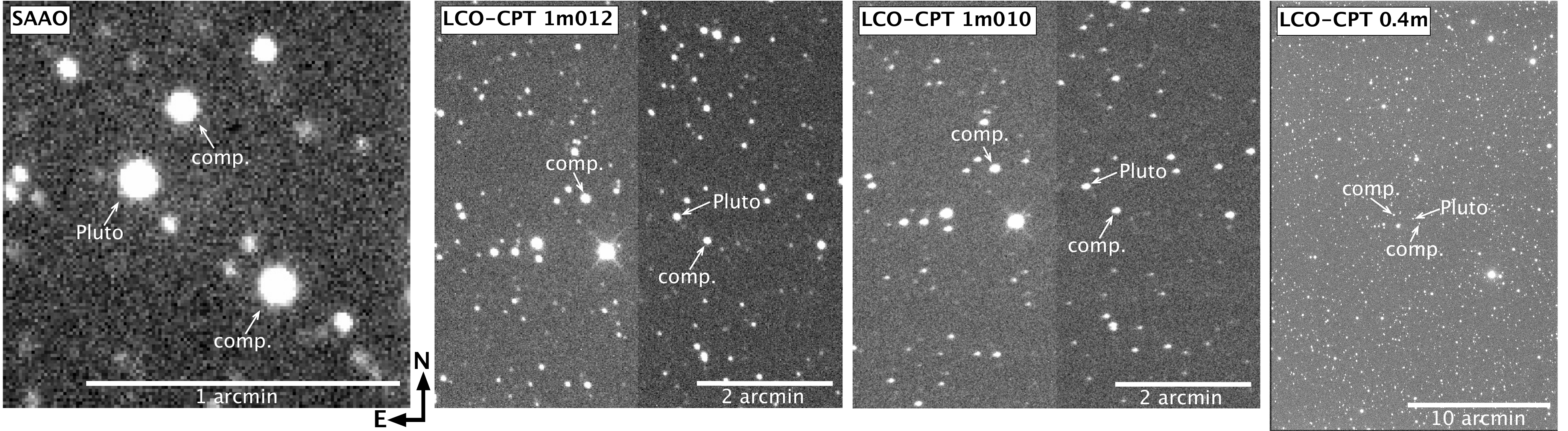}
\centering
\caption{Example images from the successful telescopes for the 2018 October 01 event (details are provided in Table~\ref{tab:scopes}). Reduced images from the two LCO-CPT 1-m telescopes show a low-level difference (a few tens of ADUs background) between the two simultaneous readout channels, which is taken into account during the background subtraction process when generating light curves. The Pluto system and the occultation star (merged) are labeled as well as the comparison stars used to derive the light curves.  \label{fig:20181001image}}
\end{figure}

\begin{figure}[ht!]
\includegraphics[width=0.3\textwidth,clip,trim=0mm 0mm 0mm 0mm]{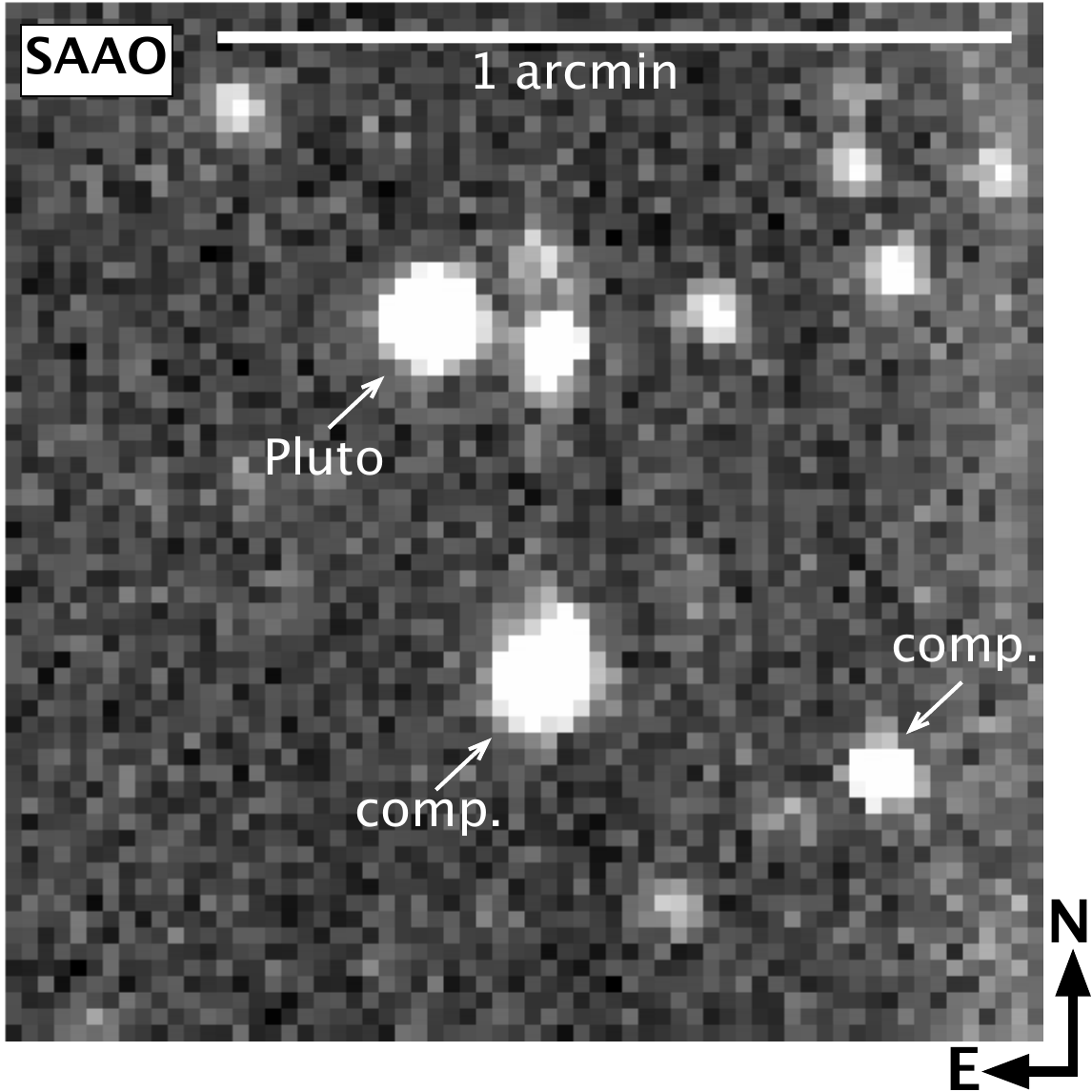}
\centering
\caption{Example image from the single successful telescope for the 2018 November 01 event (details are provided in Table~\ref{tab:scopes}). The Pluto system and the occultation star (merged) are labeled as well as the comparison stars used to derive the light curve.  \label{fig:20181101image}}
\end{figure}

\begin{figure}[ht!]
\includegraphics[width=0.3\textwidth,clip,trim=0mm 0mm 0mm 0mm]{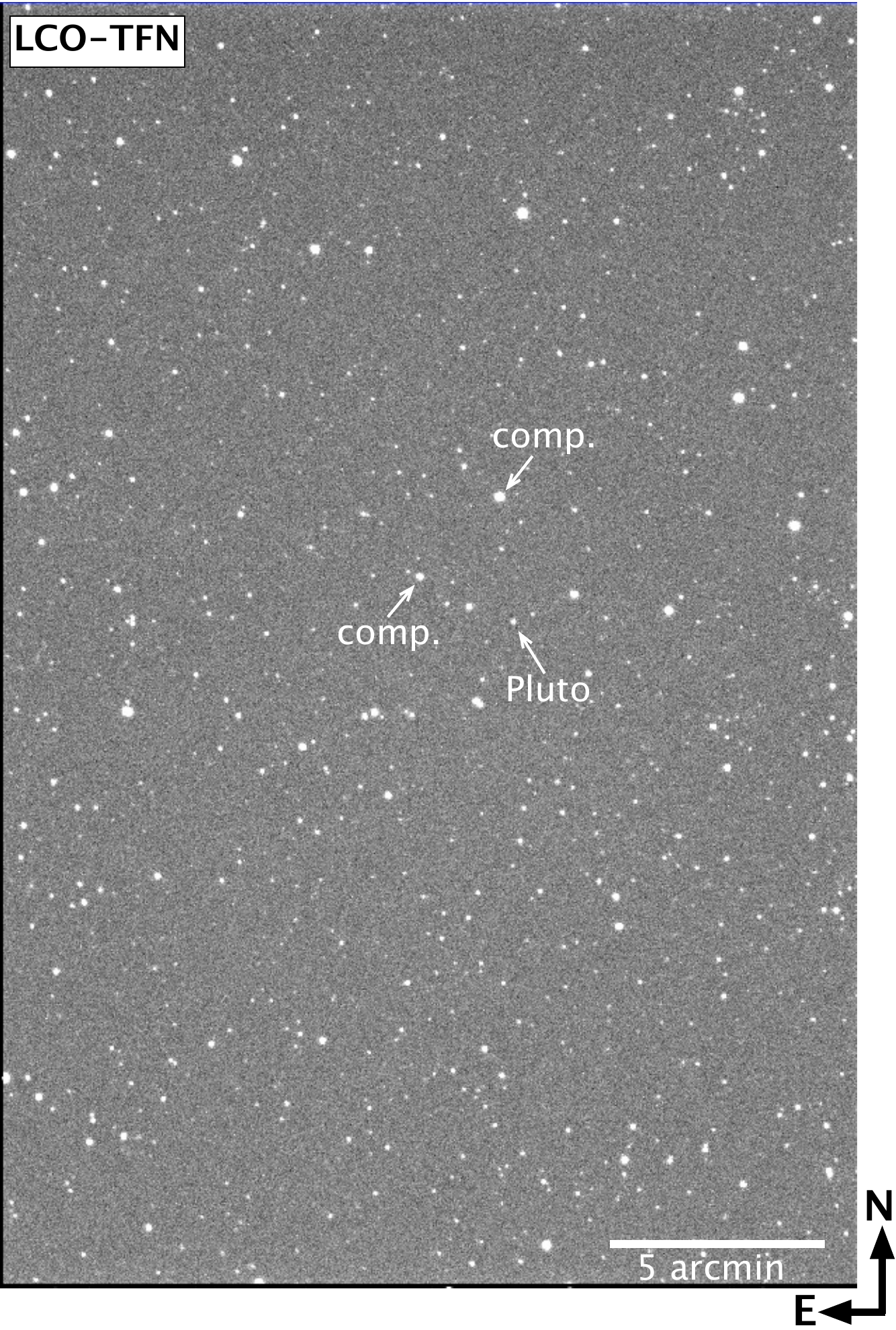}
\centering
\caption{Example image from the single successful telescope for the 2018 November 20 event (details are provided in Table~\ref{tab:scopes}). The Pluto system and the occultation star (merged) are labeled, as well as the comparison stars used to derive the light curve.  \label{fig:20181120image}}
\end{figure}

\begin{figure}[ht!]
\includegraphics[width=0.8\textwidth,clip,trim=0mm 0mm 0mm 0mm]{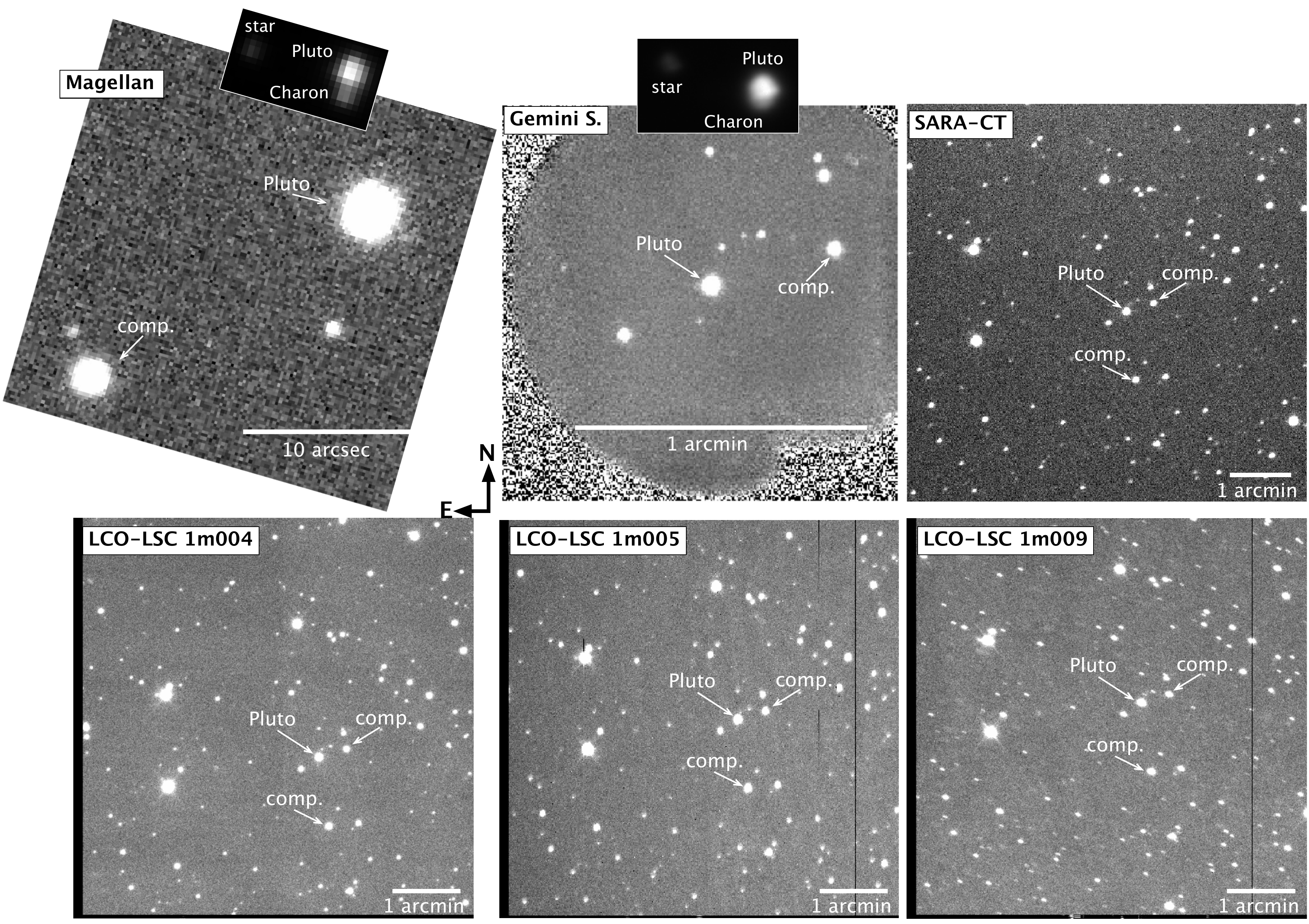}
\centering
\caption{Example images from the successful telescopes for the 2021 August 06 event (details are provided in Table~\ref{tab:scopes}). Reduced images from the three LCO-LSC 1-m telescopes show faint differences between the two simultaneous readout channels, less than those in Fig. \ref{fig:20181001image}. The Pluto system and the occultation star (merged) are labeled, as well as the comparison stars used to derive the light curves. The inset images for Magellan and Gemini S. are well-separated data taken roughly an hour after the occultation, shown at a different display scale, in which the occultation star, Pluto, and Charon, are apparent.  \label{fig:2021images}}
\end{figure}

\begin{figure}[ht!]
\includegraphics[width=1\textwidth,clip,trim=0mm 0mm 0mm 0mm]{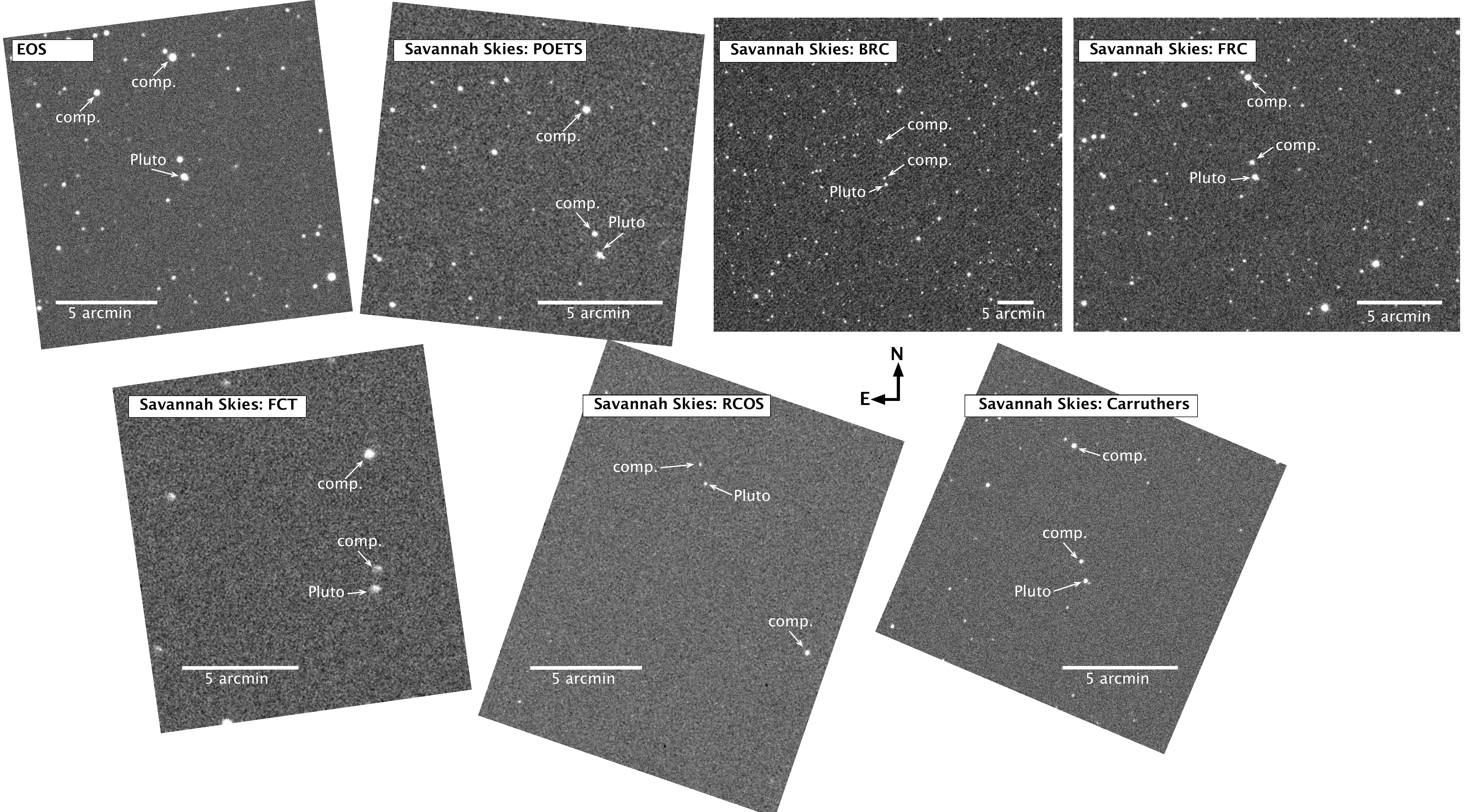}
\centering
\caption{Example images from each of the successful telescopes for the 2022 June 01 event (details are provided in Table~\ref{tab:scopes}). The Pluto system and the occultation star (merged) are labeled, as well as the comparison stars used to derive the light curves.  \label{fig:2022images}}
\end{figure}

\begin{figure}[ht!]
\includegraphics[width=0.6\textwidth,clip,trim=0mm 0mm 0mm 0mm]{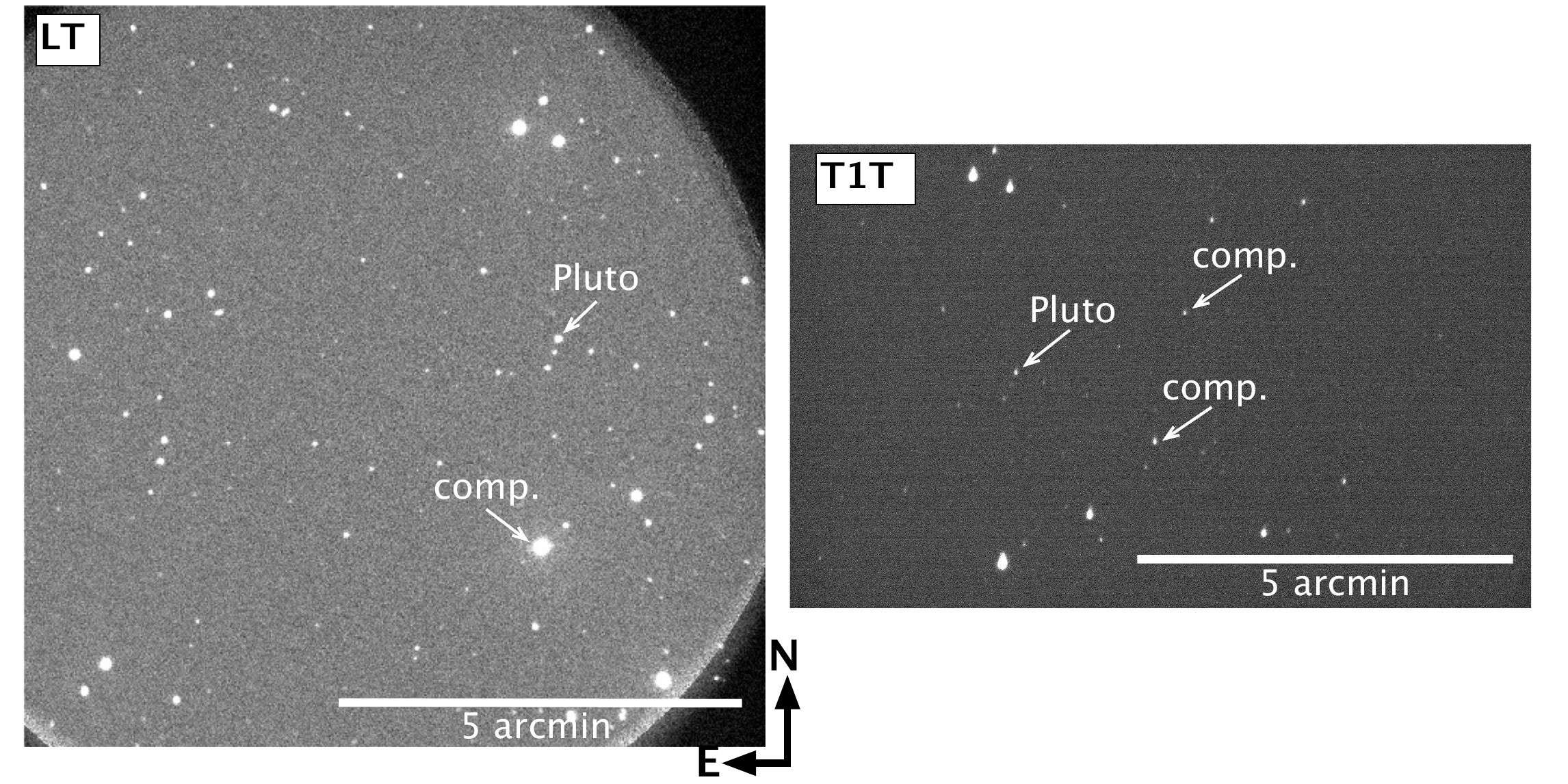} 
\centering
\caption{Example images from each of the successful sites for the 2022 August 23 event (details are provided in Table~\ref{tab:scopes}). The Pluto system and the occultation star (merged) are labeled as well as the comparison stars used to derive the light curves.  
\label{fig:20220823images}}
\end{figure}

\begin{figure}[ht!]
\includegraphics[width=0.3\textwidth,clip,trim=0mm 0mm 0mm 0mm]{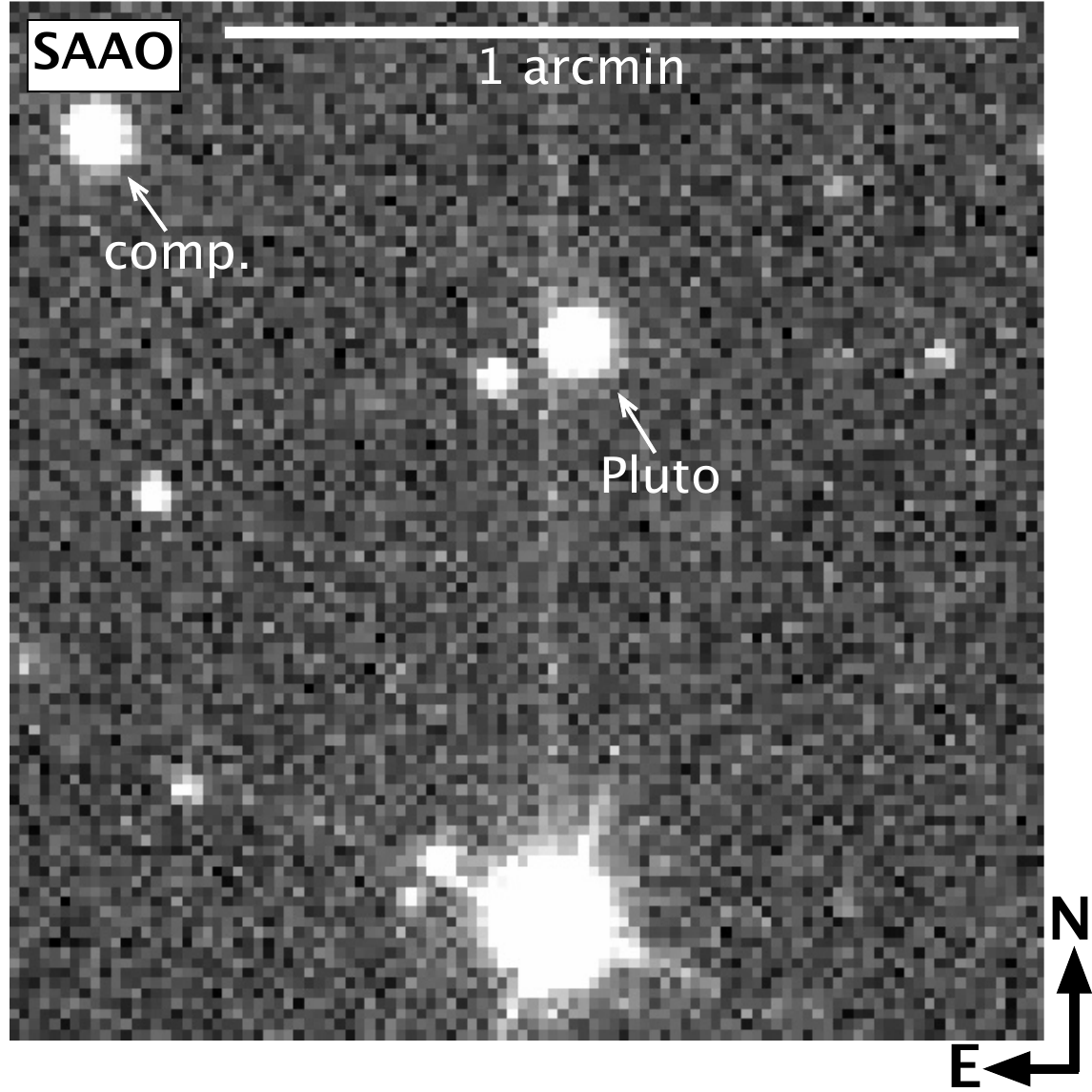}
\centering
\caption{Example image from the single successful site for the 2023 July 17 event (details are provided in Table~\ref{tab:scopes}). The Pluto system and the occultation star (merged) are labeled as well as the comparison star used to derive the light curve. The bright star to the south is saturated. 
\label{fig:20230717images}}
\end{figure}

Information about the observing sites is provided in Table~\ref{tab:sites}, and characteristics of the telescopes and instruments are listed in Table~\ref{tab:scopes}. Details of the observations are provided in Appendix \ref{subsec:appendix} for each occultation by date and site. Reconstructed globes showing the shadow paths and site positions are provided in Fig.~\ref{fig:shadowpaths}. For these postdictions, the star positions are from Gaia DR3 and the reference ephemerides for Pluto are from Numerical Integration of the Motion of an Asteroid (NIMA) v9\footnote[1]{https://lesia.obspm.fr/lucky-star/obj.php?p=818}. The NIMA code has been demonstrated to predict Pluto's ephemeris with accuracy at the milliarcsecond level \citep{RN4091}, so combining these ephemerides with Gaia should return the most accurate event geometry. Figure~\ref{fig:globes} shows reconstructions of Pluto's orientation in the sky-plane view, along with the locations at Pluto of the successful chords for each event. Figures \ref{fig:20170807images}-\ref{fig:20230717images} are example images from each telescope.
\begin{deluxetable*}{llcccccc}
%\tablenum{1}
\tablecaption{Occultation star information.\label{tab:star}}
\tablewidth{0pt}
\tablehead{
\colhead{Midtime$^a$}& \colhead{Designation} & \colhead{$\alpha$} & \colhead{$\delta$} & \colhead{G$^b$} & \colhead{BP$^b$} & \colhead{RP$^b$} & \colhead{Velocity$^c$} \\
\colhead{(UT)}& \colhead{(GDR3$^b$)} & \colhead{(ICRS; hh:mm:ss.ss)}& \colhead{(ICRS; $\arcdeg$:$\arcmin$:$\arcsec$)} & \colhead{(mag)} & \colhead{(mag)} & \colhead{(mag)} &\colhead{(km/sec)}
}
\startdata
%2010 July 04 01:59:38& 4095644224662776960 & 18:15:42.11& –18:16:41.2 & 14.7 & 16.3 & 13.5& 23.6\\% Not in Gaia on Aladin Lite map, 15.3 UCAC2 from prediction. TESS catalog has it as a giant, R=18.114 R_sun, dist 2971.1900±518.0400 pc
2017 August 07 10:37:52& 4081283090964124288 & 19:14:15.39& –21:37:03.2 & 15.11 & 15.63 & 14.41& 21.1\\
2018 April 09 11:21:29& 6772904048429765632 & 19:30:34.81& –21:29:30.5 & 17.76 & 18.36 & 17.10& 6.4\\
2018 August 15 05:33:39& 6772629170525258240 &19:22:10.46& –21:58:49.0 & 12.98 & 13.31 &12.48 &19.3\\
2018 October 01 18:05:06& 6772646861485572224 & 19:20:06.79& –22:07:27.9 & 17.42 & 17.86 & 16.79& 1.7\\
2018 November 01 19:51:37& 6772602266848686976 &19:21:13.06& –22:08:17.1 & 14.98 & 15.67 & 14.18&16.0\\
2018 November 20 19:51:31&  6772612437332651136 &19:22:50.16& –22:06:54.7& 14.43 & 15.06 & 13.67&24.5\\
2021 August 06 02:12:13 & 6864654853491878528 & 19:48:11.58  & -22:44:14.2 & 17.76 & 18.15 & 17.23&22.7\\
2022 June 01 16:16:36 & 6852184815383389824 & 20:02:18.45  & -22:33:02.3 & 12.97 & 13.62 & 12.19&14.9\\
2022 August 23 21:20:46 &6863752296183940608 & 19:54:52.56 & -23:01:53.1  & 16.61 & 17.03 & 16.02 & 19.5\\
2023 July 17 22:52:50 &6851930652101295616 & 20:06:29.65 & -22:59:55.83  & 18.40 & 18.78 & 17.86 & 24.3
\enddata
\tablecomments{$^a$Geocentric, predicted.$^b$Gaia Data Release 3 \citep{RN4103,RN3687}. $^c$Relative to Pluto, geocentric. }
\end{deluxetable*}
\begin{deluxetable*}{llccc}
%\tablenum{2}
\tablecaption{Site information.\label{tab:sites}}
\tablewidth{0pt}
\tablehead{
\colhead{Site Name}& \colhead{Site Location$^a$} & \colhead{N Latitude$^b$} & \colhead{E Longitude$^b$} & \colhead{Altitude$^b$} \\
\colhead{}& \colhead{} & \colhead{($\degr$ $\arcmin$ $\arcsec$)}& \colhead{($\degr$ $\arcmin$ $\arcsec$)} & \colhead{(m)} 
}
\startdata
CTIO & Cerro Tololo Inter-American Obs., Chile& & & \\
 & \hspace{2mm}LCO-LSC$^c$& -30 10 02 &-70 48 17& 2198\\ 
 & \hspace{2mm}SARA-CT& -30 10 19 &-70 47 57& 2012\\
EOS (Electro Optic Systems) & near Learmonth, WA, Australia& -22 14 22 &114 05 50 & 6\\
ETS$^d$ (Experimental Test Site) & White Sands Missile range, NM& 33 49 05 & -106 39 36& 1511\\
Happy Jack & near Flagstaff, AZ& 34 44 39 & -111 25 21 & 2337\\
LCO & Las Campanas Obs., Chile& -29 00 51 & -70 41 33 & 2519\\
MAO & Michael Adrian Obs., Trebur, Germany& 49 55 32 & 8 24 41& 103\\
McDonald Obs. & near Fort Davis, TX& 30 40 12 & -104 01 12& 2070\\
MKO & Mauna Kea Obs., HI& 19 49 34 & -155 28 19 & 4168\\
OAN-SPM$^e$ & Observatorio Astr\'{o}nomico Nacional, Sierra San Pedro M\'{a}rtir, Mexico &31 02 39 & -115 27 49 & 2800\\
ORM & Observatorio del Roque de los Muchachos, Spain& -17 53 41& 28 45 49 & 2396\\
Pach\'{o}n & Cerro Pach\'{o}n, Chile & -30 14 27 &	-70 44 12& 2722 \\
SAAO & South African Astronomical Obs., South Africa& & & \\
& \hspace{2mm}74 in& -32 22 44 & 20 48 42& 1822\\%from Coppejans thesis 
 & \hspace{2mm}LCO-CPT$^c$& -32 22 48 & 20 48 36& 1810\\ %altitude from 1-m Coppejans thesis, LCO website has 1760, which can't possibly be correct
Savannah Skies  & near Chillagoe, QLD, Australia& -17 03 41 & 144 21 28 & 360\\
SRO & Sierra Remote Obs., Auberry, CA& 37 04 14 & -119 24 45& 1405\\
Teide Obs. & Tenerife, Spain&-16 30 35&28 18 00& 2330\\
\enddata
\tablecomments{$^a$Subdivided by telescope, for sites at which multiple telescopes were used that were spaced farther apart than one second in latitude or longitude. $^b$Relative to WGS-84. $^c$The average of the three Las Cumbres 1-m telescopes at CTIO (LCO-LSC) and the two 1-m telescopes at SAAO (LCO-CPT). The 0.4-m LCO-CPT telescope is in close proximity to the 1-m telescopes. The LCO-CPT telescopes are within a kilometer of the 74 in.$^d$Run by MIT's Lincoln Laboratory.$^e$From table 1 in \cite{RN3990}.}
\end{deluxetable*}

\begin{deluxetable*}{cclllcccc}
%\tablenum{3}
\tablecaption{Pluto occultation observation details.\label{tab:scopes}}
\tablewidth{0pt}
\tablehead{
\nocolhead{} & \nocolhead{}  & \nocolhead{}  &\nocolhead{}  &  \nocolhead{}&
\colhead{Exposure} & \colhead{Cycle} & \nocolhead{} & \colhead{Postdicted} \\
\nocolhead{} & \nocolhead{}  & \nocolhead{}  &\nocolhead{}& \nocolhead{}  &
\colhead{Time} & \colhead{Time$^d$} &\nocolhead{} & \colhead{Closest}\\
\colhead{Date} & \colhead{Site$^a$} & \colhead{Telescope$^b$} & \colhead{Instrument$^c$} & \colhead{Filter} &
\colhead{(s)} & \colhead{(s)} & \colhead{SNR$^e$} & \colhead{Approach$^f$ (km)}
}
\startdata
20170807 & MKO          & 3.2 m IRTF      & MORIS             & open      & 2.4966    & 2.5           & 27        & 1026\\ %939 \\
20180409 & Happy Jack   & 4.3 m LDT     & POETS             & open      & 0.0983    & 0.1           & 11        & 465\\ %600 \\
20180815 & SRO          &  0.6 m ATUS   & Andor iXon DU-888 & open      & 0.6       & 0.6018        & 42        & 1106\\ %$1089.0\pm3.5^g$ \\
         & OAN-SPM$^g$  & 2.1 m         &  Andor iXon DU888 & Gunn $i$  & 0.0966    & 0.1           & 158       & 669\\ %$651.6\pm3.5$$^g$ \\
         & ETS          & 31 in (0.8 m) ETS-B   &  Andor iXon DU-888& clear     & 1         & $\sim1$& 18        & 641\\ %$624.5\pm3.5^g$ \\
         & McDonald Obs.& 1 m LCO-ELP   &  FLI MicroLine ML4720& clear  & 1.5       & $\sim$3       & 71        & 365\\ %$348.3\pm3.5^g$ \\
20181001 & SAAO         & 74 in (1.9 m)         & SHOC              & open      & 2.4966    & 2.5           & 41        & 534\\ %590  \\
         & SAAO         & 2x1 m LCO-CPT &  FLI MicroLine ML4720& clear  & 4,5       & $\sim$5.4,6.4 & 10        & 534\\ %590 \\
         & SAAO         & 0.4 m LCO-CPT & SBIG 6303         & open      & 10.28     & $\sim$20.2    & 5         & 534\\ %590 \\
20181101 & SAAO         & 74 in (1.9 m)        & SHOC              & open      &0.2966     & 0.3           & 58 
& 942\\ %$\sim$910 \\
20181120 & Teide Obs.   & 0.4 m LCO-TFN & SBIG 6303          & open      & ~5.34     & ~7.34         & 19        & 735\\ %$\sim$861xx \\
20210806 & LCO          & 6.5 m Magellan Clay& POETS        & open      & 0.4983    & 0.5           & 21        & 1201\\ %1219 \\
         & Pachon         & 8.1 m Gemini S.& Zorro            & clear     & 0.5       & 0.5034        & 17        & 1335\\ %1353 \\
         & CTIO         & 3x1 m LCO-LSC  & FLI  MicroLine ML4720 & clear& 10,11,12  & $\sim$11.4,12.4,13.4 & 6  & 1326\\ %1353xx \\
         & CTIO         & 0.6 m SARA-CT   & Andor iKon-L 936  & open      & 15        & $\sim$16.9    & 2         & 1327\\ %1353xx \\
20220601 & Savannah Skies & 20 in (0.5 m)       & POETS             & open      & 0.7483    & 0.75          & 86        & 468\\ %$454.7\pm11.4^g$\\
        & Savannah Skies & 25 cm BRC    &Apogee Alta U16    & clear      & 10        & $\sim$11.34   & 12        & 468\\ %$454.7\pm11.4^g$\\
        & Savannah Skies & 30 cm FRC    &SBIG STL 6303      & clear     & 10        & $\sim$15.36   & 36        & 468\\ %$454.7\pm11.4^g$\\
        & Savannah Skies & 15 cm FCT    &SBIG STF 8300      & clear     & 10        & $\sim$10.91   & 8         & 468\\ %$454.7\pm11.4^g$\\
        & Savannah Skies & 12.5 in (0.3 m) RCOS &SBIG STXL 6303     & clear     & 10        & $\sim$14.03   & 9         & 468\\ %$454.7\pm11.4^g$\\
        & Savannah Skies & 20 in (0.5 m) Carruthers        & SBIG STX16803     & clear     & 6         & $\sim$9.4     & 41        & 468\\ %$454.7\pm11.4^g$\\
        & EOS           & 1 m           & FLI ProLine PL4240& open      & 2.0       & $\sim$3.8     & 94        & 812\\ %$799.6\pm11.4^g$\\
20220823& ORM           & 2.3 m LT      & RISE              & V+R       & 1.5       & 1.535         & 9         & 453\\ %612xx  \\
        & MAO           & 1.2 m T1T     & QHY174M-GPS       & open      & 1.5       & $\sim$1.5    & 3         & 1117\\ %1112xx\\
20230717& SAAO          & 74 in (1.9 m)         & SHOC              & open      & 0.797     & 0.8           & 6         & 474\\ %400xx \\
\enddata
\tablecomments{$^a$ See Table \ref{tab:sites}. $^b$Telescope key: ATUS – Astronomical Telescope of the Univ. of Stuttgart; BRC - Takahashi Baker-Ritchey-Chr\'{e}tien; ETS - Experimental Test Site; FCT -  fluorite apochromatic triplet; FRC - Takahashi flat-field Ritchey-Chr\'{e}tien; IRTF – NASA's InfraRed Telescope Facility; LCO-CPT - Las Cumbres SAAO; LCO-ELP - Las Cumbres McDonald Obs.; LCO-LSC - Las Cumbres CTIO; LCO-TFN - Las Cumbres Teide Obs.; LDT – Lowell Discovery Telescope; RCOS - Takahashi RC Optical Systems; SARA-CT – Southeastern Association for Research in Astronomy; LT - Liverpool Telescope; T1T- Trebur 1.2~m Telescope. $^c$Instrument key: FLI – Finger Lakes Instruments; MORIS – MIT Optical Rapid Imaging System \citep{RN3782}; POETS – Portable Occultation, Eclipse, and Transit System \citep{RN3080}; RISE - Rapid Imaging Search for Exoplanets \citep{RN3986}; SBIG – Santa Barbara Instrument Group; SHOC – Sutherland High-speed Optical System \citep{RN3585}; Zorro \citep{RN3944}. $^d$Values with "$\sim$" are from instruments for which the cycle time varied: the value listed is the median. $^e$Signal-to-noise ratio per 60-km atmospheric scale height. The Las Cumbres data are combined when there are multiple telescopes of the same size. $^f$Closest approach values are calculated from the NIMA Pluto ephemeris and the GAIA DR3 star catalog as described in \S~\ref{sec:obs}}. $^g$This dataset was published in \citet{RN3990} and supplies the light curve used in the atmospheric fits to supplement our data.
% CA values here are used to generate Fig. 1 from Bastian and Fig. 2, in PlutoGlobes2021.# notebook.
\end{deluxetable*}
%
%% The "ht!" tells LaTeX to put the figure "here" first, at the "top" next
%% and to override the normal way of calculating a float position

\section{Analyses} \label{sec:analyses}

\subsection{Light-curve extraction} \label{subsec:lc}
This section contains details of how light curves were extracted from the data from each telescope. In general, calibrations frames (biases and flat fields) were applied when available. The optimal photometric aperture size was chosen by testing a series of aperture sizes and selecting the aperture that returned the highest signal-to-noise ratio (SNR) in the light-curve baseline. Each light curve was then generated by (i) dividing the sky-subtracted aperture occultation star counts by those of the sky-subtracted aperture comparison star(s), (ii) normalizing to one by dividing by the mean value of the baseline (out of occultation), and (iii) calibrating from zero to unit flux level using the background fraction (defined as $f_b=\frac{S_{P}}{S_{P}+S_{*}}$, where $S_{P}$ and $S_{*}$ are the full-scale signals from Pluto plus its satellites and the occultation star, respectively). Finally, a third-order polynomial was fit to the data outside the occultation and subtracted in order to remove the baseline variation. 

\begin{figure}[ht!]
\centering
\includegraphics[width=0.75\textwidth,clip,trim=0mm 0mm 0mm 0mm]{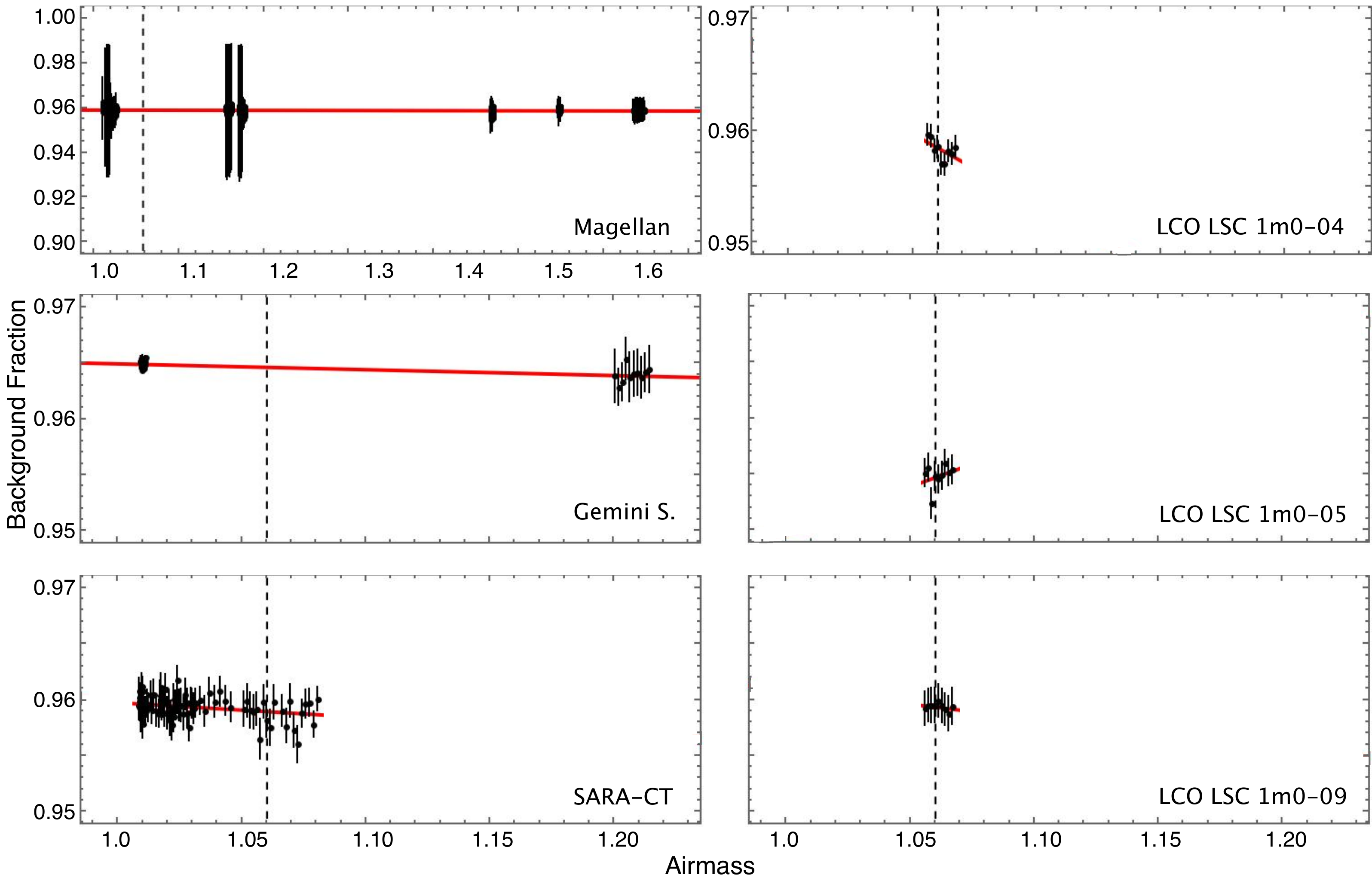}
\caption{Background fractions versus airmass for each telescope for the 2021 August 06 occultation. The dashed black lines indicate the airmasses at the times of the occultations and the red lines are least-squares, linear fits to the data. Data from Magellan and Gemini S. were taken before and after the occultation on the same night, while the other datasets were taken at similar airmasses on an adjacent night (which allows the SARA-CT and LCO data to span the airmass of the occultation). Each of the three, 1-m Las Cumbres telescopes are measured separately here, and the different error bars reflect differing image quality.  These plots demonstrate the technique for determining the background fraction when sufficient well-separated data are available. \label{fig:bfPlot}}
\end{figure}

When images were available in which Pluto and the star are well separated, we derived the background fraction using point-spread-function fitting with standard functions in Astropy's Photutils \citep{RN4082}. Because the star and Pluto are not necessarily expected to have the same spectra throughout the observed wavelengths, this ratio can change as a function of airmass. We use a linear fit to the background fractions over time, including error bars, to calculate the expected value at the airmass of the occultation midtime. Example plots illustrating this technique are shown for 2021 August 06 in Fig.~\ref{fig:bfPlot}. This event was chosen as an example because we had well-separated data for each telescope and were able to fit for background fractions; however, we note that the background fractions did not change significantly over the tested airmasses for these sites. Background fractions are functions of the response of each telescope and instrument, so the optimal result would be a well-fit value for each site for each event. However, we do not always have well-separated data. An additional complicating factor can be when well-separated data were taken one or more days before or after an occultation, since Pluto's magnitude can change tenths of magnitudes on daily timescales due to its rotation. For datasets that were part of multi-chord events where a background fraction could not be calculated, the background fraction was determined from initial atmospheric fitting to all the data (see \S~\ref{subsec:multichordatmfits}). For single-chord events where a background fraction could not be calculated, the calculated background fractions had large errors, or the calculated value returned unrealistic results (e.g. the light curve dropped below zero flux), the background fraction was set to that of the expected flux ratio between Pluto and the star at the midtime based on the star's photometric catalog magnitude and Pluto's apparent magnitude. As noted below, the expected flux ratios are largely consistent with the well-fit background fractions that we adopt per telescope and event.

Unless otherwise noted, circular-aperture photometry was carried out on the target star and two bright comparisons (bright comparisons are preferred, as comparison stars fainter than Pluto increase noise when combined during reduction). Three hand-selected boxes well outside the apertures and without background stars were used to determine the sky value. The position of the target star was determined by centroiding the brightest comparison star and using a fixed offset that was determined from the occultation- and comparison-star centroids at the start of the dataset. When multiple comparison stars were available, we used the mean of the stars and their ratio was confirmed to be a flat line, indicating no stellar variations.  We provide all light curves as normalized flux versus time in Figs.~\ref{fig:20170807LC}-\ref{fig:20230717LC}. Offsets in the occultation midtimes between different sites for a given event are a result of the spread of the locations of the observing sites along the shadow path. As a measure of data quality, the resulting SNRs per atmospheric scale height are provided in Table~\ref{tab:scopes} for every light curve that we generated.

\subsubsection{2017 August 07} \label{subsubsec:lc20170807}
\begin{figure}[ht!]
\centering
\includegraphics[width=0.4\textwidth,clip,trim=0mm 0mm 0mm 13.5mm]{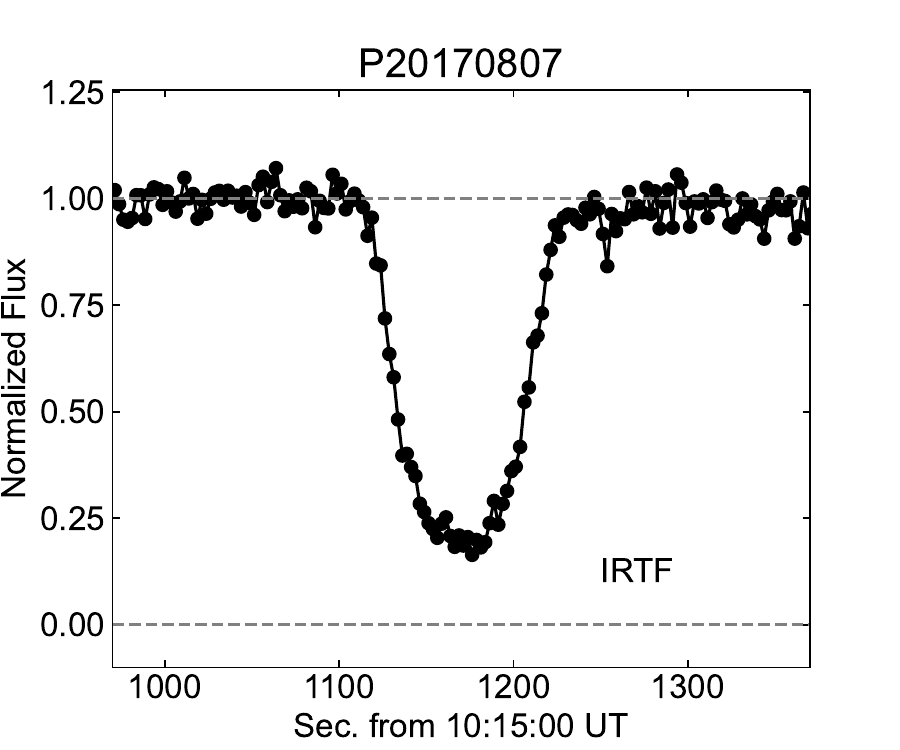}
\caption{Light curve from the 2017 August 07 occultation. Each black dot represents one data point. Dashed gray lines are provided for reference at zero and one flux levels. \label{fig:20170807LC}}
\end{figure}

A master bias was created and subtracted from the raw data. No flatfields were available. Circular-aperture photometry was carried out as described above. We note the nearby star to the northernmost comparison (see Fig. \ref{fig:20170807images}), the center of which fell within the comparison-star aperture: that nearby star is a fraction of the flux of the brighter star, and we confirmed that it did not affect the final light curve. The highest SNR was achieved with an aperture of diameter 21 pixels (4.6 arcsec). Thirty images taken before and twenty images taken after the occultation, when Pluto and the star were well separated, were analyzed to determine a background fraction of $0.684\pm0.002$ at the airmass of the event (as an example for comparison with the flux-ratio method, the expected background fraction was 0.681). The resulting light curve is shown in Fig.~\ref{fig:20170807LC}.

% Consider plotting on same x and y scales.

\subsubsection{2018 April 09} \label{subsubsec:lc20180409}
\begin{figure}[ht!]
\centering
\includegraphics[width=0.4\textwidth,clip,trim=0mm 0mm 0mm 15mm]{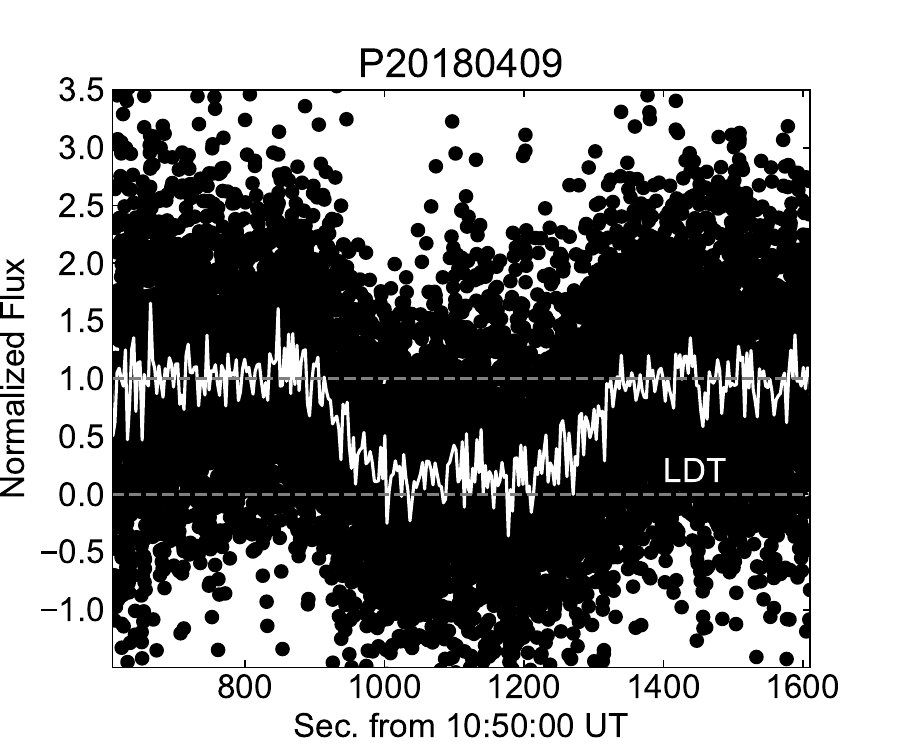}
\caption{Light curve from the 2018 April 09 occultation. Each black dot represents one data point.  Due to the low SNR, the points are not connected by lines. Plotted in white is the data binned by 30 points, to better see the light curve trends. Dashed gray lines are provided for reference at zero and one flux levels. \label{fig:20180409LC}}
\end{figure}

Master biases and flats were made for the LDT, and the data were bias-subtracted and flatfielded. Circular-aperture photometry was carried out as described above, except that only one bright comparison star was available in the frame. The highest SNR was achieved with an aperture of diameter 12 pixels (6.1 arcsec). This event was slower than usual, and the star on the fainter end (see Table~\ref{tab:star}): Pluto and the star were not separable in images taken within a few hours of the occultation. We assumed a background fraction of 0.960, based on the expected flux ratio between Pluto and the star. The resulting light curve is shown in Fig.~\ref{fig:20180409LC}. The full-resolution data are very noisy, so in Fig.~\ref{fig:20180409LC} we have also plotted a light curve binned by average by 30 points (to 3-s resolution, or 19.1-km spatial resolution), which has a SNR per scale height of 25.

\subsubsection{2018 August 15} \label{subsubsec:lc20180815}
At SRO, master bias and flat images were constructed and the data were bias-subtracted and flatfielded. The occultation data were aligned to within a full pixel (i.e. shifted without interpolation) to compensate for the telescope drift. Circular-aperture photometry was performed as described above. An aperture of diameter 8 pixels (4.8 arcsec) produced the highest SNR.  Flux values were measured on the night after the occultation from nearly four hundred well-separated images of the star and Pluto taken within $\Delta z=0.06$ of the airmass at the occultation midtime, where $z$ is airmass. A linear fit was used to determine the background fraction of $0.238\pm0.004$ at the airmass of the event. Pluto's apparent magnitude changed by only 0.002 mag between the two nights, and this calculated background fraction is consistent with the expected flux ratio value of 0.237.

The data from LCO-ELP were not run through the standard Las Cumbres pipeline \citep{2018BANZAI}%\footnote{https://lco.global/documentation/data/BANZAIpipeline/}
; therefore, we analyzed raw data. Circular-aperture photometry was carried out as described above. An aperture of diameter 20 pixels (13.5 arcsec) resulted in the highest SNR. The background fraction of $0.257\pm0.001$ was determined from nine frames taken roughly two hours before the midtime and extrapolated to the airmass at the time of the occultation.

\begin{figure}[ht!]
\centering
% Trim here is set to remove the plot title.
\includegraphics[width=0.4\textwidth,clip,trim=0mm 10mm 0mm 30mm]{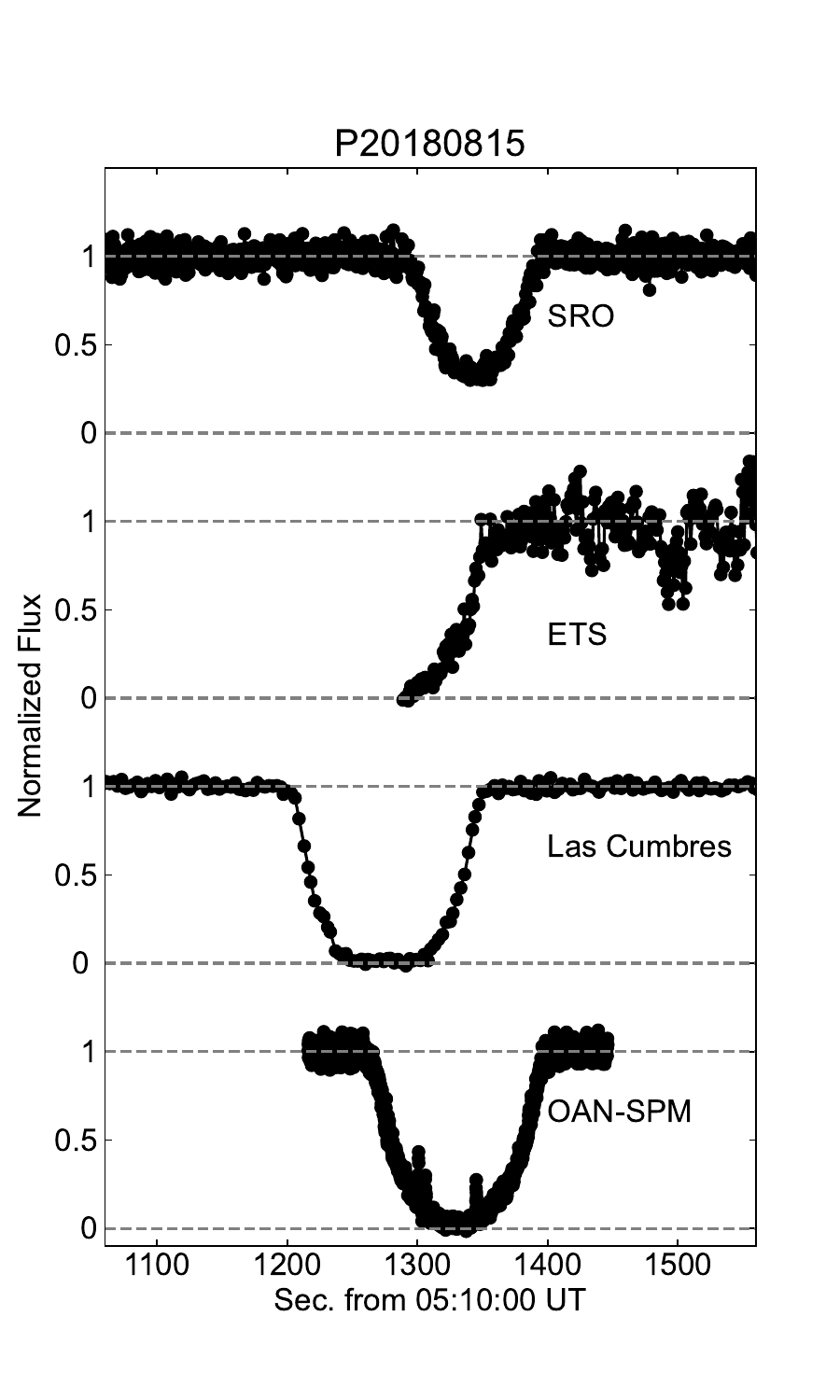}
\caption{Light curves from the 2018 August 15 occultation. Each black dot represents one data point. Dashed gray lines are provided for reference at zero and one flux levels. The ETS light curve is a partial due to clouds obscuring the ingress. Note that the OAN-SPM data are not from this work but were published in \citet{RN3990} and are included here as being a very high quality dataset that improves the atmospheric fit. %The light curves are vertically offset slightly more than in the other plots in order to prevent the Las Cumbres and OAN-SPM data from overlapping.
\label{fig:20180815LC}}
\end{figure}

The ETS data were provided with a flatfield image, from which a median-normalized flatfield was constructed and divided from the data. Circular-aperture photometry was carried out as described above; however, (i) only the brightest comparison star was used (as it returned the highest SNR light curve) and (ii) the location of the occulted star was determined differently. The comparison stars disappeared throughout the dataset due to intermittent clouds. When the comparison star signal was too low to centroid, the location of the occulted star was fixed to that from the previous frame. An aperture of diameter 18 pixels (12.2 arcsec) returned the highest SNR. No well-separated images were taken. Based on the initial atmospheric fits to all the data from this event (see \S~\ref{subsec:multichordatmfits}), the background fraction was set to $0.20$. The resulting light curves are shown in Fig.~\ref{fig:20180815LC}.

\subsubsection{2018 October 01} \label{subsubsec:lc20181001}
Raw data from SAAO were used to generate the light curve. No calibrations were done because they increased noise: (i) there was low-level structure in the bias images, the location of which varied between frames, which was not apparent in the occultation dataset and (ii) no flatfields were obtained, as it was cloudy at the start of the night and the observing window ended before morning twilight. Circular-aperture photometry was performed as described above. An aperture 11 pixels (6.7 arcsec) in diameter  returned a light curve with the highest SNR.

The data from the two LCO-CPT 1-m telescopes were processed through the Las Cumbres pipeline \citep{2018BANZAI}. The data from the 0.4-m Las Cumbres telescope were provided and analyzed raw. Photometry was carried out identically to the SAAO data. The highest-SNR light curves were achieved with 9-pixel (6.1 arcsec; telescope 1m012), 13-pixel (8.8 arcsec; 1m010), and 5-pixels (5.8 arcsec; 0m407) diameter apertures. Telescope 1m010 was slightly out of focus and thus required a particularly large aperture. 

\begin{figure}[ht!]
\centering
\includegraphics[width=0.4\textwidth,clip,trim=0mm 10mm 0mm 30mm]{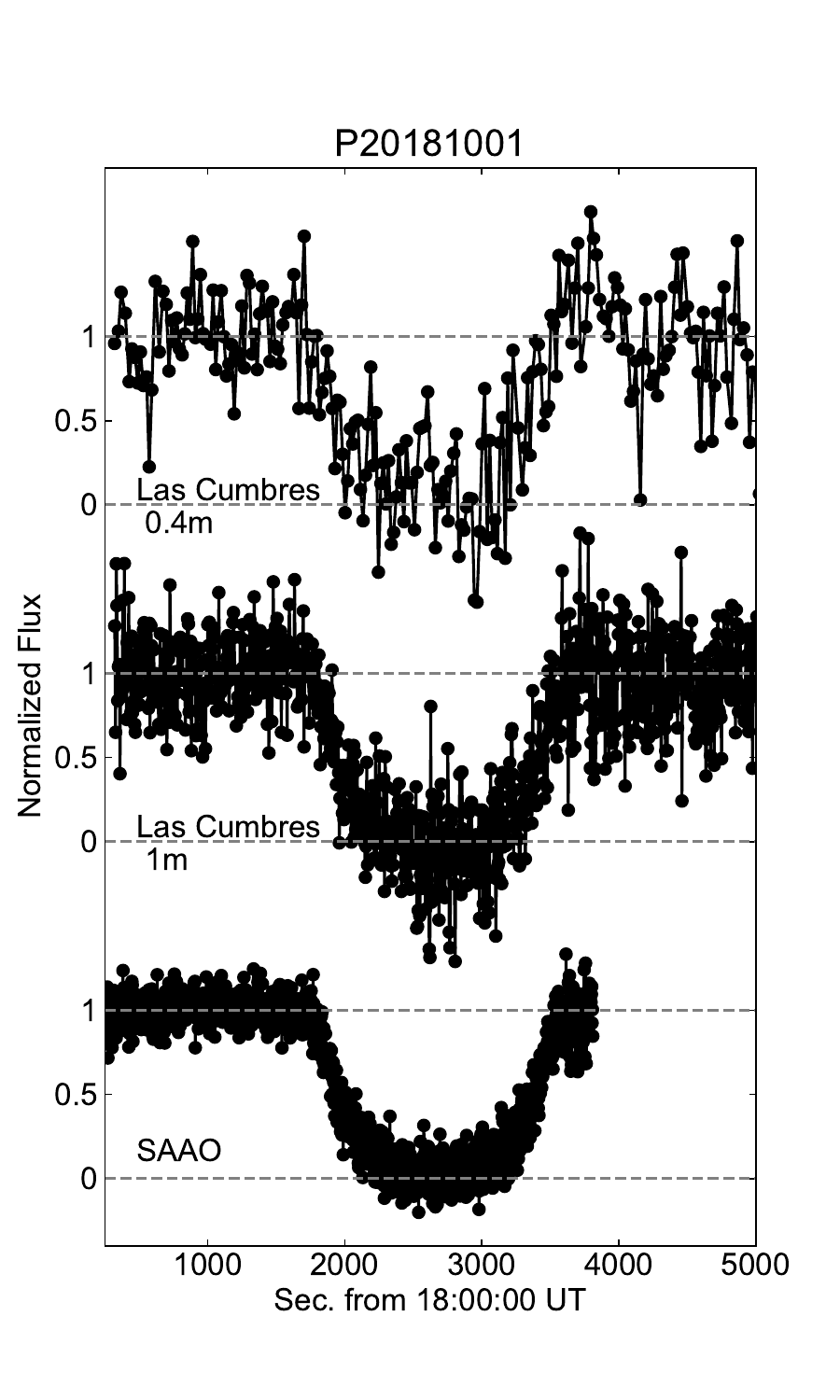}
\caption{Light curves from the 2018 October 01 occultation. Each black dot represents one data point. Dashed gray lines are provided for reference at zero and one flux levels. The data from the two 1-m Las Cumbres telescopes are combined. \label{fig:20181001LC}}
\end{figure}

Due to the unusually slow velocity of this occultation (see Table~\ref{tab:star}), it was not possible to take data on the same night with Pluto and the star well separated. For the SAAO telescope, a background fraction of $0.943$ was determined from the initial atmospheric fits. For LCO-CPT 1m010, ten frames were taken each on 2018 October 04 and 05, within 0.05 airmasses of the airmass at the occultation midtime. For LCO-CPT 1m012, ten frames were taken on 2018 October 04, also within $\Delta z=0.05$ of the airmass at occultation midtime. For LCO-CPT 0m407, fifteen frames were taken on 2018 October 05, within $\Delta z=0.09$ of the airmass at occultation midtime. Linear fits were used to determine background fractions of $0.949\pm0.002$, $0.947\pm0.002$, and $0.947\pm0.003$ at the airmass of the event for LCO-CPT 1m010, 1m012, and 0m407, respectively. Pluto's apparent magnitude changed by only 0.006 mag over this multi-day timescale, but based on the combined 1 m light curve dropping below zero, we fixed the background fractions at the consistent value of 0.945 from the expected flux ratio between Pluto and the star. The resulting light curves, with the 1-m LCO data combined, are shown in Fig.~\ref{fig:20181001LC}.

\subsubsection{2018 November 01} \label{subsubsec:lc20181101}
\begin{figure}[ht!]
\centering
% Trim here is set to remove the plot title.
\includegraphics[width=0.4\textwidth,clip,trim=0mm 0mm 0mm 15mm]{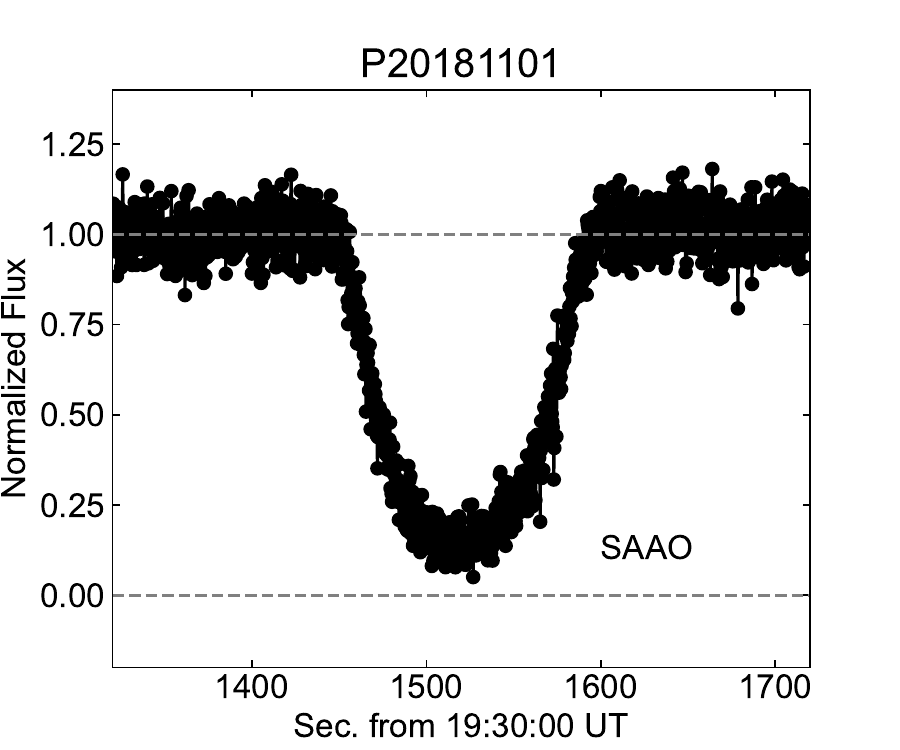}
\caption{Light curve from the 2018 November 01 occultation. Each black dot represents one data point. Dashed gray lines are provided for reference at zero and one flux levels. \label{fig:20181101LC}}
\end{figure}

Master biases and flats were created for the SAAO and used to bias subtract and flat field the data. Circular-aperture photometry was carried out as described above. The optimal aperture had a diameter of 10 pixels (12.2 arcsec). Three frames well outside the occultation (images 5721, 5816, and 6248) were excluded due to cosmic ray contamination in the background boxes. Observations of Pluto and the star separated from each other were taken both before and after the occultation. The light ratio was fit on roughly 200 frames before the event and 100 frames after the event. A linear fit returned a background fraction at the airmass of the event midtime of $0.667\pm0.005$. The resulting light curve is shown in Fig.~\ref{fig:20181101LC}.

\subsubsection{2018 November 20} \label{subsubsec:lc20181120}
\begin{figure}[ht!]
\centering
% Trim here is set to remove the plot title.
\includegraphics[width=0.4\textwidth,clip,trim=0mm 0mm 0mm 15mm]{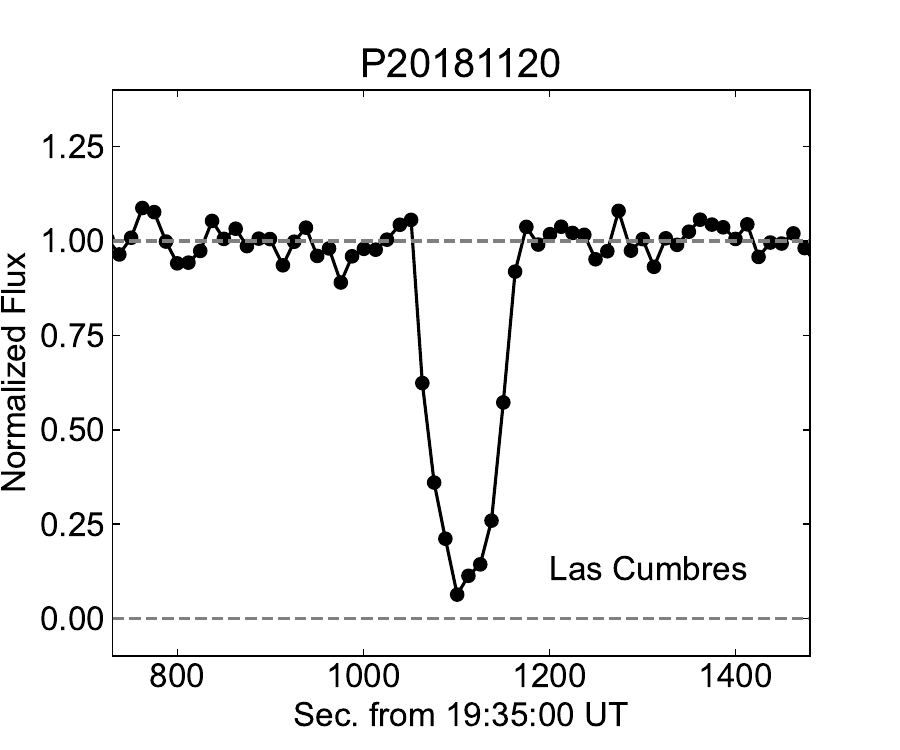}
\caption{Light curve from the 2018 November 20 occultation. Each black dot represents one data point. Dashed gray lines are provided for reference at zero and one flux levels. \label{fig:20181120LC}}
\end{figure}

The 0.4-m Las Cumbres data were provided and analyzed raw. Circular-aperture photometry was carried out as described above, and the aperture with the highest SNR had a diameter of 8 pixels (9.3 arcsec). A series of thirty, well-separated images were taken three nights after the occultation at airmasses closer than $\Delta z=0.38$ of the airmass at occultation midtime.  A linear fit was used to derive a background fraction at the time of the event of $0.507\pm0.005$. However, this fraction resulted in a light curve that was inconsistent with all the other events. We assumed the background fraction of 0.515 based on the expected flux ratio. The light curve is shown in Fig. \ref{fig:20181120LC}.

\subsubsection{2021 August 06} \label{subsubsec:lc20210806} 
For Gemini South, Magellan, and SARA-CT, master biases and flats were generated and the data were bias-subtracted and flatfielded. Circular-aperture photometry was carried out as described above, excluding Magellan and Gemini where the field of view was small and only one moderately-bright comparison was available. The LCO-LSC and SARA-CT fields were large enough to allow selection of two bright comparisons (as indicated in Fig.~\ref{fig:2021images}). The highest light-curve SNR was obtained for apertures of diameter 14 pixels (3.3 arcsec), 20 pixels (3.7 arcsec) and 18 pixels (10.9 arcsec) for Gemini, Magellan, and SARA-CT, respectively. 

The LCO-LSC data were processed through the Las Cumbres pipeline \citep{2018BANZAI}. The light curve was generated as described above. The data from all three telescopes were combined to make the final light curve. The optimal aperture diameters were 14 pixels (9.4 arcsec) for the 1m004 telescope and 16 pixels (10.8 arcsec) for the 1m005 and 1m009 telescopes.
 
\begin{figure}[ht!]
\centering
\includegraphics[width=0.4\textwidth,clip,trim=0mm 10mm 0mm 30mm]{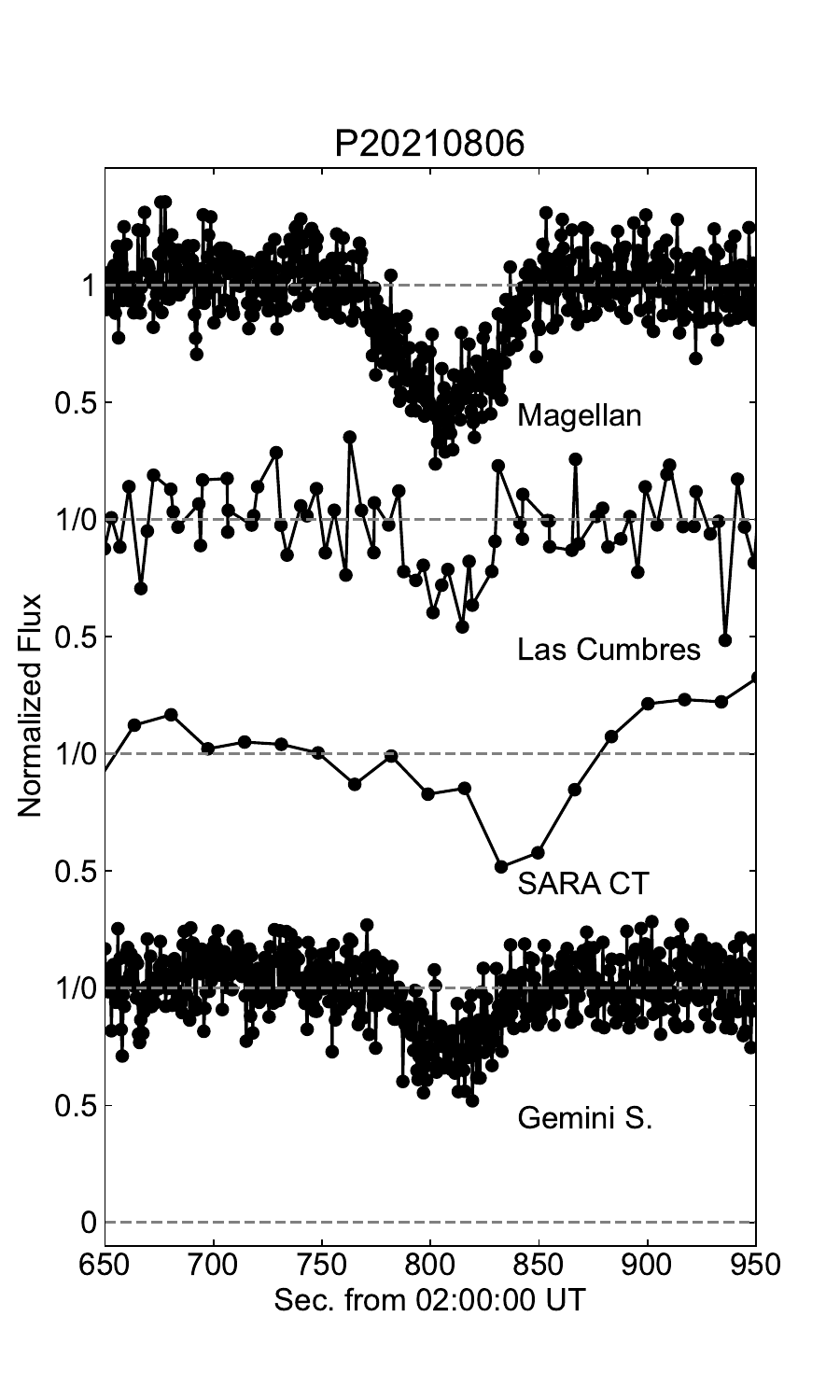}
\caption{Light curves from the 2021 August 06 occultation. Each black dot represents one data point. Dashed gray lines are provided for reference at zero and one flux levels. The data from the three 1-m Las Cumbres telescopes are combined. As apparent from the flux levels not dropping below half-light level, these chords grazed Pluto's atmosphere and were not full occultations. \label{fig:20210806LC}}
\end{figure}

As shown in Fig.~\ref{fig:bfPlot}, the background fraction was extracted from well-separated images at each site. Data were taken both before and after the event from Gemini (ten frames an hour before and twenty frames 1.5 hours after) and Magellan (nearly 900 frames 2.5 to 1 hour before and 600 frames 30 min to 1.5 hours after). Because of the high spatial resolution from these telescopes, Charon was fit as a separate source (see Fig. \ref{fig:2021images}). We suspect that the relatively large error bars on the Magellan data are due to nearby light from Charon being close to Pluto and more difficult to properly fit: Charon was roughly 10 pixels and 4 pixels away from Pluto in the Gemini and Magellan data, respectively. It may also seem counterintuitive for the Magellan error bars to be larger at lower airmasses; however, those data are closer to the occultation midtime and so the star and Pluto have a smaller separation. Linear fits were used to determine background fractions at the airmass of the event midtime of $0.965\pm0.006$ for Gemini and $0.959\pm0.0002$ for Magellan. For the Las Cumbres telescopes, background fractions of $0.958\pm0.001$, $0.955\pm0.001$, and $0.959\pm0.002$ (1m004,1m005, and 1m009 telescopes, respectively) were calculated from ten well-separated images taken on the night before the event that spanned the airmass at the occultation midtime. For SARA-CT, a background fraction of $0.960\pm0.001$ was derived based on a series of eighty images taken the previous night that spanned the airmass at the occultation midtime. Because of the unknown timing offset in the SARA-CT data, and the low SNR, we do not include it in the atmospheric fitting. For reference, the background fraction is 0.960 from the expected flux ratio between Pluto and the star. The resulting light curves are shown in Fig.~\ref{fig:20210806LC}.

\subsubsection{2022 June 01} \label{subsubsec:lc20220601} 

Biases and flatfields were provided at EOS; however, they were not used because they weren't taken using the same instrumental settings as the data.  The standard aperture photometry procedure from above was applied. Note that there was substantial field rotation, so centroiding was carried out on each frame for the comparison stars and Pluto merged with the occultation star. The highest SNR was achieved with an aperture of diameter 16 pixels (14.1 arcsec). No well-separated data were obtained. The nominal background fraction of 0.22 was determined from initial atmospheric fitting for the full dataset with haze. 

For POETS on the 20-in at Savannah Skies, raw data were used because bias subtraction and flat-fielding did not improve SNR. We find that it is not uncommon for standard data reduction, especially flat fielding, to add noise to light curves and do not think that this reflects on the data quality. Circular-aperture photometry was carried out as described above. The highest SNR was achieved with an aperture diameter of 6 pixels (8.6 arcsec). Well-separated images were only taken before the event. A linear fit was applied to the background fractions derived from 180 such images to determine the background fraction of $0.205 \pm0.005$ at the airmass of the event. 

For all other Savannah Skies telescopes, raw data were used because no calibration frames were included with the datasets. The only exception was the data from the 20-in Carruthers telescope, which were auto-dark-corrected using the feature in the TheSkyX software. TheSkyX software was used to operate the telescope and camera and automatically dark-subtracted the science images. Circular-aperture photometry was carried out as described above, for all datasets except RCOS. One of the comparison stars drifted off the RCOS frame by the end of the exposure sequence, so a different nearby comparison was used instead. The optimal aperture radius was 9 pixels for Carruthers, FRC, and FCT, and 5 pixels for BRC and RCOS. The nominal background fractions for each of these sites, of 0.25, 0.26, 0.24, 0.29 and 0.19 respectively, were determined from the initial atmospheric fits. Nominally, occultation data taken at different wavelengths can be a powerful tool to characterize haze \citep[e.g.][]{RN2820, RN3641}; however, without well-separated data to properly calculate the background fractions, we lack confidence in the normalized flux levels and cannot use them to examine flux versus wavelength trends.

\begin{figure}[ht!]
\centering
\includegraphics[width=0.65\textwidth,clip,trim=0mm 0mm 0mm 0mm]{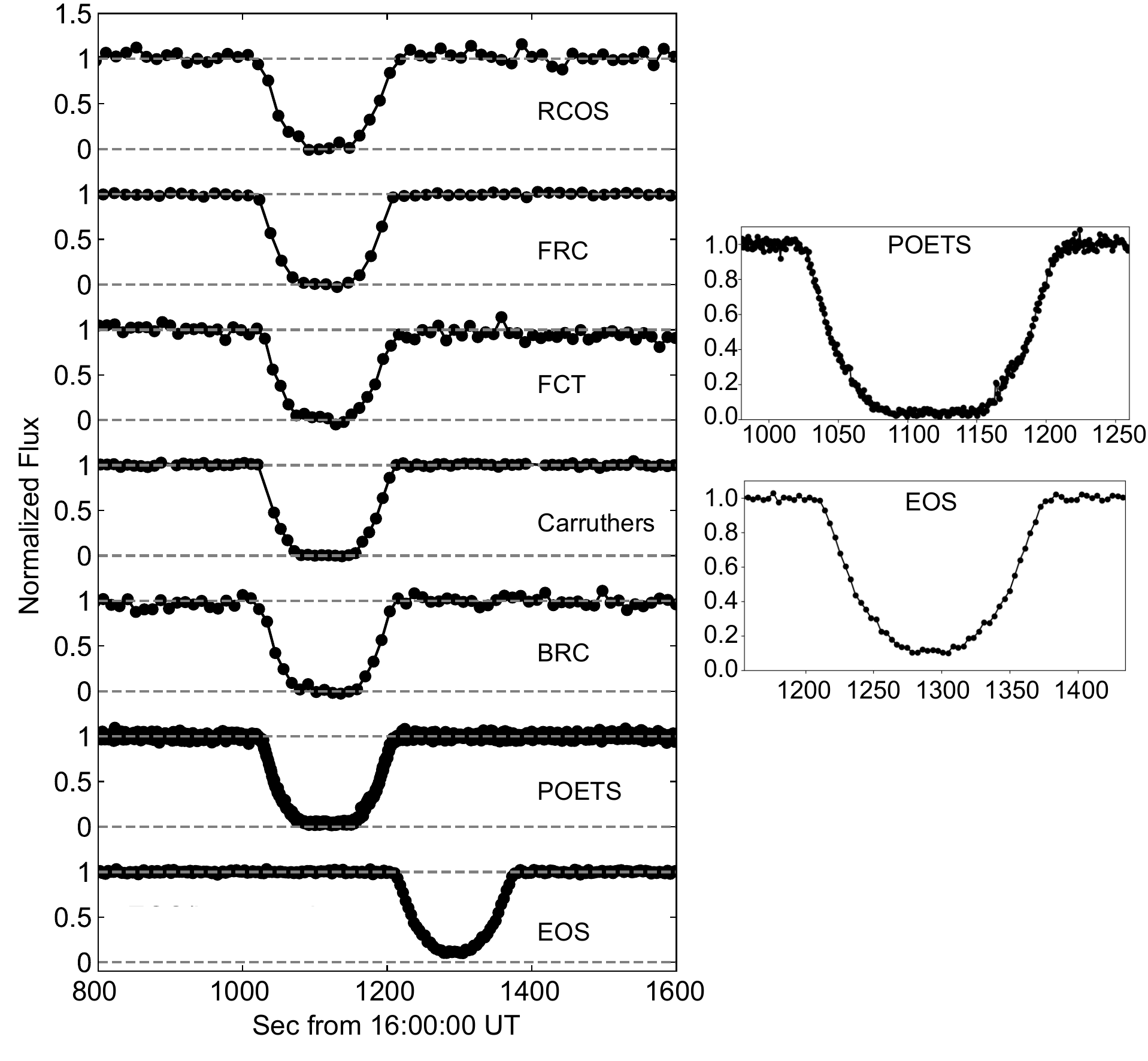}
\caption{Light curves from the 2022 June 01 occultation. Each black dot represents one data point. Dashed gray lines are provided for reference at zero and one flux levels. Zoomed extracts for the two highest-quality light curves are shown on the right, to better see the structural shapes. \label{fig:20220601LC}}
\end{figure}

The resulting light curves are shown in Fig.~\ref{fig:20220601LC}.

\subsubsection{2022 August 23} \label{subsubsec:lc20220823} 
The LT data were provided after bias-, dark- and flat-correction through LT's RISE pipeline. The light curve was generated as described above. The optimal aperture was 9 pixels (9.7 arcsec). As the event occurred roughly 17 min after evening twilight ended, well-separated images were only available on the night of the event roughly 2.5 hours after the midtime; however, images were also taken one day prior within $\Delta z\approx0.15$ of the midtime airmass and six days later. The background fractions derived from the images taken at different times were inconsistent with each other, so we fixed the background fraction at 0.886, based on the expected flux ratio between Pluto and the star. The adopted background fraction is slightly lower than that used in \cite{RN4068}, $0.897\pm0.006$, because the higher value was found to be inconsistent with the other light curves when plotted in scaled distance (see \S~\ref{subsec:lowatm}).

\begin{figure}[ht!]
\centering
\includegraphics[width=0.4\textwidth,clip,trim=0mm 10mm 0mm 30mm]{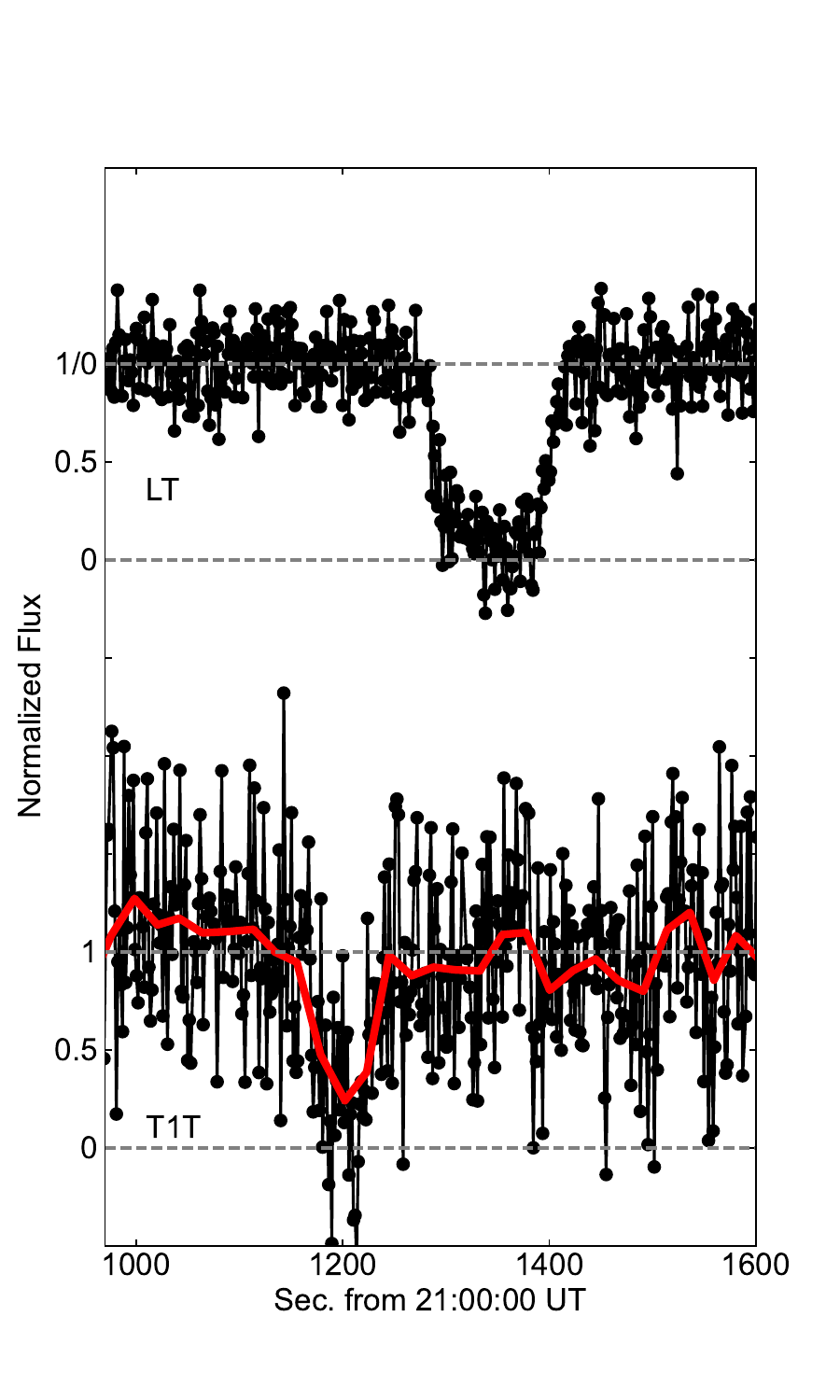}
\caption{Light curves from the 2022 August 23 occultation. Each black dot represents one data point. The red line shows the T1T data binned by 14 points to match the SNR of the LT data. Dashed gray lines are provided for reference at zero and one flux levels. \label{fig:20220823LC}}
\end{figure}

Flats, biases, and dark images were taken at T1T; however, the flats were unusable due to an error in camera settings. The data were dark-subtracted and the light curve was generated as described above. All the bright stars were saturated (see Fig.~\ref{fig:20220823images}), so two unsaturated comparison stars were used. The optimal aperture diameter was 23 pixels. Due to the noisy nature of the data, the data were binned by an average of 14 points to match the SNR=9 of the LT curve. The binned data are overplotted in Fig.~\ref{fig:20220823LC}. Due to the extremely low elevation and weather conditions in the days before and after the event, well-separated data were not available, so the background fraction was also set at 0.886.

\subsubsection{2023 July 17} \label{subsubsec:lc20230717} 
For the SAAO data, master biases and flats were created and used to bias subtract and flatfield the raw data. Circular-aperture photometry was carried out as described above. The optimal aperture had a diameter of 11 pixels (8.4 arcsec). A streak is apparent in Fig.~\ref{fig:20230717images} extending from the saturated bright star located south of Pluto, due to residual flux during frame transfer. Background selections were made to avoid this streak.

Observations of Pluto and the star when they were separated from each other were taken between one and three hours before (337 images) and after (225 images) the occultation. The light ratio fit returned a background fraction at the airmass of the event midtime of $0.967\pm0.006$. 

\begin{figure}[ht!]
\centering
\includegraphics[width=0.4\textwidth,clip,trim=0mm 0mm 0mm 0mm]{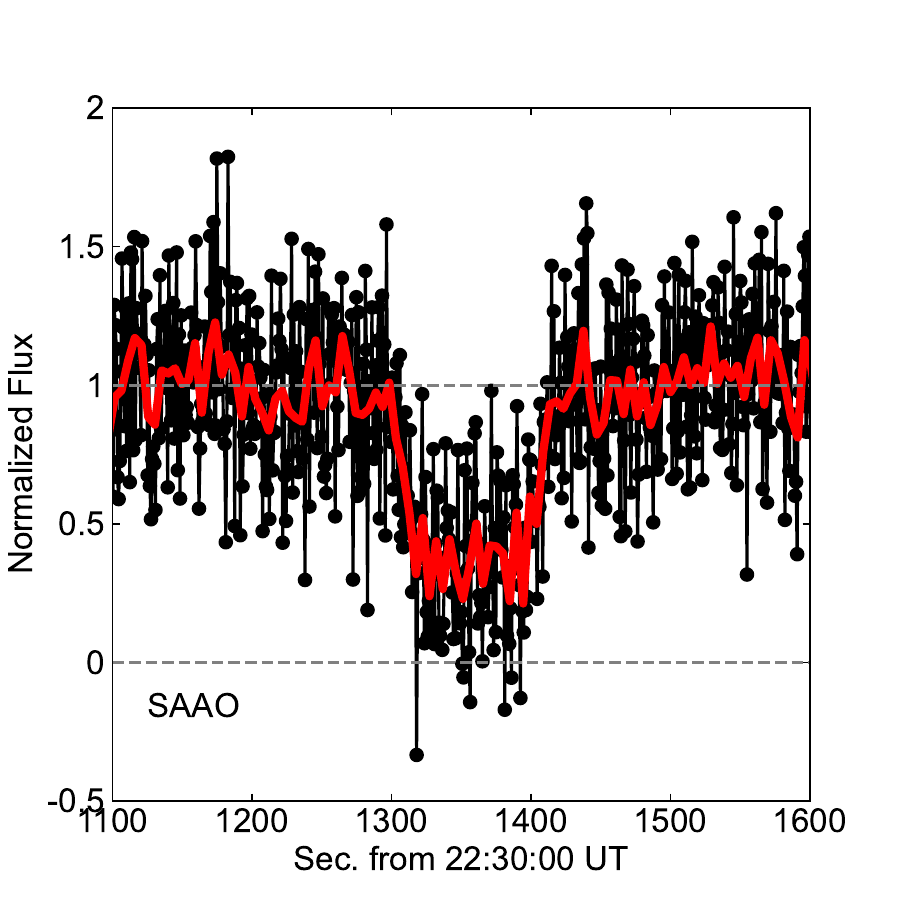}
\caption{Light curve from the 2023 July 17 occultation. Each black dot represents one data point. The red line shows the data binned by six points, to better see the light-curve structure. Dashed gray lines are provided for reference at zero and one flux levels. \label{fig:20230717LC}}
\end{figure}

The light curve is shown in Fig.~\ref{fig:20230717LC}. Because the data are noisy, a binned average of six points has also been plotted to better see the light-curve structure. The SNR per scale height for the binned data is 11.

\subsection{Fits to model atmospheres} \label{subsec:models}

To fit the data, we started by confirming that the events can be modeled with Fresnel diffraction. The spatial resolution is unique to each observation and is based on the angular size of the star and the Fresnel scale length, $\sqrt{\lambda D/2}$, where $\lambda$ is the wavelength of the observation and $D$ is the distance from the observer to Pluto. These scales are provided in Table~\ref{tab:scales}. The observations are in the regime of Fresnel diffraction when the star diameter is smaller than the Fresnel length and the spatial resolution of the observation is larger, which is the case for all data here excluding the LDT observations on 2018 April 09, which are particularly low SNR (see Fig.~\ref{fig:20180409LC}).

\begin{deluxetable*}{lccc}
%\tablenum{4}
\tablecaption{Relevant occultation distance scales.\label{tab:scales}}
\tablewidth{0pt}
\tablehead{
\colhead{Date} & \colhead{Fresnel Length$^a$} & \colhead{Maximum Stellar Angular Diameter$^b$} & \colhead{Minimum Integration$^c$}\\
\colhead{(UT)} & \colhead{(km)} & \colhead{(km)} & \colhead{(km)}
}
\startdata
2017 August 07  & 1.30 & 0.36 (giant)& 52.8 \\
2018 April 09 & 1.32 & 0.20& 0.6\\
2018 August 15 & 1.31 &0.49& 1.9\\
2018 October 01 & 1.32 & 0.16& 4.3\\
2018 November 01 & 1.33 & 0.42 & 4.8\\
2018 November 20 & 1.34 & 0.50& 178.9\\
2021 August 06 & 1.34 & 0.13 & 11.4 \\
2022 June 01 & 1.33 & 1.16 (giant)& 11.2 \\
2022 August 23 & 1.33& 0.15& 29.9\\
2023 July 17 & 1.33& 0.29$^d$ & 19.4\\
\enddata
\tablecomments{$^a$Assuming observational wavelength of 700~nm and the Earth-Pluto distances at the midtimes of the occultations from JPL Horizons. The exact observational wavelength for each dataset is based on the detector response, the stellar profile, and the transmission of the optics and atmosphere as functions of wavelength. $^b$Based on the stellar sizes and minimum distances in the TESS Input Catalog, including the error bar on distance \citep{RN3976}. All of the stars are designated as luminosity class dwarf except for 2022 June 01 and 2017 August 07, which are designated giant.$^c$Using the geocentric velocities listed in Table \ref{tab:star}.$^d$No size is given in the TESS catalog for this star. We assume the maximum size is that of the Sun, in line with the sizes of the other dwarf stars in this table. As a check, the projected stellar diameter is only 0.12 km using $B$ and $K$ magnitudes from the NOMAD catalog and the main-sequence equations from \citet{RN2265}.}
\end{deluxetable*}

For atmospheric fitting to the occultation data, we use an analytic model based on \citet{RN1023} and \citet{RN2787} and as used in previous work \citep[e.g.][]{RN3171, RN3641, RN3616,RN3469}. This model allows for a constant thermal gradient throughout the atmosphere with the optional addition of a haze layer having a defined turn-on radius for the top of the haze as well as a haze scale height. Derived model parameters include the half-light radius in Pluto’s atmosphere $r_h$, half-light pressure $P_h$, half-light temperature $T_h$, temperature gradient $dT/dr$, the thermal gradient exponent $b$, and the ratio of gravitational to thermal energy $\lambda_h$. Atmospheric model parameters include the shadow radius $\rho_h$ and atmospheric pressure scale height $H_p$. The temperature as a function of radius $r$ from the center of Pluto is defined as $T(r)=T_h(r/r_h)^b$: for an isothermal atmosphere $b=0$. For haze models, we have the additional variables of the radii at which the haze optical depth is 0 and 1, $r_{\tau0}$  and $r_{\tau1}$, as well as the haze scale height, $H_\tau$.

While fitting the datasets, we realized that the model fits are more stable when recontextualized to make the pressure scale height at half light $H_p$ the primary fitted quantity rather than the thermal energy ratio $\lambda_h$ used previously. This stability is likely due to the interconnectedness of $\lambda_h$, $b$, and the half-light radius $r_h$. As pressure scale height and half-light radius are the quantities in which we are primarily interested, we inverted the model using the formulation of \citet{RN1023} and calculated a new $\lambda_h$ at each step (based upon the instantaneous values of $b$ and $r_h$), while treating $H_p$ as the fitted quantity. This method resulted in stable fitting, and the scale height did not oscillate between iterations as in previous works \citep[e.g.][]{RN3641}.  

The background fractions for each of the light curves can also be fitted by the model. As noted in \S~\ref{subsec:lc}, for multi-site light curves that did not have sufficient data to calculate a background fraction, we used the best-fit values from initial atmospheric models for a hazy atmosphere and then reran the fits with the updated background fractions. The light curves were fit simultaneously to generate preliminary solutions for $r_h$ and the event geometry. We fit for the event geometry using $(f_0,g_0)$, which is the offset between the predicted and observed locations of Pluto in the plane of the sky, in the \textit{fgh} coordinate system \citep[c.f. Section 5.5 of][]{RN1191}. The \textit{fgh} system is defined to be in the shadow plane and to originate at the center of the planet's shadow: \textit{f} is positive in the direction of increasing right ascension, \textit{g} is positive in the direction of in increasing declination, and \textit{h} is positive in the direction of the occulted star. Single-chord events had fixed $(f_0,g_0)$ (see the next paragraph). After these initial, global fits, the atmospheric parameters were fit while being carefully monitored to ensure convergence. The closest approach values for each station were calculated based on the best fits for $(f_0,g_0)$ and the known station separation distances (see Table \ref{tab:sites}). 

The event geometry plays an important role in fitting atmospheric models to light curves. There is a degeneracy in fitting for $(f_0,g_0)$ and the shadow radius: for an occultation with a single chord, these parameters do not converge to a single solution to match the data. For example, a light curve could be equally well fit by a chord located closer to Pluto's center with a smaller atmosphere or a chord farther away with a bigger atmosphere. While a single chord can be used to constrain Pluto's atmospheric properties, multiple chords are needed to provide tighter constraints. Additionally, astrometric data before and after the occultation can be interpolated to provide an independent measure of the closest approach distance; however, this method often leads to large uncertainties \citep[e.g. Fig. 8 in][]{RN3689}. As indicated in Table~\ref{tab:scopes}, closest approach distances for this work were calculated after the occultations using NIMA and the GAIA DR3 star catalog. These postdiction values can be compared to the fitted close-approach values for multi-chord events below. Note that the fitted $(f_0,g_0)$ values are with respect to stellar reference positions that we used at the times of planning the observations, and they do not include any ephemeris corrections.

Discussions of Pluto's atmosphere have typically been split into upper and lower, with the dividing point being roughly at the occultation half-light level. The structure of the upper part of the atmosphere has been fairly constant over time and is generally well-matched by isothermal, clear models \citep[noting that a trend of slight cooling with altitude has been reported, e.g.][]{RN3366, RN3757}. The lower part of the atmosphere requires a steep temperature gradient and/or haze and its observed profile has evolved temporally \citep[e.g.][]{RN3366}. For these reasons, some previous modeling efforts have considered only the upper atmosphere when studying bulk properties \citep[e.g.][]{RN3171, RN3616}. Here, we consider clear atmosphere fits using only data above the half-light levels as well as hazy atmospheres fitted to the full datasets. 

The highest-SNR dataset from recent epochs, and the one with the most complete understanding of the star-planet geometry, is the SOFIA occultation observed in 2015, which was nearly coincident with and consistent with the \textit{New Horizons} dataset \citep{RN3641}. This dataset, comprising multiple high-SNR chords with well-constrained geometry, enabled full atmospheric modeling, including haze components and thermal gradients. Using the \citet{RN1023} model applied to the 2015 data, the thermal-gradient (power-law) exponent, $b$, was fitted to a value of $-2.3 \pm 0.2$. For the present work, this parameter was fixed at $b=-2.3$ to enable more direct comparisons across epochs and to permit the inclusion of lower-SNR datasets, for which fitting $b$ is not feasible. Determination of $b$ depends critically on differential slope measurements through the steepest portion of the occultation light curve, precisely the information that is first lost as photometric SNR decreases.

Occultation data do not probe all the way to Pluto's surface. Therefore, we provide the pressure at half-light level, $P_h$, as well as at the reference height of 1275~km from the center, $P_{1275}$. To facilitate comparison with other publications, we also report pressure at 1250 km $P_{1250}$, and 1215 km, $P_{1215}$. The pressure as a function of height depends on the detailed atmospheric structure (especially temperature gradients) and whether or not haze is included --- factors that are thought to be significant and evolving below half-light level. Therefore, extrapolation of pressure values to the surface, $P_{\text{surface}}$, and even to 1215~km, requires assumptions about atmospheric structure and is significantly less reliable than measuring $P_h$ and extrapolating to $P_{1275}$. For example, \citet{RN3741} assumes a fixed-profile atmosphere to convert pressures at 1215~km to surface pressures, $P_{\text{surface}}=1.84 P_{1215}$, rather than integrating through the full atmosphere. This simple conversion can be used to compare pressures that have been reported at different heights, but we must be aware of the assumptions involved.

\section{Results} \label{sec:results}
\subsection{Atmospheric fits for multi-site observations} \label{subsec:multichordatmfits}

%We started by fitting data from the multi-site events, given the degeneracy in geometry and atmospheric size described above. 
Four of the events reported here have data from multiple sites. We fit the data simultaneously for each of these events, considering four different cases: (i) a clear atmosphere with fixed parameters $H_p$ and \textit{b}, fit to data only above 0.5 flux; (ii) a clear atmosphere fit to data only above 0.5 flux with fixed \textit{b}; (iii) an atmosphere with a layer of haze and fixed parameters $H_p$ and \textit{b}; (iv) and an atmosphere with haze and fixed \textit{b}. We include a fixed atmosphere because fitting for \textit{b} is not feasible for low-SNR data and to allow direct comparison with previously published results. Data from multiple sites can be sufficient to break the degeneracy between geometry and atmospheric size described above, if they are far enough apart. The most accurate results are based on data from sites both north and south of the centerlines (which we do not have for any of our datasets, see Figs.~\ref{fig:shadowpaths} and \ref{fig:globes}). 

\subsubsection{Atmospheric fits: 2018 August 15}\label{subsubsec:20180815atmfits}

Observations of 2018 August 15 were previously published in \citet{RN3990}, and we have included the highest-quality light curve from that publication in our atmospheric fits, the Andor iXon light curve from the 2.1-m telescope at the Sierra San Pedro M\'{a}rtir Observatory (OAN-SPM) (as shown in Table~\ref{tab:scopes} and Fig.~\ref{fig:20180815LC}). Inclusion of the OAN-SPM data was particularly useful in order to have chords that spanned more of Pluto. The resulting best-fit model parameters are listed in Table~\ref{tab:20180815fits}. The hundreds of well-separated images taken at SRO allowed calculation of background fraction and closest approach distance; however, the location at the time of the event could not be derived with sufficient accuracy to determine a usable closest approach, so that value is fitted. 

\begin{figure}[ht!]
\centering
\includegraphics[width=1\textwidth,clip,trim=0mm 0mm 0mm 0mm]{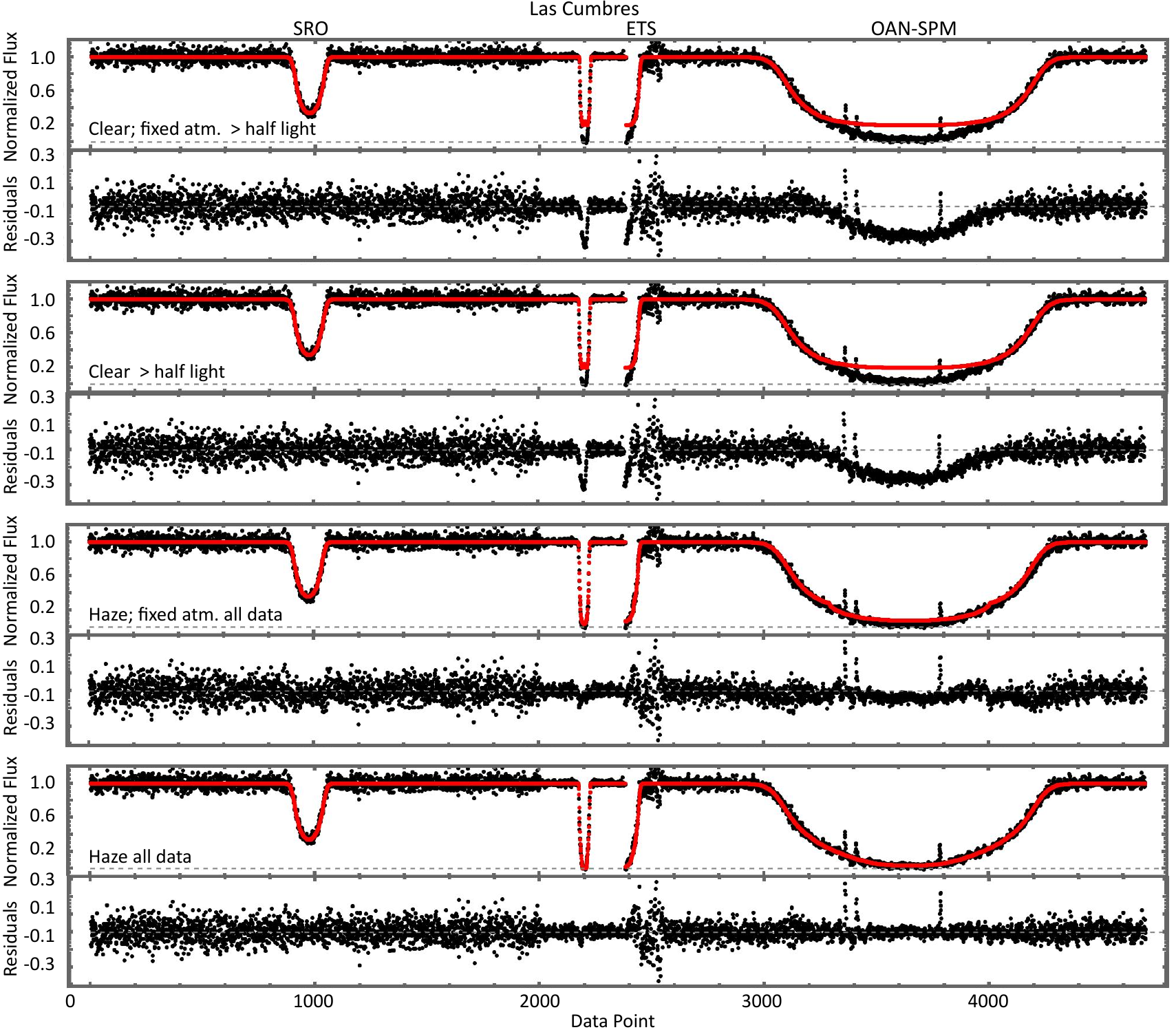}
\caption{Atmospheric fits for the 2018 August 15 dataset. The occultation data are plotted as black points and the best-fit models are shown as red lines, with the residuals plotted below each fit. The station names are noted at the top of the figure. The model parameters are listed in Table \ref{tab:20180815fits}. A clear, isothermal atmosphere does not accurately reproduce the flux at the bottom of the light curve, while an atmosphere with haze returns the best fit overall. The three spikes discussed in \S~\ref{subsec:spikes} are apparent. The results for each of the multi-site events show similar trends, and these data are provided as an example of the quality of fits. \label{fig:20180815fits}}
\end{figure}

As an example of the atmospheric fitting technique, the light curve data are shown along with the best-fit models in Fig.~\ref{fig:20180815fits}. The clear models do not accurately match the lowest part of Pluto's atmosphere (as they do not in any of the following analyses), indicating the presence of haze and/or a steep thermal gradient. Our simple exponential haze layer model provides a good match to all the data points, although we note that the steep thermal gradient for the atmospheric temperature to reach that of the surface is not included in this model (see discussion in \S~\ref{subsec:spikes}).

\begin{deluxetable*}{lcccc}
\tablecaption{Best-fit model atmosphere parameters for 2018 August 15.}
\tablewidth{0pt}
\tablehead{ \colhead{}  & \colhead{Clear; fixed atm.$^a$} & \colhead{Clear} & \colhead{Haze; fixed atm.$^a$} & \colhead{Haze}  \\
\colhead{Fit parameter}  & \colhead{$>$ half light} & \colhead{$>$ half light} &\colhead{all data} & \colhead{all data} 
\label{tab:20180815fits}}
\startdata
 \text{Reduced $\chi^2$} & 0.98 & 0.98 & 1.16 & 0.93 \\
 \text{SRO CA$^b$ (km)} & \text{1098.4 $\pm$ 2.9} & \text{1098.4 $\pm$ 2.9} & \text{1098.4 $\pm$ 2.9} & \text{1098.4 $\pm$ 2.9} \\
 \text{LCO CA$^b$ (km)} & \text{357.7 $\pm$ 2.9} & \text{357.7 $\pm$ 2.9} & \text{357.7 $\pm$ 2.9} & \text{357.7 $\pm$ 2.9} \\
 \text{ETS CA$^b$ (km)} & \text{633.9 $\pm$ 2.9} & \text{633.9 $\pm$ 2.9} & \text{633.9 $\pm$ 2.9} & \text{633.9 $\pm$ 2.9} \\
 \text{OAN-SPM CA$^b$ (km)} & \text{661.1 $\pm$ 2.9} & \text{661.1 $\pm$ 2.9} & \text{661.1 $\pm$ 2.9} & \text{661.1 $\pm$ 2.9} \\
 \text{Shadow Radius, $\rho_h$} & \text{1207.9 $\pm$ 1.0} & \text{1199.6 $\pm$ 2.9} & \text{1202.9 $\pm$ 0.5} & \text{1220.1 $\pm$ 1.2} \\
 \text{Pressure at half-light, $P_h$ ($\mu$bar)} & \text{1.36 $\pm$ 0.01} & \text{1.55 $\pm$ 0.07} & \text{1.41 $\pm$ 0.01} & \text{1.44 $\pm$ 0.03} \\
 \text{Pressure at 1215 km, $P_{1215}$ ($\mu $bar)} & \text{5.50 $\pm$ 0.09} & \text{5.70 $\pm$ 0.11} & \text{4.01 $\pm$ 0.03} & \text{5.61 $\pm$ 0.10} \\
 \text{Pressure at 1250 km, $P_{1250}$ ($\mu $bar)} & \text{2.86 $\pm$ 0.05} & \text{3.05 $\pm$ 0.08} & \text{2.08 $\pm$ 0.02} & \text{2.95 $\pm$ 0.07} \\
 \text{Pressure at 1275 km, $P_{1275}$ ($\mu $bar)} & \text{1.79 $\pm$ 0.03} & \text{1.95 $\pm$ 0.06} & \text{1.29 $\pm$ 0.01} & \text{1.86 $\pm$ 0.05} \\
 \text{Half-light temperature, $T_h$ (K)} & \text{93.1 $\pm$ 0.1} & \text{97.8 $\pm$ 1.5} & \text{96.0 $\pm$ 0.1} & \text{95.1 $\pm$ 0.8} \\
 \text{Temperature gradient, $dT/dr$ (K/km)} & \text{-0.17 $\pm$ 0.01} & \text{-0.17 $\pm$ 0.01} & \text{-0.17 $\pm$ 0.01} &\text{-0.17 $\pm$ 0.01} \\
 \text{Half-light radius, $r_h$ (km)} & \text{1289.6 $\pm$ 1.0} & \text{1287.6 $\pm$ 1.3} & \text{1270.3 $\pm$ 0.5} & \text{1289.0 $\pm$ 0.9} \\
 \text{Pressure Scale Height, $H_P$ (km)} & 52.8 & \text{55.3 $\pm$ 0.8} & 52.8 & \text{53.9 $\pm$ 0.4} \\
 \text{Thermal gradient exponent, $b$ } & -2.3 & -2.3 & -2.3 & -2.3 \\
 \text{Haze onset radius (km)} & \nodata & \nodata & 1200 & \text{1247.3 $\pm$ 4.7} \\
 \text{Haze radius at $\tau=1$ (km)} & \nodata & \nodata &  1150 & \text{1176.7 $\pm$ 0.8} \\
 \text{Haze scale height (km)} & \nodata &  \nodata & 50 & \text{19.3 $\pm$ 0.8} \\
 \text{$f_0^{c}$ (km)} & \text{-1570.8 $\pm$ 2.0} & \text{-1570.8 $\pm$ 2.0} & \text{-1570.8 $\pm$ 2.0} & \text{-1570.8 $\pm$ 2.0} \\
 \text{$g_0^{c}$ (km)} & \text{308.0 $\pm$ 2.8} & \text{308.0 $\pm$ 2.8} & \text{308.0 $\pm$ 2.8} & \text{308.0 $\pm$ 2.8} \\
 \enddata
\tablecomments{$^a$Assuming fixed atmosphere values from \cite{RN3641}. $^b$CA – closest approach distance based on model fit. $^c$With respect to stellar reference positions at the times of planning the observations, neglecting ephemeris corrections.}

\end{deluxetable*}

\subsubsection{Atmospheric fits: 2021 August 06} \label{subsubsec:20210806fits}

For the 2021 August 06 dataset, background fractions were determined at each site and no initial fitting was required (see \S~\ref{subsubsec:lc20210806}). However, as can be seen in Table~\ref{tab:scopes}, the SNRs for these light curves were significantly lower than the data from 2018 August 15. Note that we did not include the SARA-CT data in the fits due to the unknown timing offset and the low SNR. In addition, all the chords grazed the atmosphere, not reaching below $\sim0.5$ flux, and they were on the same side of Pluto (see discussion of geometry in \S~\ref{subsec:models}). Thus, there were difficulties in the atmospheric fitting for this event and the results were poorly constrained. In fact, fitting this dataset is what forced the recontextualization of the model from $\lambda_h$ to $H_p$ as described in \S~\ref{subsec:models}.  The atmospheric-fit results are shown in Table~\ref{tab:20210806fits}. There are larger error bars for values such as $r_h$, $3-12 ~\text{km}$, compared to better than $\sim1 \text{km}$ for 2018 August 15 (see Table~\ref{tab:20180815fits}). Nonetheless, notable values such as $H_p$ are stable between fits, with smaller error bars overall, than when fitting using the previous $\lambda_{\text{h}}$-based fitting methodology. %The fitted closest approach values are consistent with that measured for Magellan of $1113\pm387 \, \text{km}$ from well-separated images taken before and after the event.

\begin{deluxetable*}{lcccc}[h]
\tablecaption{Best-fit model atmosphere parameters for 2021 August 06.}
\tablewidth{0pt}
\tablehead{ \colhead{}  & \colhead{Clear; fixed atm.$^a$} & \colhead{Clear} & \colhead{Haze; fixed atm.$^a$} & \colhead{Haze}  \\
\colhead{Fit parameter}  & \colhead{$>$ half light} & \colhead{$>$ half light} &\colhead{all data} & \colhead{all data} 
\label{tab:20210806fits}}
\startdata
 \text{Reduced $\chi^2$} & 0.81 & 0.80 & 0.81 & 0.82 \\
 \text{Magellan CA$^b$ (km)} & \text{1218.9 $\pm$ 3.7} & \text{1218.9 $\pm$ 3.7} & \text{1218.9 $\pm$ 3.7} & \text{1218.9 $\pm$ 3.7} \\
 \text{Gemini CA$^b$ (km)} & \text{1353.7 $\pm$ 3.7} & \text{1353.7 $\pm$ 3.7} & \text{1353.7 $\pm$ 3.7} & \text{1353.7 $\pm$ 3.7} \\
 \text{LCO-LSC CA$^b$ (km)} & \text{1344.9 $\pm$ 3.7} & \text{1344.9 $\pm$ 3.7} & \text{1344.9 $\pm$ 3.7} & \text{1344.9 $\pm$ 3.7} \\
 \text{Shadow Radius, $\rho_h$} & \text{1213.4 $\pm$ 3.6} & \text{1181.2 $\pm$ 11.0} & \text{1227.9 $\pm$ 3.6} & \text{1234.8 $\pm$ 21.0} \\
 \text{Pressure at half-light, $P_h$ ($\mu$bar)}  & \text{1.32 $\pm$ 0.01} & \text{2.15 $\pm$ 0.28} & \text{1.32 $\pm$ 0.01} & \text{1.40 $\pm$ 0.46} \\
 \text{Pressure at 1215 km, $P_{1215}$ ($\mu $bar)}& \text{5.90 $\pm$ 0.34} & \text{6.96 $\pm$ 0.50} & \text{5.90 $\pm$ 0.34} & \text{7.09 $\pm$ 0.62} \\ 
 \text{Pressure at 1250 km, $P_{1250}$ ($\mu $bar)}& \text{3.07 $\pm$ 0.18} & \text{4.00 $\pm$ 0.36} & \text{3.07 $\pm$ 0.18} & \text{3.76 $\pm$ 0.32} \\
 \text{Pressure at 1275 km, $P_{1275}$ ($\mu $bar)}& \text{1.92 $\pm$ 0.11} & \text{2.68 $\pm$ 0.28} & \text{1.92 $\pm$ 0.11} & \text{2.38 $\pm$ 0.27} \\ 
 \text{Half-light temperature, $T_h$ (K)}& \text{92.4 $\pm$ 0.5} & \text{110.1 $\pm$ 5.1} & \text{92.4 $\pm$ 0.5} & \text{93.7 $\pm$ 11.6} \\
 \text{Temperature gradient, $dT/dr$ (K/km)} & \text{-0.16 $\pm$ 0.01} & \text{-0.20 $\pm$ 0.02} & \text{-0.16 $\pm$ 0.01} & \text{-0.17 $\pm$ 0.02} \\
 \text{Half-light radius, $r_h$ (km)} & \text{1295.0 $\pm$ 3.5} & \text{1288.9 $\pm$ 4.5} & \text{1295.0 $\pm$ 3.5} & \text{1304.0 $\pm$ 11.7} \\
 \text{Pressure Scale Height, $H_P$ (km)} & 52.8 & \text{62.3 $\pm$ 2.6} & 52.8 & \text{54.3 $\pm$ 5.8} \\
 \text{Thermal gradient exponent, $b$ } & -2.3 & -2.3 & -2.3 & -2.3 \\
 \text{Haze onset radius (km)} & \nodata & \nodata & 1200 & \text{1358.2 $\pm$ 17.3} \\
 \text{Haze radius at $\tau=1$ (km)}  & \nodata & \nodata & 1150 & \text{1175.2 $\pm$ 191.3} \\
 \text{Haze scale height (km)} & \nodata & \nodata & 50 & \text{61.9 $\pm$ 92.5} \\
 \text{$f_0^c$ (km)}& \text{686.6 $\pm$ 11.2} & \text{686.6 $\pm$ 11.2} & \text{686.6 $\pm$ 11.2} & \text{686.6 $\pm$ 11.2} \\
 \text{$g_0^c$ (km)} & \text{438.7 $\pm$ 4.5} & \text{438.7 $\pm$ 4.5} & \text{438.7 $\pm$ 4.5} & \text{438.7 $\pm$ 4.5} \\
\enddata
\tablecomments{$^a$Assuming fixed atmosphere values from \citep{RN3641}. $^b$CA – closest approach distance based on model fit. $^c$With respect to stellar reference positions at the times of planning the observations, neglecting ephemeris corrections.}
\end{deluxetable*}

\subsubsection{Atmospheric fits: 2022 June 01} \label{subsubsec:20220601fits}

Results for the atmospheric fits for 2022 June 01 are listed in Table~\ref{tab:20220601fits}. Initial fits for all data with haze were used to determine the background fractions for EOS and all light curves from Savannah Skies, excluding that from POETS (as noted in \S~\ref{subsubsec:lc20220601}). The results have error bars more similar to 2018 August 15 than 2021 August 06 as a result of the high SNR data (see Table~\ref{tab:scopes}) and the two successful sites being full occultations located more centrally on Pluto.% The well-separated images from POETS were sufficient for deriving a background fraction but did not return a usable closest approach.

\begin{deluxetable*}{lcccc}
\tablecaption{Best-fit model atmosphere parameters for 2022 June 01.}
\tablewidth{0pt}
\tablehead{ \colhead{}  & \colhead{Clear; fixed atm.$^a$} & \colhead{Clear} & \colhead{Haze; fixed atm.$^a$} & \colhead{Haze}  \\
\colhead{Fit parameter}  & \colhead{$>$ half light} & \colhead{$>$ half light} &\colhead{all data} & \colhead{all data} 
\label{tab:20220601fits}}
\startdata
 \text{Reduced $\chi^2$} & 1.14 & 1.13 & 1.28 & 1.11 \\
 \text{POETS CA$^b$ (km)} & \text{455.0 $\pm$ 4.1} & \text{455.0 $\pm$ 4.1} & \text{455.0 $\pm$ 4.1} & \text{455.0 $\pm$ 4.1} \\
 \text{EOS CA$^b$ (km)} & \text{799.9 $\pm$ 4.1} & \text{799.9 $\pm$ 4.1} & \text{799.9 $\pm$ 4.1} & \text{799.9 $\pm$ 4.1} \\
 \text{Shadow Radius, $\rho_h$} & \text{1202.1 $\pm$ 1.6} & \text{1179.0 $\pm$ 5.4} & \text{1198.9 $\pm$ 0.8} & \text{1221.2 $\pm$ 1.5} \\
 \text{Pressure at half-light, $P_h$ ($\mu$bar)} & \text{1.33 $\pm$ 0.01} & \text{1.85 $\pm$ 0.12} & \text{1.38 $\pm$ 0.01} & \text{1.29 $\pm$ 0.04} \\
 \text{Pressure at 1215 km, $P_{1215}$ ($\mu $bar)} & \text{4.86 $\pm$ 0.12} & \text{5.26 $\pm$ 0.16} & \text{3.63 $\pm$ 0.05} & \text{5.22 $\pm$ 0.14} \\
 \text{Pressure at 1250 km, $P_{1250}$ ($\mu $bar)} & \text{2.53 $\pm$ 0.06} & \text{2.92 $\pm$ 0.11} & \text{1.89 $\pm$ 0.03} & \text{2.70 $\pm$ 0.09} \\
 \text{Pressure at 1275 km, $P_{1275}$ ($\mu $bar)} & \text{1.58 $\pm$ 0.04} & \text{1.91 $\pm$ 0.09} & \text{1.17 $\pm$ 0.02} & \text{1.68 $\pm$ 0.06} \\
 \text{Half-light temperature, $T_h$ (K)} & \text{93.9 $\pm$ 0.2} & \text{105.9 $\pm$ 2.5} & \text{96.6 $\pm$ 0.1} & \text{92.7 $\pm$ 0.9} \\
 \text{Temperature gradient, $dT/dr$ (K/km)} & \text{-0.17 $\pm$ 0.01} & \text{-0.19 $\pm$ 0.01} & \text{-0.17 $\pm$ 0.01} & \text{-0.17 $\pm$ 0.01} \\
 \text{Half-light radius, $r_h$ (km)}  & \text{1284.0 $\pm$ 1.5} & \text{1277.1 $\pm$ 2.2} & \text{1266.4 $\pm$ 0.8} & \text{1288.9 $\pm$ 1.3} \\
 \text{Pressure Scale Height, $H_P$ (km)} & 52.8 & \text{58.9 $\pm$ 1.3} & 52.8 & \text{52.5 $\pm$ 0.5} \\
 \text{Thermal gradient exponent, $b$ } & -2.3 & -2.3 & -2.3 & -2.3 \\
 \text{Haze onset radius (km)} & \nodata & \nodata & 1200 & \text{1214.4 $\pm$ 1.2} \\
 \text{Haze radius at $\tau=1$ (km)} & \nodata & \nodata & 1150 & \text{1184.6 $\pm$ 1.6} \\
 \text{Haze scale height (km)} & \nodata & \nodata & 50 & \text{129.5 $\pm$ 33.0} \\
 \text{$f_0^c$ (km)} & \text{741.1 $\pm$ 2.6} & \text{741.1 $\pm$ 2.6} & \text{741.1 $\pm$ 2.6} & \text{741.1 $\pm$ 2.6} \\
 \text{$g_0^c$ (km)} & \text{415.9 $\pm$ 3.9} & \text{415.9 $\pm$ 3.9} & \text{415.9 $\pm$ 3.9} & \text{415.9 $\pm$ 3.9} \\
\enddata
\tablecomments{$^a$Assuming fixed atmosphere values from \cite{RN3641}. $^b$CA – closest approach distance based on model fit. $^c$With respect to stellar reference positions at the times of planning the observations, neglecting ephemeris corrections.}
\end{deluxetable*}

\subsubsection{Atmospheric fits: 2022 August 23} \label{subsubsec:20220823fits}

2022 August 23 was challenging to fit as a result of the low SNR of the data (see Table~\ref{tab:scopes}) and the event geometry. We proceeded as before with clear atmosphere fits using only data above the $50\%$ flux level and haze fits using all the data, and with the atmosphere fixed as well as allowing the atmosphere to be free. The fits have significant error bars, with $r_h$ having errors of $\approx10 \text{--} 30$ km. Even so, the reduced $\chi^2$ values for all fits are acceptable, indicating that the dispersions caused by the low SNRs are appropriately modeled by the error bars. Therefore, with the error bars included, the results for this event may be treated as reliable (even if not particularly useful).

\begin{deluxetable*}{lcccc}
\tablecaption{Best-fit model atmosphere parameters for 2022 August 23.}
\tablewidth{0pt}
\tablehead{ \colhead{}  & \colhead{Clear; fixed atm.$^a$} & \colhead{Clear} & \colhead{Haze; fixed atm.$^a$} & \colhead{Haze}  \\
\colhead{Fit parameter}  & \colhead{$>$ half light} & \colhead{$>$ half light} &\colhead{all data} & \colhead{all data} 
\label{tab:20220823fits}}
\startdata
 \text{Reduced $\chi^2$} & 1.07 & 1.07 & 1.08 & 1.07 \\
 \text{LT CA$^b$ (km)} & \text{508.5 $\pm$ 27.7} & \text{508.5 $\pm$ 27.7} & \text{508.5 $\pm$ 27.7} & \text{508.5 $\pm$ 27.7} \\
 \text{T1T CA$^b$ (km)} & \text{1172.42 $\pm$ 27.7} & \text{1172.42 $\pm$ 27.7} & \text{1172.42 $\pm$ 27.7} & \text{1172.42 $\pm$ 27.7} \\
 \text{Shadow Radius, $\rho_h$} & \text{1183.2 $\pm$ 19.1} & \text{1153.2 $\pm$ 72.5} & \text{1236.6 $\pm$ 9.9} & \text{1215.6 $\pm$ 15.8} \\
 \text{Pressure at half-light, $P_h$ ($\mu$bar)} & \text{1.39 $\pm$ 0.06} & \text{2.12 $\pm$ 1.73} & \text{1.28 $\pm$ 0.03} & \text{1.34 $\pm$ 0.05} \\
 \text{Pressure at 1215 km, $P_{1215}$ ($\mu $bar)} & \text{3.63 $\pm$ 1.10} & \text{4.28 $\pm$ 1.80} & \text{6.74 $\pm$ 1.08} & \text{4.80 $\pm$ 1.23} \\
 \text{Pressure at 1250 km, $P_{1250}$ ($\mu $bar)} & \text{1.88 $\pm$ 0.58} & \text{2.41 $\pm$ 1.30} & \text{3.51 $\pm$ 0.57} & \text{2.50 $\pm$ 0.64} \\
 \text{Pressure at 1275 km, $P_{1275}$ ($\mu $bar)} & \text{1.17 $\pm$ 0.36} & \text{1.60 $\pm$ 1.01} & \text{2.20 $\pm$ 0.36} & \text{1.56 $\pm$ 0.40} \\
 \text{Half-light temperature, $T_h$ (K)} & \text{96.7 $\pm$ 2.8} & \text{112.5 $\pm$ 33.1} & \text{91.2 $\pm$ 1.4} & \text{94.1 $\pm$ 2.3} \\
 \text{Temperature gradient, $dT/dr$ (K/km)} & \text{-0.2 $\pm$ 0.1} & \text{-0.2 $\pm$ 0.1} & \text{-0.2 $\pm$ 0.1} & \text{-0.2 $\pm$ 0.1} \\
 \text{Half-light radius, $r_h$ (km)} & \text{1265.9 $\pm$ 18.4} & \text{1257.8 $\pm$ 29.0} & \text{1303.5 $\pm$ 9.8} & \text{1282.9 $\pm$ 15.5} \\
 \text{Pressure Scale Height, $H_P$ (km)} & 52.8 & \text{60.7 $\pm$ 15.8} & 52.8 & \text{61.0 $\pm$ 19.4} \\
 \text{Thermal gradient exponent, $b$ } & -2.3 & -2.3 & -2.3 & -2.3 \\
 \text{Haze onset radius (km)} & \nodata & \nodata & 1250 & \text{1243.8 $\pm$ 36.7} \\
 \text{Haze radius at $\tau=1$ (km)}  & \nodata & \nodata & 1200 & \text{1157.2 $\pm$ 20.9} \\
 \text{Haze scale height (km)} & \nodata & \nodata & 50 & \text{49.8 $\pm$ 45.1} \\
 \text{$f_0^c$ (km)} & \text{739.7 $\pm$ 19.1} & \text{739.7 $\pm$ 19.1} & \text{739.7 $\pm$ 19.1} & \text{739.7 $\pm$ 19.1} \\
 \text{$g_0^c$ (km)} & \text{366.9 $\pm$ 27.2} & \text{366.9 $\pm$ 27.2} & \text{366.9 $\pm$ 27.2} & \text{366.9 $\pm$ 27.2} \\
\enddata
\tablecomments{$^a$Assuming fixed atmosphere values from \cite{RN3641}. $^b$CA – closest approach distance based on model fit. $^c$With respect to stellar reference positions at the times of planning the observations, neglecting ephemeris corrections.}
\end{deluxetable*}

\subsection{Atmospheric fits for single-site observations} \label{subsubsec:singlechordatmfits}

As described in \S~\ref{subsec:models}, single-chord atmospheric fits are far more challenging than those with multiple sites. Without having more than one chord to allow the fitting to determine the geometric offset for the event, the $(f_0,g_0)$ values {\em must} be fixed for any reliable results. The extreme correlation between atmospheric radius and closest approach distance in a single chord event almost guarantees divergent fit results. However, the correlation is not total, and fitting a single chord without a fixed geometry {\em is} possible, but only for chords having SNR of at least $\sim250$ per scale height, a constraint that none of our data reached.

To deal with this problem, an outside geometric solution is necessary for single-chord fits. It is possible to derive a solution from astrometric data. However, events that result in a single chord are typically also those events with the fewest observational resources, frequently just a single telescope in the path taking only an hour or two of queued data.  These events often also involve fainter stars -- and for Pluto recently, crowded fields -- which leads to lower SNR light curves.

To provide a consistent set of $(f_0,g_0)$ values in order to stabilize the single-chord fits, we use the closest approach values from the postdictions (in Table \ref{tab:scopes}). As a check on their accuracy, we compared the postdiction closest approaches to the fitted values from the multi-chord solutions (the values in Tables~\ref{tab:20180815fits}, \ref{tab:20210806fits}, \ref{tab:20220601fits}, and \ref{tab:20220823fits}). These are two, independent methods to determine the closest approaches. The differences are between eight to twenty kilometers; the values diverge at $\sim55 \,\text{km}$ ($<3 \text{mas}$) for 2022 August 23, which has the lowest SNR data for the multi-chord events. Such differences can be due to a star position that is slightly off, and they are within expected GAIA catalog variation. For single-site events, we assumed the postdiction closest-approach values and carried out a sanity check by plotting the light curves versus distance from the shadow center in units of half-light radii (see \S~\ref{subsec:lowatm}). In these units, the overall atmospheric size should be consistent between datasets. %One single-chord event \textbf{(2017 August 07)} required a shift closer in closest approach of \textbf{70} km in order to be consistent. As the deviations between planetary ephemerides and the star positions are combined, they are difficult to disentagle. In this case, since all of the postdiction values are consistent but for the one outlier, it is likely a star position that is slightly off. A \textbf{70} km shift at the distance of Pluto equates to a position error of $<5 \, \text{mas}$, within expected GAIA catalog variation.

%Note that the differences between postdicted and fitted clos\textbf{est} approaches are identical for all telescopes of a given event, as even photon noise can not cause the observation sites (and hence the chords) to move with respect to each other. While we did not have sufficient well-separated data to accurately determine the closest approaches for the majority of events, we got a value of $950\pm15 \,\text{km}$ on 2017 August 07, which is consistent with the postdiction values in Table \ref{tab:scopes} and demonstrates that the methodology can work.

% Updated postdiction 20170807 shifted that event so that it became consistent.
%Date   Site    Fitted  Postdicted Difference
%220823 LT  508  CA 453  (55 km)
%220601 POETS 455   468 (13 km)
%210806 Mag 1219    1201 (18 km)
%180815 SRO 1098   1106 (8 km)

Using the postdictions to fix the geometry of each single-chord event, the fits were stable. Rather than create individual tables for each single chord event, the results of the single-chord fits are given along with the fitted and derived atmospheric parameters from the multi-site occultations (for completeness) in Tables~\ref{tab:cleartable} and \ref{tab:hazetable} for clear and hazy atmospheres, respectively. 

\begin{longrotatetable}
\begin{deluxetable*}{lcccccccccl}
\tablecaption{Pluto's atmospheric characteristics from CLEAR model fits to the stellar occultation data in this work. \label{tab:cleartable}}
\tablewidth{0pt}
\tablehead{
\colhead{} &\colhead{}  & \colhead{}& \colhead{}&\colhead{} & \colhead{}& \colhead{}& \colhead{}& \colhead{}& \colhead{}& \colhead{Temp. Gradient}\\
\colhead{Date} & \colhead{Type} &\colhead{$\rho_h$} & \colhead{$r_h$}& \colhead{$H_p$}&\colhead{$P_h$}& \colhead{$P_{1215}$} & \colhead{$P_{1250}$}&\colhead{$P_{1275}$}&  \colhead{$T_h$}& \colhead{at $r_h$}\\
\colhead{(UT)} & \colhead{} &\colhead{(km)} & \colhead{(km)}&  \colhead{(km)} &\colhead{($\mu$bar)} &\colhead{($\mu$bar)} & \colhead{($\mu$bar)}&\colhead{($\mu$bar)}&   \colhead{(K)}& \colhead{(K/km)}
}
\startdata
\text{2017 August 07}&Single & \text{1276 $\pm$ 8}& \text{1356 $\pm$ 5} & \text{53 $\pm$ 3}& \text{1.2 $\pm$ 0.2}& \text{16.3 $\pm$ 1.3} & \text{8.5 $\pm$ 0.8}& \text{5.4 $\pm$ 0.6}& \text{84 $\pm$ 5}& \text{-0.14 $\pm$ 0.01} \\
\text{2018 April 09$^a$}&Single & \text{1268 $\pm$ 21}& \text{1374 $\pm$ 16}& \text{63 $\pm$ 6}& \text{1.9 $\pm$ 0.5} & \text{21.8 $\pm$ 5.0} & \text{12.8 $\pm$ 3.3}& \text{8.7 $\pm$ 2.4}& \text{98 $\pm$ 9} & \text{-0.16 $\pm$ 0.02} \\
\text{2018 August 15}&Multi & \text{1200 $\pm$ 3}& \text{1288 $\pm$ 1}& \text{55 $\pm$ 1}& \text{1.6 $\pm$ 0.1} & \text{5.7 $\pm$ 0.1} & \text{3.1 $\pm$ 0.1}& \text{1.9 $\pm$ 0.1}& \text{98 $\pm$ 1} & \text{-0.17 $\pm$ 0.01} \\
%\text{2018 October 01}&Single& \text{1292.8 $\pm$ 7.0} & \text{1365.3 $\pm$ 4.4} & \text{49.7 $\pm$ 2.5} & \text{0.95 $\pm$ 0.13} & \text{18.55 $\pm$ 1.44} & \text{9.36 $\pm$ 0.77} & \text{5.73 $\pm$ 0.54}& \text{77.9 $\pm$ 3.9} & \text{-0.13 $\pm$ 0.01} \\
\text{2018 October 01} & Single& \text{1220 $\pm$ 11} & \text{1301 $\pm$ 5} & \text{53 $\pm$ 3} & \text{1.3 $\pm$ 0.2} & \text{6.4 $\pm$ 0.6} & \text{3.3 $\pm$ 0.4} & \text{2.1 $\pm$ 0.3} & \text{91 $\pm$ 6} & \text{-0.16 $\pm$ 0.01} \\
%\text{2018 November 01}&Single& \text{1272.5 $\pm$ 3.8}& \text{1332.7 $\pm$ 2.5}& \text{43.3 $\pm$ 1.5}& \text{0.68 $\pm$ 0.06} & \text{9.95 $\pm$ 0.51}& \text{4.52 $\pm$ 0.27}& \text{2.56 $\pm$ 0.18} & \text{71.2 $\pm$ 2.5}& \text{-0.12 $\pm$ 0.01} \\
\text{2018 November 01} & Single& \text{1231 $\pm$ 5} & \text{1308 $\pm$ 2} & \text{51$\pm$ 2} & \text{1.1 $\pm$ 0.1} &\text{6.8 $\pm$ 0.3} & \text{3.4 $\pm$ 0.2} & \text{2.1 $\pm$ 0.2} & \text{87 $\pm$ 3} & \text{-0.15 $\pm$ 0.01} \\
\text{2018 November 20$^a$}&Single& \text{1090 $\pm$ 22}& \text{1168 $\pm$ 13} & \text{50 $\pm$ 8} & \text{1.4 $\pm$ 0.7}& \text{0.6 $\pm$ 0.4}& \text{0.3 $\pm$ 0.2}& \text{0.2 $\pm$ 0.2} & \text{106 $\pm$ 17}& \text{-0.21 $\pm$ 0.03} \\
\text{2021 August 06}&Multi& \text{1181 $\pm$ 11}  & \text{1289 $\pm$ 5} & \text{62 $\pm$ 3}& \text{2.2 $\pm$ 0.3}& \text{7.0 $\pm$ 0.5} & \text{4.0 $\pm$ 0.4}& \text{2.7 $\pm$ 0.3}& \text{110 $\pm$ 5}  & \text{-0.20 $\pm$ 0.01} \\
\text{2022 June 01}&Multi& \text{1179 $\pm$ 5}  & \text{1277 $\pm$ 2} & \text{59 $\pm$ 1}& \text{1.8 $\pm$ 0.1}& \text{5.3 $\pm$ 0.2} & \text{2.9 $\pm$ 0.1}& \text{1.9 $\pm$ 0.1}& \text{106 $\pm$ 3}  & \text{-0.19 $\pm$ 0.01} \\
%\text{2022 August 23}&Multi& \text{1173.8 $\pm$ 61.8}  & \text{1267.4 $\pm$ 25.0} & \text{57.1 $\pm$ 15.1}& \text{1.74 $\pm$ 1.40}& \text{4.33 $\pm$ 1.81} & \text{2.36 $\pm$ 1.27}& \text{1.52 $\pm$ 0.97}& \text{104.2 $\pm$ 30.5}  & \text{-0.19 $\pm$ 0.06} \\
\text{2022 August 23}&Multi& \text{1153 $\pm$ 73}  & \text{1259 $\pm$ 29} & \text{61 $\pm$ 16}& \text{2.1 $\pm$ 1.7}& \text{4.3 $\pm$ 1.8} & \text{2.4 $\pm$ 1.3}& \text{1.6 $\pm$ 1.0}& \text{112 $\pm$ 33}  & \text{-0.20 $\pm$ 0.10} \\ 
%\text{2023 July 17}&Single& \text{1297.9 $\pm$ 44.5}& \text{1353.7 $\pm$ 32.7}& \text{41.1 $\pm$ 18.0}& \text{0.57 $\pm$ 0.68}& \text{15.77 $\pm$ 11.45}& \text{6.90 $\pm$ 5.03}& \text{3.80 $\pm$ 3.18}& \text{65.6 $\pm$ 28.6}& \text{-0.11 $\pm$ 0.05} \\
\text{2023 July 17$^a$}&Single& \text{1228 $\pm$ 58} & \text{1337 $\pm$ 29} & \text{60 $\pm$ 17} & \text{1.7 $\pm$ 1.4} & \text{10.9 $\pm$ 5.2} & \text{6.1 $\pm$ 3.5} & \text{4.0 $\pm$ 2.7} & \text{99 $\pm$ 29} & \text{-0.17 $\pm$ 0.05} \\
\enddata
\tablecomments{$^a$These datasets are low SNR and are plotted in lighter shades in Figs. \ref{fig:shadowRad}-\ref{fig:masterpressureplot} See the discussion in \S~\ref{subsec:atmevolution}.}

\end{deluxetable*}

\end{longrotatetable} 

\begin{longrotatetable}
\footnotesize
\begin{deluxetable*}{lccccccccccccl}
\tablecolumns{14}
\tablecaption{Pluto's atmospheric characteristics from HAZE model fits to the stellar occultation data in this work. \label{tab:hazetable}}
\tablewidth{0pt}
\tablehead{
\colhead{} &\colhead{}  & \colhead{}& \colhead{}&\colhead{} & \colhead{}& \colhead{}& \colhead{}& \colhead{}& \colhead{}& \colhead{Temp. Gradient}& \colhead{}& \colhead{}& \colhead{}\\
\colhead{Date} & \colhead{Type} &\colhead{$\rho_h$} & \colhead{$r_h$}& \colhead{$H_p$}&\colhead{$P_h$}& \colhead{$P_{1215}$} & \colhead{$P_{1250}$}&\colhead{$P_{1275}$}&  \colhead{$T_h$}& \colhead{at $r_h$}&\colhead{$r_{\tau_0}$}&  \colhead{$r_{\tau_1}$}& \colhead{$H_\tau$}\\
\colhead{(UT)} & \colhead{} &\colhead{(km)} & \colhead{(km)}&  \colhead{(km)} &\colhead{($\mu$bar)} &\colhead{($\mu$bar)} & \colhead{($\mu$bar)}&\colhead{($\mu$bar)}&   \colhead{(K)}& \colhead{(K/km)}&\colhead{(km)}&   \colhead{(km)}& \colhead{(km)}
}
\startdata
\text{2017 August 07}&Single& \text{1263 $\pm$ 5}  & \text{1327 $\pm$ 5} & \text{52 $\pm$ 2}& \text{1.2 $\pm$ 0.1} & \text{10.1 $\pm$ 0.8}& \text{5.2 $\pm$ 0.5}& \text{3.2 $\pm$ 0.4}& \text{86 $\pm$ 4}& \text{-0.15 $\pm$ 0.01}& \text{1275 $\pm$ 12}& \text{1207 $\pm$ 54}& \text{52 $\pm$ 117} \\
\text{2018 April 09$^a$}&Single& \text{1252 $\pm$ 19}  & \text{1334 $\pm$ 12} & \text{62 $\pm$ 7}& \text{1.9 $\pm$ 0.6}& \text{12.7 $\pm$ 2.7} & \text{7.3 $\pm$ 1.8}& \text{4.9 $\pm$ 1.4}& \text{102 $\pm$ 11}  & \text{-0.18 $\pm$ 0.02} & \text{1267 $\pm$ 12}& \text{1140 $\pm$ 63}& \text{200 $\pm$ 200}\\
\text{2018 August 15}&Multi& \text{1220 $\pm$ 1}  & \text{1289 $\pm$ 1} & \text{54 $\pm$ 1}& \text{1.4 $\pm$ 0.1}& \text{5.6 $\pm$ 0.1} & \text{3.0 $\pm$ 0.1}& \text{1.9 $\pm$ 0.1}& \text{95 $\pm$ 1}  & \text{-0.17 $\pm$ 0.01} & \text{1247 $\pm$ 5}& \text{1177 $\pm$ 1}& \text{19 $\pm$ 1}\\
%\text{2018 October 01}&Single & \text{1277.1 $\pm$ 4.7}& \text{1338.6 $\pm$ 4.3} & \text{49.7 $\pm$ 1.8}& \text{1.00 $\pm$ 0.10}& \text{11.64 $\pm$ 0.92} & \text{5.86 $\pm$ 0.54}& \text{3.57 $\pm$ 0.37}& \text{81.0 $\pm$ 2.8} & \text{-0.14 $\pm$ 0.01} \\
\text{2018 October 01}&Single &  \text{1230 $\pm$ 5}  & \text{1296 $\pm$ 4} & \text{52 $\pm$ 2}  & \text{1.2 $\pm$ 0.1} & \text{5.7 $\pm$ 0.5} & \text{2.9 $\pm$ 0.3} & \text{1.8 $\pm$ 0.2} & \text{90 $\pm$ 3} & \text{-0.16 $\pm$ 0.01} & \text{1248 $\pm$ 7}& \text{1184 $\pm$ 4}& \text{33 $\pm$ 6}\\
%\text{2018 November 01}&Single& \text{1266.7 $\pm$ 2.7}& \text{1318.4 $\pm$ 2.5}&\text{43.0 $\pm$ 1.1} & \text{0.69 $\pm$ 0.05}& \text{7.42 $\pm$ 0.40}& \text{3.34 $\pm$ 0.21}& \text{1.88 $\pm$ 0.14}& \text{72.3 $\pm$ 1.9} & \text{-0.13 $\pm$ 0.01} \\
\text{2018 November 01} &Single&\text{1236 $\pm$ 3} & \text{1299 $\pm$ 2} & \text{50 $\pm$ 1} & \text{1.1 $\pm$ 0.1} & \text{5.8 $\pm$ 0.3} & \text{2.9 $\pm$ 0.2} & \text{1.8 $\pm$ 0.1} & \text{87 $\pm$ 2} & \text{-0.15 $\pm$ 0.01}& \text{1253 $\pm$ 3}& \text{1194 $\pm$ 8}& \text{33 $\pm$ 13} \\
\text{2018 November 20$^a$}&Single& \text{1141 $\pm$ 23}& \text{1211 $\pm$ 17}& \text{54 $\pm$ 8}& \text{1.6 $\pm$ 0.7}& \text{1.5 $\pm$ 0.8}& \text{0.8 $\pm$ 0.5} & \text{0.5 $\pm$ 0.3}& \text{107 $\pm$ 16}& \text{-0.20 $\pm$ 0.03} & \text{1172 $\pm$ 288}& \text{1104 $\pm$ 17}& \text{25 $\pm$ 36}\\
\text{2021 August 06}&Multi& \text{1235 $\pm$ 21}  & \text{1304 $\pm$ 12} & \text{54 $\pm$ 6}& \text{1.4 $\pm$ 0.5}& \text{7.1 $\pm$ 0.6} & \text{3.8 $\pm$ 0.3}& \text{2.4 $\pm$ 0.3}& \text{94 $\pm$ 12}  & \text{-0.17 $\pm$ 0.02}& \text{1358 $\pm$ 17}& \text{1175 $\pm$ 119}& \text{62 $\pm$ 92} \\
\text{2022 June 01}&Multi& \text{1222 $\pm$ 1}  & \text{1289 $\pm$ 1} & \text{52 $\pm$ 1}& \text{1.3 $\pm$ 0.1}& \text{5.2 $\pm$ 0.1} & \text{2.7 $\pm$ 0.1}& \text{1.7 $\pm$ 0.1}& \text{93 $\pm$ 1}  & \text{-0.17 $\pm$ 0.01} & \text{1214 $\pm$ 1}& \text{1185 $\pm$ 2}& \text{97 $\pm$ 20}\\
%\text{2022 August 23}&Multi& \text{1232.0 $\pm$ 16.5}  & \text{1284.7 $\pm$ 14.0} & \text{43.5 $\pm$ 7.1}& \text{0.78 $\pm$ 0.35}& \text{3.80 $\pm$ 1.39} & \text{1.72 $\pm$ 0.79}& \text{0.97 $\pm$ 0.52}& \text{77.3 $\pm$ 12.6}  & \text{-0.14 $\pm$ 0.02} \\
\text{2022 August 23}&Multi& \text{1216 $\pm$ 16}  & \text{1283 $\pm$ 16} & \text{61 $\pm$ 19}& \text{1.3 $\pm$ 0.1}& \text{4.8 $\pm$ 1.2} & \text{2.5 $\pm$ 0.7}& \text{1.6 $\pm$ 0.4}& \text{94 $\pm$ 2}  & \text{-0.20 $\pm$ 0.01}& \text{1244 $\pm$ 37}& \text{1157 $\pm$ 20}& \text{50 $\pm$ 45} \\
%\text{2023 July 17}&Single& \text{1246.7 $\pm$ 47.8} & \text{1310.5 $\pm$ 66.5}& \text{50.9 $\pm$ 19.0}& \text{1.13 $\pm$ 1.19}& \text{7.21 $\pm$ 5.21}& \text{3.67 $\pm$ 3.23}& \text{2.26 $\pm$ 2.30} & \text{86.6 $\pm$ 32.7}& \text{-0.15 $\pm$ 0.06} \\
\text{2023 July 17$^a$} & Single & \text{1213 $\pm$ 45} & \text{1297 $\pm$ 98} & \text{62 $\pm$ 16} & \text{2.1 $\pm$ 1.5}& \text{7.6 $\pm$ 4.0}& \text{4.4 $\pm$ 2.7}& \text{2.9 $\pm$ 2.1}& \text{108 $\pm$ 28}& \text{-0.19 $\pm$ 0.05} & \text{1261 $\pm$ 46} & \text{1067 $\pm$ 98} & \text{119 $\pm$ 348}\\
\enddata
\tablecomments{$^a$These datasets are low SNR and are plotted in lighter shades in Figs. \ref{fig:shadowRad}-\ref{fig:masterpressureplot} See the discussion in \S~\ref{subsec:atmevolution}.}

\end{deluxetable*}

\end{longrotatetable} 

\subsection{Comparison of atmospheric properties over time} \label{subsec:atmevolution}

Because $r_h$ depends on atmospheric parameters, the most direct size measurement from an occultation is $\rho_h$, defined as the radius in Pluto's shadow at which a light curve reaches the half-light level. Both values are dependent upon the geometric solution, but $\rho_h$ is otherwise independent of the atmospheric solution. For an isothermal atmosphere, the shadow radius is smaller than the half-light radius by one scale height. Following \citet{RN3616} and \citet{RN3469}, we therefore focus on shadow radii and the pressure in the upper atmosphere when examining changes in Pluto's atmosphere over time. 

\begin{figure}[ht!]
\centering
\includegraphics[width=0.8\textwidth,clip,trim=0mm 0mm 0mm 0mm]{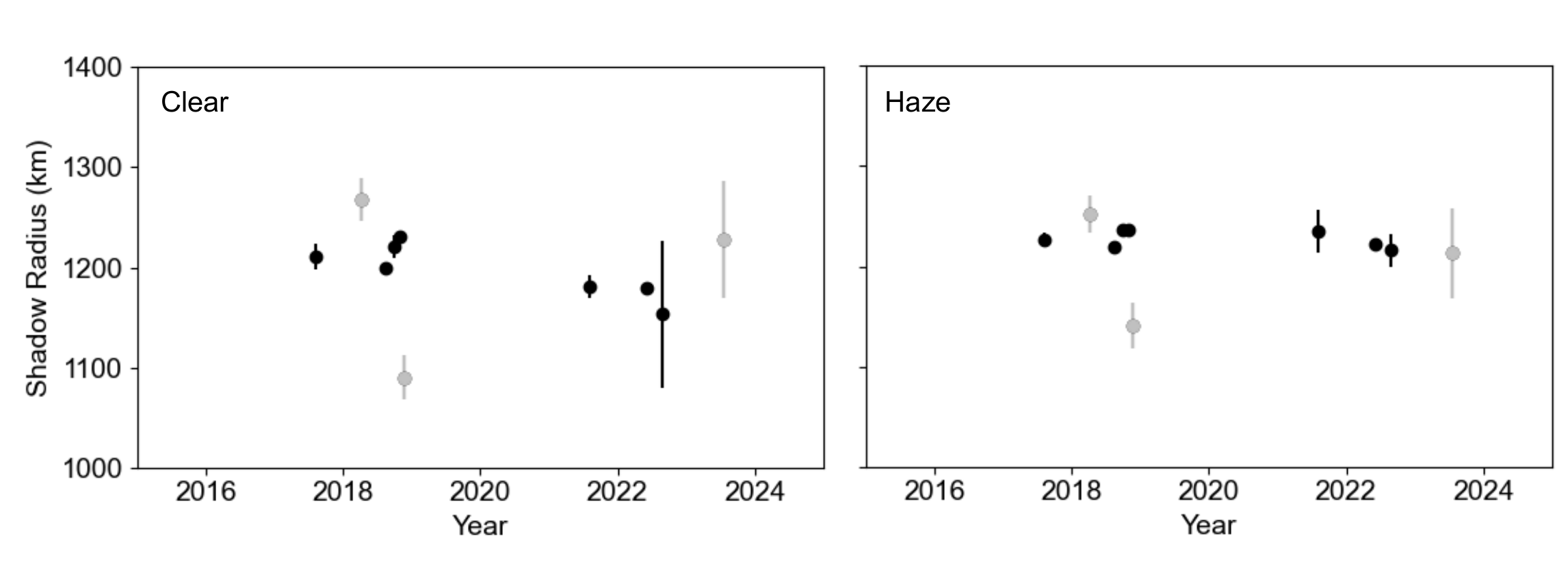}
\caption{Fitted shadow radii for the datasets presented in this work, with values from Tables \ref{tab:cleartable} and \ref{tab:hazetable}. The clear atmosphere results are on the left and the haze atmosphere results are on the right. The three low-quality datasets are plotted in a lighter shade. \label{fig:shadowRad}}
\end{figure}

\begin{figure}[ht!]
\centering
\includegraphics[width=0.4\textwidth,clip,trim=0mm 0mm 0mm 0mm]{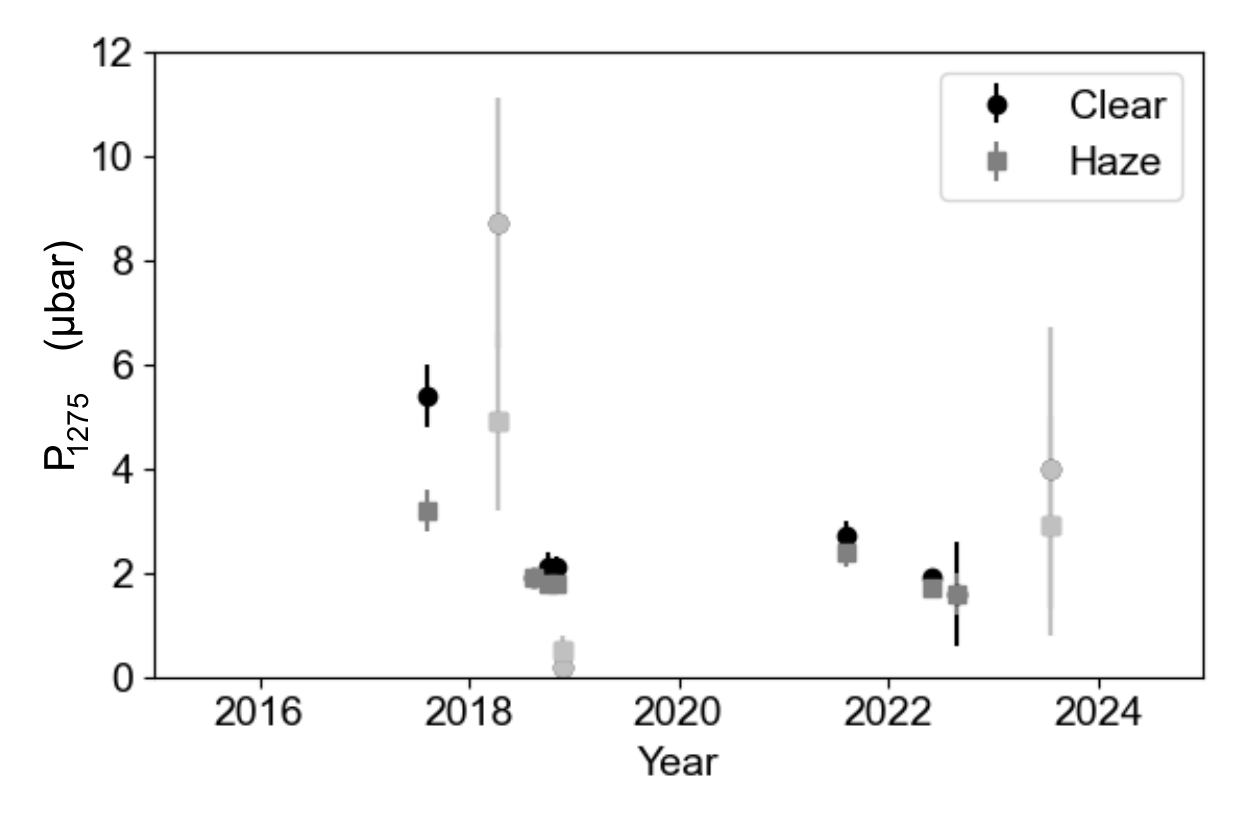}
\caption{Fitted atmospheric pressures at 1275 km radius, $P_{1275}$, for the datasets presented in this work, with values from Tables \ref{tab:cleartable} and \ref{tab:hazetable}. The three low-quality datasets are plotted in a lighter shade. \label{fig:p1275}}
\end{figure}

Tables~\ref{tab:cleartable} and \ref{tab:hazetable} respectively contain all the fitting results for clear atmospheres using only high-altitude data and haze atmospheres using all data. These fits are selected for temporal comparison because (i) the former considers the part of Pluto's atmosphere that has remained relatively unchanged over time, allowing better comparison to previous results, and (ii) the latter returns the best overall fits to the data. The $\rho_h$ values from these tables are plotted in Fig.~\ref{fig:shadowRad}, and $P_{1275}$ are plotted in Fig.~\ref{fig:p1275}. We opt for $P_{1275}$ rather than $P_{h}$ because the half-light radius varies. By eye, we see that the haze shadow radius has stayed roughly constant over time, while there is an overall downward trend for the clear shadow radius and $P_{1275}$. The high-quality results from the 2022 June 01 event, in particular, pin the later epochs to lower than the earlier typical values. The chords that possibly break this trend (2018 April 09, 2018 November 20, and 2023 July 17) are single chords and have the large error bars (see Tables~\ref{tab:cleartable} and \ref{tab:hazetable}). This is due to low SNRs. As shown in Fig.~\ref{fig:20181120LC}, the 2018 November 20 chord has only nine measurements during the occulted part of the light curve. Worse, when fitting for a clear atmosphere using only data above half-light (as in Table~\ref{tab:cleartable}), fewer than three data points are left in the occultation range. While better sampled, Figs.~\ref{fig:20180409LC} \& \ref{fig:20230717LC} likewise show that the 2018 April 09 and 2023 July 17 data are noisy, with SNRs per scale height of 11 and 6, respectively. The exceptionally low SNRs of these datasets suggests that they are unreliable for detecting atmospheric parameters that vary with altitude, and we have plotted the results for these dates in lighter shades in all the figures. From 2017-2022, excluding the low-SNR datasets, the haze shadow radius has stayed roughly constant over time (fitted slope of $-0.5\pm1.5 \, {\text{km/yr}}$), with a weighted mean of $1222.7\pm0.7 \, {\text{km}}$. The clear shadow radius changed by $-10.8\pm5.4 \, {\text{km/yr}}$, and $P_{1275}$ shows a $8\pm6\%$ change pre- and post-2021.

\begin{figure}[ht!]
\centering
\includegraphics[width=0.8\textwidth,clip,trim=0mm 0mm 0mm 0mm]{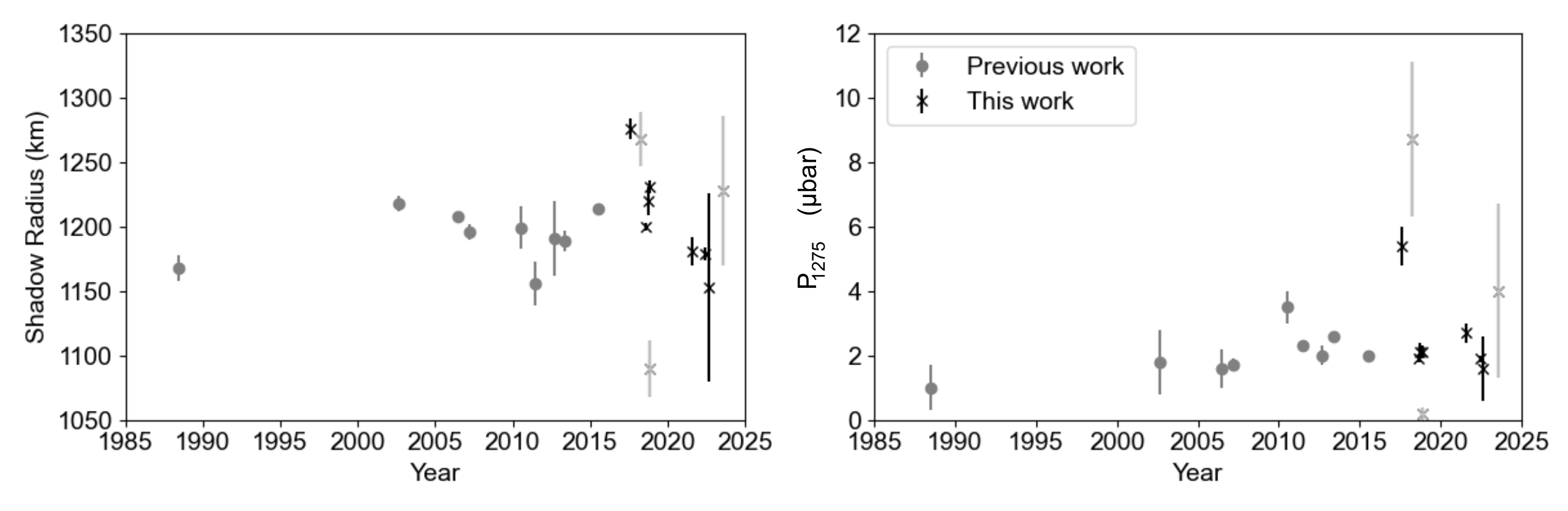}
\caption{Results for Pluto occultation datasets since 1988 that have been fitted using the same clear atmospheric model and fitting technique employed here, with values from Tables \ref{tab:atmvstime} and \ref{tab:cleartable}. The shadow radii are on the left and the atmospheric pressures at 1275 km radius are on the right. The three low-quality datasets from this work are plotted in a lighter shade. \label{fig:radpressAll}}
\end{figure}

To span a longer time period, we compiled results for Pluto occultation data since 1988 for which clear atmospheric fits were carried out with similar methods to this work (see Table~\ref{tab:atmvstime}). These results should be the most consistent way to compare bulk atmospheric characteristics over time, since they measure the relatively stable upper atmosphere. As shown in Fig.~\ref{fig:radpressAll}, there is an increase in shadow radius ($\rho_h$) and pressure ($P_{1275}$) from 1988 to roughly 2015, relative stability, and then what appears to be a decrease in the post-2021 measurements. 

\begin{figure}[ht!]
\centering
\includegraphics[width=0.8\textwidth,clip,trim=0mm 0mm 0mm 0mm]{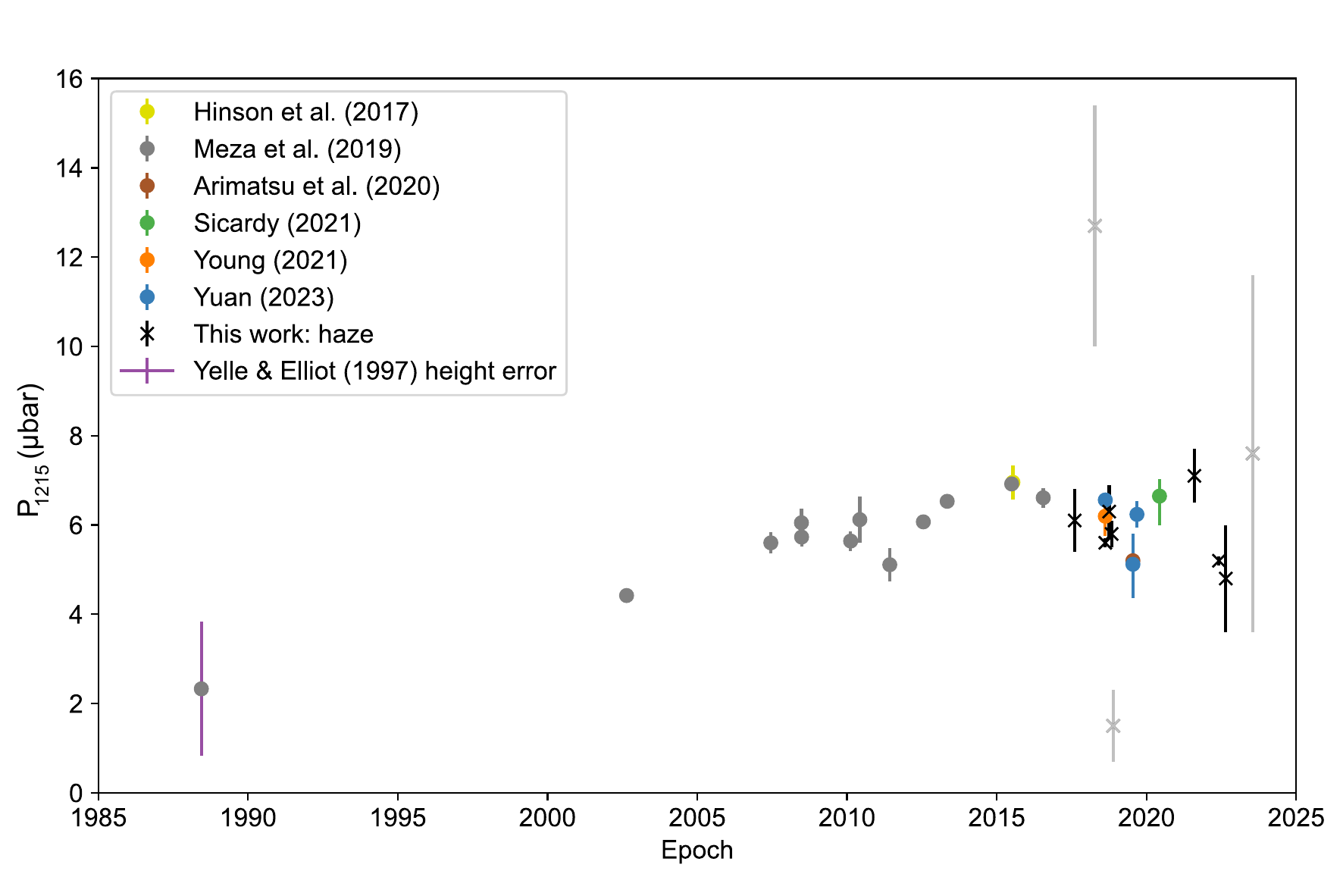}
\caption{Pluto's atmospheric pressure at 1215~km over time for published datasets from \citet{RN2137,RN3744,RN3823,RN3947,RN3918, RN4067} and with \citet{RN3741} including references therein. Publications quoting surface pressure were scaled to this height using the factor of 1.84 from \cite{RN3741}. The results from this work for the best-fitted atmospheres with haze are shown as black x's: the lowest-quality datasets are plotted in a lighter shade. The error bars on the 1988 data point are increased here from those shown in the given reference in order to properly include the error bar on the height (from Table 1 in \cite{RN2137}, see \S~\ref{subsec:pressref}). Pressure measurements were reported for ingress and egress from the \textit{New Horizons} occultation in \cite{RN3744}: the ingress value is plotted here as it was measured at the adopted surface radius, rather than the egress value that was measured 5 km higher.} \label{fig:masterpressureplot}
\end{figure}

Comparison with other published Pluto atmospheric parameters is not straightforward, due to the differing models and assumptions (see \S~\ref{sec:discussion}). As a representative compilation of results to date, we have plotted the estimated pressure at 1215~km in Fig.~\ref{fig:masterpressureplot}. We selected this height because it has served as the reference in many publications, and we note that a scaling factor has been assumed to convert from reported surface pressures to $P_{1215}$ \citep[following][]{RN3741}. We stress that the atmospheric profiles differ between analyses and thus these pressure values are not completely comparable --- rather, the plot serves as a check on whether results are roughly consistent between methods/publications and to try to identify broad temporal trends. The pressures at this lower altitude are larger than those at $P_{1275}$, and the changes in pressure are more visually apparent. The trend remains of increasing pressure from 1988 to roughly 2015, a plateau through 2021, and then a more recent pressure drop.

\begin{longrotatetable}
\begin{deluxetable*}{lcccccccccl}
%\tablenum{6}
\tablecaption{Pluto's atmospheric characteristics from published clear model fits to stellar occultation data using a similar methodology to this work.  
\label{tab:atmvstime}}
\tablewidth{0pt}
\tablehead{
\colhead{} & \colhead{} & \colhead{}& \colhead{}&\colhead{} & \colhead{}& \colhead{}& \colhead{}& \colhead{}& \colhead{Temp.}&\\
\colhead{} & \colhead{} & \colhead{}& \colhead{}&\colhead{} & \colhead{}& \colhead{}& \colhead{}& \colhead{}& \colhead{Gradient}&\\
\colhead{Date} & \colhead{$\rho_h$} & \colhead{$r_h^a$}& \colhead{$H_p$}&\colhead{$P_h$} & \colhead{$P_{1275}$}& \colhead{$\lambda_h^a$}& \colhead{$b^a$}& \colhead{$T_h$}& \colhead{at $r_h$}&\colhead{Ref.$^b$}\\
\colhead{(UT)} & \colhead{(km)} & \colhead{(km)}&  \colhead{(km)} &\colhead{($\mu$bar)} & \colhead{($\mu$bar)}& & & \colhead{(K)}& \colhead{(K/km)}& 
}
\startdata
%+ means double checked
29 June 2015    &   $1213.8\pm1.1$&  $1295.0\pm0.5$&   $52.8\pm0.7$&   $1.37\pm0.03$& $2.0\pm0.1^c$& $18.8\pm0.2$ & $-2.3\pm0.2$  &   $92.4\pm1.2$&   $-0.16\pm0.01$& 6 \\
04 May 2013    &   $1188.7\pm7.7$&  $1299.2\pm3.8$&   $54.4\pm0.5$&   $1.66\pm0.03$& $2.6\pm0.1^c$& $15.6\pm0.3$ & $-3.2\pm0.2$  &   $94.6\pm1.0$&   $-0.24\pm0.01$& 2,6 \\
&   $1212.8\pm4.3$&  $1304.0\pm3.8$&   $57.4\pm0.3$&   $1.70\pm0.03$& $2.8\pm0.3$& $17.2\pm0.2$ & $-2.2$  &   $99.1\pm0.9$&   $-0.17\pm0.01$& 2 \\
09 Sept. 2012    &   $1191\pm29$&  $1282\pm14$&   $56.7\pm7.6$&   $1.72\pm0.68$& $2.0\pm0.3^c$& $17.1\pm3.1$ & $-2.2$  &   $101\pm15$&   $-0.17\pm0.03$& 2 \\
   &   $1198\pm38$&  &  &  & $2.0\pm2.0$& & &   &  & 2 \\
23 June 2011     &   $1156\pm17$&  $1273.1\pm4.0$&   $61.2\pm1.0$&   $2.39\pm0.12$& $2.3\pm0.1^c$& $14.0\pm0.9$ & $-2.7\pm0.4$  &   $110.7\pm1.7$&   $-0.24\pm0.03$& 2 \\
&   $1206.5\pm1.0$&  $1290.3\pm1.1$&   $54.2\pm0.1$&   $1.52\pm0.01$& $2.0\pm0.1$& $18.3$ & $-2.2$  &   $95.5\pm0.1$&   $-0.16\pm0.01$& 2,5 \\
04 July 2010   &   $1199.2\pm16.5$&  $1303.5\pm14.8$&   $62.4\pm2.2$&   $2.23\pm0.24$& $3.5\pm0.5^c$& $15.4\pm0.8$ & $-2.2$  &   &   & 5 \\
 &   $1211.9.\pm15.5$&  $1304.2\pm14.6$&   $71.7\pm8.7$&   $2.57\pm0.5$& $3.9\pm0.2^c$& $17.7\pm1.6$ & $-0.2\pm1.4$  &   &   & 4 \\
&   $1226\pm12$&  $1311.2\pm13.1$&   $55.1\pm0.5$&   $1.53\pm0.01$& $3.0\pm0.7^c$& $18.3$ & $-2.2$  &   &   & 4 \\
18 March 2007   &   $1196\pm6$&  $1276.1$&   $53.5\pm0.3$&   $1.51\pm0.12$& $1.7\pm0.2^c$& $17.9\pm1.1$ & $-1.9\pm0.8$  &   $95\pm1$&   $-0.13\pm0.02$& 3 \\
 &   $1207\pm4$&  $1291.1\pm4.6$&   $54.2\pm0.2$&   $1.51\pm0.10$& $2.0\pm0.3$& $18.3$ & $-2.2$  &   $95\pm1$&   $-0.16\pm0.01$& 2,3 \\
12 June 2006    &   $1208\pm4$&  $1276.1\pm3.5$&   $53.8\pm2.9$&   $1.58\pm0.14$& $1.6\pm0.1^c$& $18.3\pm0.8$ & $-2.2\pm0.7$  & $97\pm5$  &  $-0.17\pm0.05$  & 1 \\
 &   $1208\pm4$&  &   &   & $1.6\pm0.6$&  &   &  &  & 2 \\
21 Aug 2002    &   $1218\pm6$&          &          &   & $1.8\pm1.0$&  &   &   &   & 2 \\
                &     &      $1279\pm5$&    $61\pm4$&   &   &  &   & $108\pm9$& $0.05\pm0.08$  & 1 \\
09 June 1988    &   $1168\pm10$&          &          &   & $1.0\pm0.7$&  &   &   &   & 2\\
 &              &    $1233\pm4$&   $56\pm5$ &   &   &  &   & $114\pm10$& $-0.29\pm0.56$& 1
\enddata
\tablecomments{The first line for each date is the preferred solution. $^a$Values without error bars were fixed to the best-fit values from previous years. $^b$Reference key: 1 – \citet[][Tables 3, 5, and 10]{RN3171}; 2 – \citet[][Tables 3 and 6]{RN3616}; 3 – \citet[][Tables 3 and 5]{RN3286}; 4 –\citet[][]{RN3431}; 5 – \citet[][Table 5]{RN3469}; 6 – \citet[][Table 4]{RN3641}.$^c$Pressures at this height were calculated based on the model atmosphere using the listed values for $r_h$, $\lambda_h$, and $b$.} 

\end{deluxetable*}
\end{longrotatetable}

\section{Discussion} \label{sec:discussion}
We have presented ten new stellar occultation datasets for Pluto, spanning 2017-2023, with consistently fitted parameters for derived atmospheric properties such as pressure, temperature, scale height, and the most directly observable value, shadow radius. The decrease in pressure and $\rho_h$ between pre- and post-2021 events suggests that the atmospheric contraction in response to declining solar insolation could be underway \citep[e.g.][]{RN3365, RN3594, RN3485, Bertrand2016, RN3978, Johnson2021}. However, differing occultation data quality, analytic methods, and interpretations strongly color our understanding of Pluto's atmospheric evolution.

\subsection{Atmospheric pressure reference radii}\label{subsec:pressref}
For this work, we calculate the fitted atmospheric pressure at half-light level, $P_h$, and we use the best-fit atmospheric structure from each epoch to extrapolate to the pressure at 1275~km radius. This is the reference height used in \citet{RN3641} and \citet{RN3616} and was selected to be in the stable upper atmosphere.  
%calculated a pressure at this height of $1\pm0.7$ for the 1988 data. 
The reference radius of 1215~km selected by \citet{RN3741} and carried forward by others \citep[e.g.][]{RN3947} was presumably chosen because the pressure at this height was reported for the 1988 data in \citet{RN2137}. It's important to note that the true pressure reference height for that dataset, there labeled $r_0$, was at 1250~km with pressure of $1.23\pm0.26$ $\mu$bar for an $N_2$ atmosphere \citep{RN2137}. The reference radius $r_1=1215\pm11 \, \text{km}$ was the fitted location of the light-curve "knee", at which the pressure was reported to be $2.33\pm0.24$ $\mu$bar \citep{RN2137}. Caution should be used when quoting the latter pressure: the extrapolation between reference radii required assumptions about atmospheric structure and profile for the 1988 dataset, and the radius error bar also needs to be taken into consideration (as we have plotted in Fig.~\ref{fig:masterpressureplot}). 

It's useful to report surface pressure because this is a common atmospheric characteristic and it can be compared with results from \textit{New Horizons}. However, for the majority of Pluto occultations, occultation light curves do not reach the surface and this parameter is not fit. \citet{RN3741} used their best-fit atmospheric profile to determine a constant scaling factor (1.84) to extrapolate from fitted $P_{1215}$ values to $P_{\text{surface}}$. The same profile and scaling factor was used in \citet{RN3947} and \citet{RN3823}, and it is used here to convert published $P_{\text{surface}}$ values in Fig.~\ref{fig:masterpressureplot}. If the atmospheric structure is evolving, then this scaling factor would not necessarily be expected to remain constant over time. 

\subsection{Pressure drop versus plateau}

\cite{RN3823} and \cite{RN3918} proposed that there had been a pressure drop in Pluto's atmosphere. From a single chord taken on 2019 July 17, \cite{RN3823} reported $P_{1215}=5.20_{-0.19}^{+0.28} \, \mu \text{bar}$ (surface pressure of $P_{\text{surface}}=9.56_{-0.34}^{+0.52} \, \mu \text{bar}$). While this pressure is lower than any previous measurements, the degeneracy in fitting pressure and closest-approach distance for a single chord is a concern and the authors clearly stated that the finding required confirmation.  Subsequently, results were presented from multiple, high-quality chords taken on 2018 August 15 \citep[an event observed from different sites than reported in this work; ][]{RN3918}. Central flashes were detected for this event, which are particularly sensitive to surface pressure and haze opacity. The best-fit surface pressure of $P_{\text{surface}}=11.4\pm0.8 \, \mu \text{bar}$ was found to be consistent with that measured by \textit{New Horizons} in 2015 \citep{RN3918}. The lack of a continued monotonic pressure increase seemed to be interpreted as the onset of a pressure drop (as in the "freezing out" title of \citet{RN3918}), although the pressure had quantitatively leveled off.

More recently, observations were reported for the occultation on 2020 June 06, which returned a pressure of $P_{1215}=6.665_{-0.21}^{+0.35} \, \mu \text{bar}$ \citep{RN3947}. (Note that data were taken from a different site for this event in \cite{RN3943}: see the discussion in \cite{RN3947} regarding those results.) Assuming the constant scaling factor from \cite{RN3741}, the surface pressure was $P_{\text{surface}}=12.23_{-0.38}^{+0.65} \,\mu \text{bar}$ \citep{RN3947}. This pressure was consistent with those measured prior to 2018, and the interpretation was that of an ongoing pressure plateau \citep{RN3947}. Our results are consistent with a pressure plateau through approximately 2021 and thereafter a decrease in pressure. The possible recent change in pressure depends on which type of fit and atmospheric height is used. Considering consistent clear atmospheric fits to the upper atmosphere only, and excluding the three low-SNR datasets, the weighted percent drop in $P_{1275}$ between 2015-2021 and 2022 was $7\pm6\%$ (from $2.04\pm0.06 \, {\mu\text{bar}}$ to $1.90\pm0.10 \, {\mu\text{bar}}$, from Tables \ref{tab:cleartable} and \ref{tab:atmvstime}). Considering consistent haze fits and the pressure at a lower altitude, $P_{1215}$, the percent drop is $16\pm2\%$ (from $6.17\pm0.01 \, {\mu\text{bar}}$ to $5.20\pm0.10 \, {\mu\text{bar}}$, from Tables \ref{tab:hazetable} and \ref{tab:atmvstime}).

\subsection{Clear atmosphere vs. haze}\label{subsec:clearvhaze}
The shape of the occultation light curves for Pluto's upper atmosphere has remained fairly constant over time, being clear and predominantly isothermal with a slightly negative thermal gradient \citep[e.g.][]{RN3171}. However, the lower atmosphere has exhibited a variable, steeper or shallower profile, indicating temporal changes in thermal gradient and/or haze \citep[sometimes distinguished by a "knee" in the light curve, e.g.][]{RN77,RN3171,RN3366,RN3641}. Indeed, lower-atmosphere haze particles were inferred from stellar occultation data \citep[e.g.][]{RN77,RN2820,RN3641,RN3614} before their detection by \textit{New Horizons} in 2015 \citep[e.g.][]{RN3742}. Here, we find that clear atmosphere models alone do not fit the data well and some haze is likely present. We know that there must be a strong thermal gradient to connect the $\sim100 \, {\text{K}}$ mesosphere to the surface-ice temperature, which is not included in our haze model. To check whether the light-curve structure could be explained by thermal changes without haze, we carried out inversions to extract temperature profiles as functions of altitude \citep{RN2787} for the 2018 August 15 OAN-SPM data. The inversions extended more than 10 km below the known surface, demonstrating that the inversion technique is inappropriate for these data and that some haze is needed to explain the light-curve structure (as was found for e.g. the 2015 SOFIA data in \citet{RN3641}).

Many reported occultation results are based on models that assume a transparent, clear, pure $\text{N}_2$ atmosphere \citep[e.g.][]{RN3741,RN3641,RN3823}. This assumption is unlikely, given the changes in the structure of the lower atmosphere and variable hazes observed by \textit{New Horizons} (at up to $\sim500 \, \text{km}$ altitude, with the number density decreasing by nearly a factor of ten in the first 100~km above the surface; \citet{RN3977}). The best comparison of bulk atmospheric properties between different epochs is thus for clear atmosphere models only for data with $\geq0.5$ normalized flux (e.g. Fig.~\ref{fig:radpressAll}), which is above any discontinuities in observed light curves and should contain only the clear and constant upper atmosphere. The altitude of the half-light flux level is historically $\sim45-120 \, {\rm km}$, depending on the half-light radius of the atmosphere. As also noted by \cite{RN3613}, this selection should result in the most directly-comparable set of atmospheric parameters across datasets with the fewest complicating assumptions. Note that the assumption of a pure nitrogen atmosphere is acceptable here, since the expected mixing ratio of atmospheric constituent $\text{CO}$ does not affect occultation light curves and $\text{CH}_4$ (i) doesn't significantly change the bulk shape of the light curves and (ii) is only detectable in high SNR data \citep[although noting that the atmospheric temperature structure is sensitive to these mixing ratios, e.g. ][]{RN3554}.

\subsection{Changes in light-curve structure}\label{subsec:lowatm}
\begin{figure}[ht!]
\centering
\includegraphics[width=0.8\textwidth,clip,trim=0mm 0mm 0mm 0mm]{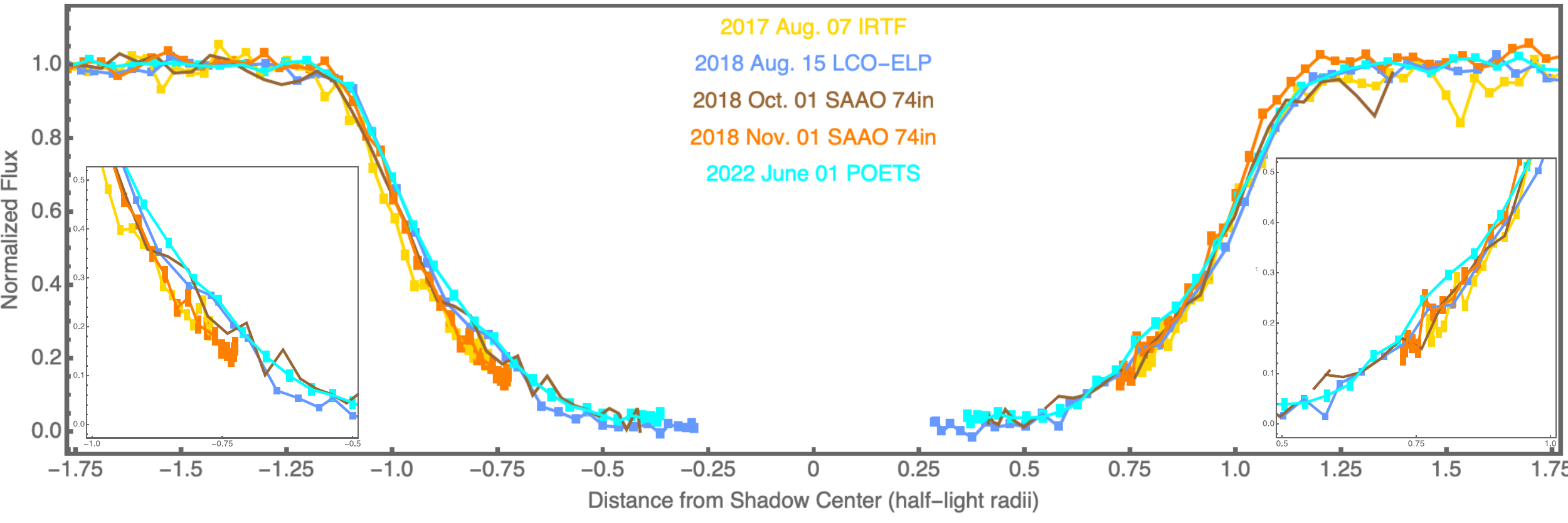}
\caption{A comparison of light-curve profiles over time, with the highest SNR light curves from this work overplotted versus distance from the center of the shadow, in units of half-light radii. Inset panels show the lower part of the atmosphere during immersion and emersion in greater detail.} For clarity, the data are binned at roughly one scale height. Filled-in error bars are shown, based on the errors in background fractions and noting that 2018 October 01 had a fixed background fraction. The overall atmospheric shape has not changed substantially between 2017 and 2022; however, there are variations in the slopes of the light curves in the lower atmosphere, with the 2022 data having the shallowest slope. \label{fig:rhoPlot}
\end{figure}

%Comparisons of the profile of Pluto occultation light curves over time have indicated that the upper atmosphere (above 0.5 flux) has been relatively stable. The lower atmosphere has exhibited changes in slope that have been attributed to varying temperature gradients and/or extinction \cite[e.g. Fig. 6 in][]{RN3171}. 
To look for variations in Pluto's atmospheric structure over the timescale of observations in this work, we plot the highest SNR light curves ($\gtrapprox30$), using only the best from a given epoch, on the common scale of distance from the shadow center in units of half-light radius (Fig.~\ref{fig:rhoPlot}). The data are binned as close as possible to one scale height, $\sim60 \, {\rm km}$, since we are looking for broader trends. Unlike the ``knee'' and steep lower-atmosphere slope in the 1988 data, the 2017-2022 light curves exhibit the bowl-like shape seen since 2006, with the normalized flux never reaching zero \cite[e.g.][]{RN3171,RN3741}. As noted in previous occultations, the upper atmospheric profiles (above half-light level) in Fig.~\ref{fig:rhoPlot} are consistent over time. However, there are variations in the light curve slope of the lower atmosphere, more notably on immersion than emersion, with 2022 having a shallower slope than the earlier years. Changes in the slope of the lower light curve can be caused by different thermal gradients and/or evolving haze properties. In addition, there is a slight bump in the 2022 data near 0.3 flux seen in both ingress and egress. As shown in the zoomed light-curve extracts in Fig. \ref{fig:20220601LC}, both the Savannah Skies POETS and the EOS data for this event show a light-curve change near 0.3 flux, indicating a non-localized (possibly global) phenomenon. The lower-altitude slope changes seen here are consistent with evolving haze: haze particles settle through the lower atmosphere on timescales of $\sim 10$ to $\sim 400$ Earth days, depending on the particle sizes ($0.2 \, \mu\text{m}$ to 10 nm for the given timescales; \citet{RN3737}). The capability of JWST to study haze in Pluto's atmosphere should help increase our understanding of how the haze evolves over time \citep[c.f.][]{2025A&A...696A.147L, 2025NatAs...9.1300B}.

\subsection{Light-curve spikes and vertically-propagating waves}\label{subsec:spikes}

While most of the light curves obtained show the smooth, bowl-shaped data characteristic of Pluto occultation light curves in recent years (see e.g. Figs. \ref{fig:20181001LC}, \ref{fig:20220601LC}, \& \ref{fig:rhoPlot}), there are significant intra-occultation flux spikes visible in the 2018 August 15 data (see Figs. \ref{fig:20180815LC} \& \ref{fig:20180815fits}). These spikes are extremely common in occultations by planets with thick atmospheres, such as Jupiter, Uranus, and Titan \citep[e.g.][]{1976Icar...27..359E, 2024PSJ.....5..247S, 2007Icar..192..503Z}, but they can also occur in thin atmospheres like those of Pluto and Mars \citep[e.g.][]{2017Icar..296..305P, 1977ApJ...217..661E}. 

Such spikes are generally attributable to instabilities or inversions in otherwise smooth density gradients, due to tides, vertical waves, and/or winds. Light curves with spikes can be analyzed in cases where the atmosphere is clear (minimal extinction) via full inversions, to return perturbations in the density profiles. For the 2007 Pluto occultation, light-curve inversion revealed perturbations of approximately 40 km in extent that were primarily attributed to internal gravity waves \citep{2009Icar..204..284H}, with a small percentage of the power being attributable to planetary-scale Rossby waves \citep{2008AJ....136.1510P}, both vertically propagating.  Similarly, smaller perturbations of only a few kilometers have been attributed to global-scale thermal tides resulting from diurnal insolation on isolated nitrogen ice patches \citep{2010Icar..208..402T}.  

In this work, the 2018 data are well fit by a dense haze model (see Fig. \ref{fig:20180815fits}), noting that our model is simple and excludes the known steep temperature gradient from the lower atmosphere to the surface. Forward models can include haze \citep[e.g.][]{RN1985}, but they are computationally expensive. More simply, recent analyses of forward models including wavelet-based perturbations have revealed several general rubrics for interpreting isolated spikes in occultation light curves \citep{2025PSJ.....6..257Y}. Here, we apply these rubrics to the spikes detected in the 2018 light curve.

The first step is to determine which features can be identified as spikes. In the case of the 2018 August 15 OAN-SPM curve, we limit our initial consideration to flux variations during the occultation that rise at least three sigma above the fitted occultation model. The three obvious spikes visible in the bottom panel of Fig. \ref{fig:20180815LC} and in the residuals of Fig. \ref{fig:20180815fits} are the only perturbations that meet this criterion. These correspond to the points at approximately 1300, 1306, and 1345 s after the reference time shown in the figure. Reducing this criterion to only a two-sigma rise in flux, there are four more possible spikes at 1276, 1278, 1363, and 1377 s from the reference time. Although these meet the 2-sigma criteria (barely), we exclude them here as being too weak to clearly identify spike duration. The three primary peaks are all greater than 8-sigma variations, and thus definitive in their extent.  

Focusing on these three spikes, we can make some observations following the methodology of \citet{2025PSJ.....6..257Y}. First, spike-generating waves cannot have wavelengths significantly greater than the atmospheric scale height, as perturbations of this size instead result in large-scale quasi-oscillations in the light curve. The observed, confined spikes imply wavelengths no larger than a few scale heights. Furthermore, because the spikes do not exceed the baseline flux level, they should be even smaller, confined to approximately $0.6 \, H_p$. Most constraining, \citet{2025PSJ.....6..257Y} determined that the extent of the waves in the radial space of the light curve is the approximate limit of the vertical wavelength. We can determine this extent by converting the timing duration of the waves into the change in vertical radius in the atmosphere. Using the astrometric solution determined from the fits in Table \ref{tab:20180815fits}, we calculate that the three spikes occur at atmospheric radii of 1195, 1184, and 1169 km, respectively. Note that these radii, which appear to extend below Pluto’s 1188 km surface radius \citep{2015Sci...350.1815S} need to then be adjusted upwards by $\sim20$ km to account for the haze extinction that is not included in the \citet{RN1023} flux-radius calculations, as demonstrated by \citet{2021Icar..35613572P}.  The adjustment results in radii of 1221, 1210, and 1195 km for the three spikes. Their durations result in vertical spreads, and thus, assuming they are at the static stability limit, vertical wavelengths with upper limits of approximately 3.4, 2.6, and 1.0 kilometers, respectively.

These wavelengths are superficially consistent with the wave results of \citet{2009Icar..204..284H} and \citet{2008AJ....136.1510P}, but there are key differences. In the 2007 occultation, the vertical wavelength was measured as 35 km at a radius of 1460 km, and 25 km at a radius of 1340 km. These measurements were made using direct inversion, which was justified as the grazing occultation sampled large atmospheric radii, well above any haze. Stemming from variations in the Brunt V\"{a}is\"{a}l\"{a} frequency, the decrease in wavelength with decreasing radius can be extrapolated to the radii of the current three spikes, yielding comparable wavelengths of 6.5, 6.0, and 5.0 km. These are the same magnitude, but a few times larger than, the calculated values for the 2018 data given above.

While it is certainly possible that the specific perturbations in 2018 are smaller than those in 2007, wavelengths of $<4 \, \text{km}$ are characteristic of the diurnal thermal tidal forcing identified by \citet{2010Icar..208..402T}. The weakness of the tidal oscillation explanation is that it should be generally of global scale and persistent between occultations. In contrast, buoyancy waves (also called ''internal waves'' or ''internal gravity waves'') are location-focused based upon their generating surface topology and are thus likely to vary greatly between occultation light curves (even from the same event), being present in some curves and missing in others, as is evident in the 2018 August 15 dataset.  Other light curves in this dataset may have spikes smoothed out due to their lower cadences and thus lower spatial resolutions per point, but on 2018 August 15, at least, the SRO curve would show the spikes seen in the OAM-SPM if they were present.

We conclude that the observed spikes in the 2018 dataset represent vertically propagating buoyancy waves, with small wavelengths ranging from 1.0 to 3.5 km, at atmospheric radii ranging from $\sim 1195-1220 \, \text{km}$.  

\subsection{Comparison with volatile-transport models}
Trends in atmospheric parameters derived from these occultations can be compared with volatile transport models to better understand Pluto's overall atmospheric evolution and the physical characteristics of its surface. Detailed comparisons are beyond the scope of this work, but the data presented here should be further considered.

For example, models such as those by \citet{RN3906} and \citet{RN3978} predict Pluto's atmospheric pressure far into the future. \citet{RN3906} specifically consider different ice distributions for the unimaged southern hemisphere: bare, a south polar cap, a southern zonal band, and where Sputnik Planitia is the only $\text{N}_2$ ice deposit. Perhaps the most relevant for a pressure decrease in the modern era are the "Exchange with Persistent Plateau" (EPP) models of \citet{2013ApJ...766L..22Y}. These models feature ice caps in both hemispheres, which mostly persist across seasons, trading ice mass between them annually. This class of plateau models begins to decay around this epoch ($\sim2025$). In fact, model EPP7 from \citet{2013ApJ...766L..22Y} begins its atmospheric decay as we describe in this work, while simultaneously fitting the rapid 1988-2002 increase identified from the 2002 occultation \citep{RN2820}. The models shown in \citet{RN3947} indicate that occultation data of typical quality could only detect a pressure decrease by the late 2020s, even if it began sooner and more subtly. Therefore, continuing Pluto occultation observations are needed. 

\section{Conclusions}
We present observations of stellar occultations by Pluto for ten different epochs between 2017-2023. Results from atmospheric-model fits to the light curves are compared with results from previous epochs in order to study Pluto's atmospheric evolution. 
\begin{itemize}[nosep]
\itemsep0em 
    \item The datasets presented here are of varying quality, with SNRs per scale height ranging from 2 to 158. Atmospheric fits for particularly low SNR light curves are unreliable and/or have large error bars. 
    \item Atmospheric fitting for events with only a single successful chord is challenging due to the degeneracy between closest approach distance and atmospheric size.
    \item We fit the closest approach for events with multiple sites and find that all but one are within 20 km of the postdiction values: the outlier (2022 August 23) has a fitted value 55 km, $\sim3 \, \text{mas}$, closer than the postdiction. %For single-site events, we assume the postdiction closest-approach values and find one event (2018 October 01) that requires a shift of $\sim10 \, \text{mas}$ closer to be consistent with the other datasets. 
    \item Combining our data with previous results, we confirm that Pluto's atmospheric size and pressure increased from 1988 to $\sim2015$ and then were roughly consistent through 2021. We find that these parameters have decreased in 2022, with a percent change of $7\pm6\%$ at $P_{1275}$ for clear upper-atmosphere fits and $16\pm2\%$ at $P_{1215}$ for atmospheric fits with haze. These changes are similar to double ice-cap models such as those of \citet{2013ApJ...766L..22Y} and \citet{RN3906}.
    \item Comparing the structures of the highest-quality light curves, we find that Pluto's upper atmosphere has not changed within the errors while there are slight changes in the light-curve slope in the lower atmosphere (likely due to changes in haze and/or thermal gradients): the most recent data have the shallowest slope. This change-of-slope is consistent with haze in the lower atmosphere settling on the expected timescales of roughly 10 to 400 days for $0.2 \, \mu\text{m}$ to 10 nm particles.
    \item Pluto continues to exhibit intermittent signs of vertically-propagating waves, with light-curve spikes appearing (or not) between chords of the same event and different epochs, indicating topology-formed, buoyancy-wave-induced perturbations to the base atmosphere.We detect such waves in one dataset in 2018, with wavelengths from 1-3.5 km at radii between 1195 and 1220 km.
    \item Ongoing observations, especially of high quality and with multiple chords, are important to confirm what appears to be a drop in Pluto's shadow radius and pressure after 2021. This time of accelerating pressure change (if confirmed) will provide key differentiation between competing atmospheric models.
    \item We encourage analyses of these Pluto occultation datasets by other teams --- results from consistently applying alternative atmospheric fitting techniques to the same data can be compared with the results presented here.
\end{itemize}
\begin{acknowledgments}
We are grateful for two anonymous reviewers, whose careful reading and suggestions improved the manuscript. AAS was partly funded by NASA SSO grant 80NSSC21K0432 and the National Research Foundation of South Africa. MJP was party funded by a subaward from NASA SSO grant to 80NSSC21K0432. AC was partly funded by a grant from the MIT Undergraduate Research Opportunities Program.

AAS was a visiting Astronomer at the Infrared Telescope Facility (IRTF), which is operated by the University of Hawaii under contract 80HQTR19D0030 with the National Aeronautics and Space Administration. The authors acknowledge the very significant cultural role and reverence that the summit of Maunakea has always had within the indigenous Hawaiian community.  We are grateful to have the opportunity to conduct observations from this mountain and recognize that we are guests.

This work uses observations made at the South African Astronomical Observatory (SAAO) and telescopes operated by Las Cumbres Observatory.

This work includes data gathered with the 6.5-meter Magellan Telescopes located at Las Campanas Observatory, Chile.

This work made use of the Lowell Discovery Telescope (LDT) at Lowell Observatory. Lowell is a private, non-profit institution dedicated to astrophysical research and public appreciation of astronomy and operates the LDT in partnership with Boston University, the University of Maryland, the University of Toledo, Northern Arizona University and Yale University.

Some of the observations in the paper made use of the High-Resolution Imaging instrument Zorro. Zorro was funded by the NASA Exoplanet Exploration Program and built at the NASA Ames Research Center by Steve B. Howell, Nic Scott, Elliott P. Horch, and Emmett Quigley. Zorro was mounted on the Gemini South telescope of the international Gemini Observatory, a program of NSF NOIRLab, which is managed by the Association of Universities for Research in Astronomy (AURA) under a cooperative agreement with the U.S. National Science Foundation on behalf of the Gemini Observatory partnership: the U.S. National Science Foundation (United States), National Research Council (Canada), Agencia Nacional de Investigaci\'{o}n y Desarrollo (Chile), Ministerio de Ciencia, Tecnolog\'{i}a e Innovaci\'{o}n (Argentina), Minist\'{e}rio da Ci\^{e}ncia, Tecnologia, Inova\c{c}\~{o}es e Comunica\c{c}\~{o}es (Brazil), and Korea Astronomy and Space Science Institute (Republic of Korea). The data were taken under program GS-2021B-Q-109.

This work makes use of data taken by Electro Optic Systems (EOS) in Australia. We are grateful specifically to Ian Ritchie, Chun Morton, Alex Pollard, and James Bennett for their collaboration.

The Liverpool Telescope is operated on the island of La Palma by Liverpool John Moores University in the Spanish Observatorio del Roque de los Muchachos of the Instituto de Astrof\'isica de Canarias with financial support from the UK Science and Technology Facilities Council.

Some observations were obtained with the SARA Observatory at CTIO, which is owned and operated by the Southeastern Association for Research in Astronomy (saraobservatory.org).

This work has made use of data from the European Space Agency (ESA) mission
{\it Gaia} (\url{https://www.cosmos.esa.int/gaia}), processed by the {\it Gaia}
Data Processing and Analysis Consortium (DPAC,
\url{https://www.cosmos.esa.int/web/gaia/dpac/consortium}). Funding for the DPAC
has been provided by national institutions, in particular the institutions
participating in the {\it Gaia} Multilateral Agreement.

\end{acknowledgments}


\vspace{5mm}
\facilities{Gemini:South(Zorro), LCOGT, LDT(POETS), Magellan:Clay(POETS), SAAO:74in(SHOC), ATUS, SARA}


\software{Mathematica \citep{RN4102}, astropy \citep{RN4083} , matplotlib \citep{RN4085}, numpy \citep{RN4087}, photutils \citep{RN4082}, scipy \citep{RN4086}}
% Photutils used version 1.1.0

\appendix
\section{Observational Details}\label{subsec:appendix}
This appendix contains details for the observations for each event and each telescope listed in Table \ref{tab:scopes}.

\subsection{2017 August 07} \label{subsec:20170807obs}
NASA's 3.2-m Infrared Telescope Facility (IRTF) at Mauna Kea Observatory (MKO) was the only telescope to successfully observe the 2017 August 07 UT occultation. The MIT Optical Rapid Imaging System (MORIS) was cooled to -65\textcelsius{}, run with the 1 MHz conventional amplifier with 16-bit A/D, and binned 2x2 for a superpixel size of 0.22 arcsec. Each frame was triggered with a Spectrum TM-4 GPS for absolute times to microsecond accuracy. The weather was clear beforehand with 0.7-arcsec seeing, and there was some cirrus after the occultation. Data were taken from 10:15-10:55 UT. An example of the data is shown in Fig.~\ref{fig:20170807images}.

\subsection{2018 April 09} \label{subsec:20180409obs}
The 2018 April 09 UT occultation was observed from the 4.3-m Lowell Discovery Telescope (LDT; Happy Jack site). The weather was mostly clear with scattered clouds on the horizon. A Portable Occultation, Eclipse, and Transit System (POETS; e2v CCD-97) instrument was used. The CCD was cooled to -60\textcelsius{}, run with the 1 MHz conventional amplifier, 16-bit A/D, and binned 4x4 for a superpixel size of 0.51 arcsec.  Each frame was triggered with a Spectrum TM-4 GPS for absolute times to microsecond accuracy. Data were taken from 10:50 to 11:34 UT. An example of the data is shown in Fig.~\ref{fig:20180409images}.

\subsection{2018 August 15} \label{subsec:20180815obs}

For the 2018 August 15 UT event, data were taken from the Astronomical Telescope of the University of Stuttgart (ATUS; \citet{Schindler2026}) at Sierra Remote Observatory (SRO) in California, the Experimental Test Site (ETS) in New Mexico, and Las Cumbres Observatory in Texas (LCO-ELP). At SRO, the camera was set to the conventional output amplifier and a 1~MHz readout, with 16-bit A/D sampling. Binning 3x3 resulted in a superpixel size of 1.7~arcsec. The camera was cooled to -60\textcelsius{} and provided time-to-live pulses at the start of each exposure, which were logged by a Spectrum TM-4 GPS receiver to microsecond accuracy. The observatory's Polaris seeing monitor indicated an average zenith seeing of 1~arcsec. Flux readings showed some minor reduction in sky transparency around event time, potentially caused by wildfire smoke towards the north, but sky conditions in the southern sky appeared unaffected on an otherwise clear night with the waxing crescent moon having already set. Data were taken from 05:12 to 06:02 UT.

The ETS data were $1024 \times 1024$ pixels unbinned, with a plate scale of 0.68 arcsec per pixel. Each image had a header timestamp in UT at the middle of the exposure time. Data were taken from 05:10 to 05:50 UT. There were scattered clouds, resulting in the majority of the data prior to the midtime being unusable.  However, an occultation emersion and post-event baseline were observed.

Las Cumbres Observatory has telescopes at a number of sites worldwide (see footnotes of Table \ref{tab:scopes} for the sites used in this work). We have a long-running program to observe stellar occultations using their guider cameras: this is a non-standard observing mode. These observations are scheduled and executed remotely, in a queue. When more than one telescope is available at a given site, we stagger the start times and combine the data in order to generate a combined light curve with the highest spatial resolution. For all Las Cumbres observations presented in this work, the instrument computers are synchronized via Network Time Protocol (NTP) to the site GPS timeserver and times are reliable to better than 5~milliseconds. For this event, observations from LCO-ELP were taken with only one telescope. The camera was binned $2{\times}2$ for a superpixel size of 0.674~arcsec. The CCD was cooled to -20\textcelsius{}. Each image had a header timestamp for the start of the exposure in UT. The sky was clear, but the images were slightly out of focus. Data were taken from 05:11 to 05:40 UT.

Examples of the data are shown in Fig.~\ref{fig:20180815images}.

\subsection{2018 October 01} \label{subsec:20181001obs}
The 2018 October 01 UT occultation was detected from three telescopes co-located at the South Africa Astronomical Observatory (SAAO) site in Sutherland, South Africa: the SAAO 74-in and two 1-m plus one 0.4-m Las Cumbres telescopes (LCO-CPT). There were scattered clouds, which cleared during the occultation observing run. Seeing varied between 1.7–2 arcsec. This was an unusually slow event, approximately a factor of 15 slower than the typical relative transverse velocity (see Table~\ref{tab:star}): the full occultation lasted nearly half an hour, as opposed to the usual timescale of a few minutes. 

On the 74-in telescope, a Sutherland High-speed Optical Camera (SHOC) instrument was run in 1 MHz, 16-bit A/D with the conventional amplifier mode. The CCD was cooled to –50\textcelsius{} and images were binned 8x8 for a superpixel size of 0.61 arcsec. Each frame was triggered using a Spectrum TM-4 GPS, for absolute times to microsecond accuracy. Data were taken from 18:03 to 19:03 UT. 

Observations made on the two, 1-m LCO-CPT telescopes used the guider cameras, binned 2x2 for a superpixel size of 0.674 arcsec. The CCDs were cooled to -20\textcelsius{}. Each image had a header timestamp for the start of the exposure in UT. Data were taken from 18:05 to 19:34 UT. The camera on the 0.4-m LCO-CPT telescope was also cooled to -20\textcelsius{}. Data were taken from 18:05 to 19:25 UT in $2\times2$ binning with a superpixel size of 1.16 arcsec. 

Examples of the data are provided in Fig.~\ref{fig:20181001image}.

\subsection{2018 November 01} \label{subsec:20181101obs}
The 2018 November 01 UT occultation was also observed from the SAAO's 74-inch telescope in Sutherland. The sky was clear, but the seeing was marginal, rising from 2.5 to greater than 3 arcsec. The SHOC instrument was run in 1 MHz, 16-bit A/D conventional amplifier mode. The CCD was cooled to –60\textcelsius{} and images were binned 16x16 for a superpixel size of 1.22 arcsec. Each frame was triggered using a Spectrum TM-4 GPS, for absolute times to microsecond accuracy. Data were taken from 19:31 to 20:16 UT. An example of the data is provided in Fig.~\ref{fig:20181101image}.

\subsection{2018 November 20} \label{subsec:20181120obs}
The 2018 November 20 UT occultation was observed from one 0.4-m Las Cumbres telescope at Teide Observatory (LCO-TFN). The sky was clear and seeing was approximately 0.5 arcsec. The camera was cooled to -20\textcelsius{}. Images were binned $2\times2$ for a superpixel size of 1.16 arcsec.  Data were taken from 19:34 to 20:00 UT. An example of the data is provided in Fig.~\ref{fig:20181120image}.

\subsection{2021 August 06} \label{subsec:20210806obs}
Three sites in Chile were used to observe the 2021 August 06 UT event: Las Campanas Observatory (LCO), Cerro Pach\'{o}n, and Cerro Tololo Inter-American Observatory (CTIO). At LCO, the 6.5-m Magellan Clay telescope was used with a POETS instrument in 1 MHz, 16-bit A/D conventional amplifier mode. The CCD was cooled to -60\textcelsius{} and the images were binned 4x4 for a superpixel size of 0.184 arcsec. Each frame was triggered using a Spectrum TM-4 GPS, for absolute times to microsecond accuracy. Data were taken from 01:54 to 02:34 UT. The sky was clear and seeing was 0.5 arcsec. 

At Cerro Pach\'{o}n, the seeing was 1.5 arcsec with thin clouds. On Gemini South, the Zorro instrument was configured in wide-field mode; the Zorro blue camera was not available, and the red camera was used in 1 MHz, 16-bit A/D conventional amplifier mode. The CCD was cooled to -60\textcelsius{} and binned 4x4 for a superpixel size of 0.236 arcsec. Within the .fits headers of the image datacubes, UTC times were recorded for each frame to six or more decimal places. The absolute timing accuracy of Zorro is $167\pm0.07 \text{ms}$ with average frame-to-frame precision of 73 ns \citep{RN3944}. Data were taken from 02:07 to 02:27 UT. 

At CTIO, the guide cameras were used at each of the three 1-m Las Cumbres telescopes (LCO-LSC). The cameras were cooled to -20\textcelsius{} and binned 2x2 for a superpixel size of 0.674 arcsec and conditions for all three systems were the same. Data were taken from 01:50 to 02:30 UT.  On the Southeastern Association for Research in Astronomy telescope at this site (SARA-CT), the camera was binned 2x2 for a super pixel size of 0.61 arcsec. The CCD was cooled to -85\textcelsius{}. Data were taken from 01:55 to 02:37 UT. The computer used to control the instrument was time-synced with a network clock at the start of the observations; however, the computer that controlled the camera was not synced. Subsequent timing tests demonstrated that the file header times could be tens of seconds off from the instrument computer. 

Examples of the data from this event are provided in Fig.~\ref{fig:2021images}.

\subsection{2022 June 01} \label{subsec:20220601obs}
The 2022 June 01 UT occultation was attempted from three sites in Australia: EOS (Electro Optic Systems space situational awareness facility site), Savannah Skies Observatory, and the Earth Sanctuary. These are all private sites located outside the cities of Learmonth WA,  Chillagoe QLD, and Alice Springs NT, respectively. Data were successfully obtained at the first two sites while there were heavy clouds over the Earth Sanctuary 14" Meade telescope, on which a POETS instrument was mounted. 

At EOS, the seeing was 1.8~arcsec with fair weather. The camera was cooled to -30\textcelsius{} and unbinned images were taken with a plate scale of 0.88~arcsec/pixel.The full $2048 \times 2048 \, {\text{pixel}}$ frame was used for calibration data; a subframe of $512 \times 512 \, \text{pixel}$ was used for the occultation observations.  Start and end UTC times were recorded in each file header to six decimal places. We assume that the times are accurate to the millisecond level. Data were taken from 16:08 to 16:52 UT. % From SEL: This is a shuttered camera, most likely with an iris shutter.  CCD diagonal is 39.1mm, so the shutter is likely 45 or 50mm in diameter.  The shutter throw time would likely be on the order of 0.1sec. Not sure how that impacts the accuracy estimate. 

At Savannah Skies, there were thin clouds on the horizon. Multiple telescopes were used at this site. A POETS was mounted on the largest of them. This camera had a $1024 \times 1024 \, \text{pixel}$ CCD (e2v CCD201-20) of which a $512 \times 512$ subframe was used for the occultation observations.  All data were taken in 1 MHz, 16-bit A/D conventional amplifier mode. The CCD was cooled to -60\textcelsius{} and the images were binned $2 \times 2$. Each frame was triggered using a Spectrum TM-4 GPS, for absolute times to microsecond accuracy. Data were taken from 15:56 to 16:36 UT. Five other datasets were obtained at the Savannah Skies site. These are labeled as 20-in SBIG, BRC, FCT, FRC, and RCOS, with instrument details in Table~\ref{tab:scopes}. They each consisted of series of photometric frames taken with non-high-speed, consumer-grade cameras, and they have varying cycle times. Additionally, these five Savannah skies telescopes had smaller apertures than the telescope on which the POETS camera was mounted. As a result of these two factors, while all the Savannah Skies data were of sufficient quality to be included in the analyses below, the relative weighting (by signal-to-noise ratio of the light curves) of the non-POETS datasets reduces their impact upon the final atmospheric-fitting solution.  

Examples of the data from the largest telescope at each of the two successful sites for this event are shown in Fig.~\ref{fig:2022images}.

\subsection{2022 August 23} \label{subsec:20220823obs}
The 2022 August 23 UT occultation was attempted from three telescopes: the Aristarchos telescope, the Liverpool Telescope (LT), and the Trebur 1.2~m Telescope (T1T). The Aristarchos in the Peloponnese, Greece and the LT at Observatorio del Roque de los Muchachos (ORM) in the Canary Islands were selected for both their locations and similar instruments, RISE and RISE2 \citep{RN3986,RN4059}. LT and T1T were successful, while Aristarchos was unable to observe due to thunderstorms. 

The LT is a fully-robotic telescope. On this night, the sky was clear with good conditions. The E2V CCD 47-20 frame transfer camera was cooled to -40\textcelsius{} and we used a ``V+R'' filter (roughly $400 - 700 \, \text{nm}$ passband). Images were binned 2x2 for a superpixel size of 1.08 arcsec. Data were taken from 20:58:32 to 21:41:35 UT. The clock was synchronized to a GPS receiver, with demonstrated accuracy of 90~msec for each frame\footnote{from https://telescope.livjm.ac.uk/TelInst/Inst/RISE/}.

From the T1T at the Michael Adrian Observatory (MAO) in Germany, the pointing elevation was challenging, being 17\textdegree{} and rising at the occultation midtime. Observers used a QHY174M-GPS Complementary Metal-Oxide Semiconductor (CMOS) camera with a \texttimes0.37 focal reducer, leading to a plate scale of 0.31 arcsec/pixel. The CMOS sensor was cooled to -15\textcelsius{}. Data were taken from 20:58:29 to 21:41:01 UT. Time stamps were provided from the camera-internal GPS receiver. % timing was from the computer, which was NTP synced before the event to an expected accuracy of roughly 500 msec. 

Examples of the data from this event are provided in Fig.~\ref{fig:20220823images}. These data have been previously published, as a real-world test case to demonstrate that Gaussian process regression can be used to model occultation light curves and to verify a Pluto ephemeris correction model that was generated from publicly-available Zwicky Transit Facility (ZTF) images \citep{RN4068}.

\subsection{2023 July 17} \label{subsec:20230717obs}
The 2023 July 17 UT occultation was observed from the SAAO's 74-inch telescope in Sutherland. The sky was clear and seeing was 1.5-2 arcsec. The SHOC instrument was run in 3 MHz, 14-bit A/D conventional amplifier mode and cooled to –50\textcelsius{}. Images were binned $10 \times 10 \, \rm{pixels}$ for a superpixel size of 0.76 arcsec. Each frame was triggered using a Spectrum TM-4 GPS, for absolute times to microsecond accuracy. Data were taken from 22:30 to 23:10 UT. An example of the data from this event is provided in Fig.~\ref{fig:20230717images}.

\bibliography{references.bib}{} %Export this file from EndNote, replace & and # with \& and \#.
\bibliographystyle{aasjournalv7}



\end{document}